\documentclass[prd,nopacs,preprintnumbers,twocolumn,amsmath,nofootinbib,amssymb]{revtex4}
\usepackage{graphicx,color,dcolumn,booktabs,bm}
\usepackage{longtable,lscape}
\usepackage{txfonts}
\usepackage{overpic}
\usepackage{epstopdf}
\usepackage{indentfirst}
\usepackage{threeparttable}
\usepackage{feynmf}   
\usepackage{slashed}  
\usepackage{cases}
\usepackage{color}
\usepackage{float}
\usepackage{makecell}
\usepackage{multirow}
\usepackage{array}
\usepackage{graphicx,color,dcolumn,booktabs,bm}
\usepackage{epsfig,dsfont,amssymb,amsmath,amsfonts,amsbsy,mathrsfs}

\graphicspath{{Figures/}} %

\usepackage{hyperref}
\hypersetup{colorlinks,citecolor=blue,anchorcolor=red,menucolor=red, linkcolor=red,filecolor=red,runcolor=red,urlcolor=blue,frenchlinks=red}

\makeatletter
\@addtoreset{equation}{section}
\makeatother
\renewcommand{\theequation}{\arabic{section}.\arabic{equation}}
\allowdisplaybreaks

\newcolumntype{I}{!{\vrule width 0.9pt}}

\begin{document}

\title{Angular distribution of the FCNC process $B_{c}\to{D}_{s}^{*}(\to{D}_{s}\pi)\ell^{+}\ell^{-}$}
\author{Yu-Shuai Li$^{1,2,3,5}$}\email{liysh20@lzu.edu.cn}
\author{Xiang Liu$^{1,2,3,4,5}$\footnote{Corresponding author}}\email{xiangliu@lzu.edu.cn}

\affiliation{$^1$School of Physical Science and Technology, Lanzhou University, Lanzhou 730000, China\\
$^2$Lanzhou Center for Theoretical Physics, Key Laboratory of Theoretical Physics of Gansu Province, Lanzhou University, Lanzhou 730000, China\\
$^3$Key Laboratory of Quantum Theory and Applications of MoE, Lanzhou University,
Lanzhou 730000, China\\
$^4$MoE Frontiers Science Center for Rare Isotopes, Lanzhou University, Lanzhou 730000, China\\
$^5$Research Center for Hadron and CSR Physics, Lanzhou University and Institute of Modern Physics of CAS, Lanzhou 730000, China}

\begin{abstract}
In this work, we study the flavor-changing neutral-current process $B_{c}\to{D}_{s}^{*}(\to{D}_{s}\pi)\ell^{+}\ell^{-}$ ($\ell$= $e$, $\mu$, $\tau$). The relevant weak transition form factors are obtained by using the covariant light-front quark model, in which, the main inputs, i.e., the meson wave functions of $B_{c}$ and $D_{s}^{*}$, are adopted as the numerical wave functions from the solution of the Schr\"{o}dinger equation with the modified Godfrey-Isgur model. With the obtained form factors, we further investigate the relevant branching fractions and their ratios, and some angular observables, i.e., the forward-backward asymmetry $A_{FB}$, the polarization fractions $F_{L(T)}$, and the $CP$-averaged angular coefficients $S_{i}$ and the $CP$ asymmetry coefficients $A_{i}$. We also present our results of the clean angular observables $P_{1,2,3}$ and $P^{\prime}_{4,5,6,8}$, which can reduce the uncertainties from the form factors. Our results show that the corresponding branching fractions of the electron or muon channels can reach up to $10^{-8}$. With more data being accumulated in the LHCb experiment, our results are helpful for exploring this process, and deepen our understanding of the physics around the $b\to{s}\ell^{+}\ell^{-}$ process.
\end{abstract}

\maketitle

\section{Introduction}\label{sec01}

The flavor-changing neutral-current (FCNC) process, like the $b\to{s}\ell^{+}\ell^{-}$ ($\ell$=$e$, $\mu$, $\tau$) we are concerned with has attracted the attention of both theorists and experimentalists, and of course has been widely studied. The FCNC process is forbidden at the tree level, and can only operate through loop diagrams in the Standard Model (SM). At the lowest order, three amplitudes contribute to the decay width, i.e., the photo penguin diagram, the $Z$ penguin diagram, and the $W^{+}W^{-}$ box diagram. In all three diagrams, the virtual $t$ quark plays a dominant role, while the $c$ and $u$ quarks are the secondary contributions. The FCNC process is very sensitive to the new physical effects. This suggests that it can serve as a perfect platform to search directly for new physics (NP) beyond the SM~\cite{Altmannshofer:2014rta,Descotes-Genon:2015uva,SinghChundawat:2022ldm}.

The $b\to{s}\ell^{+}\ell^{-}$ in the bottom(-stranged) mesons sector is an attractive experimental topic. The experimental search of the FCNC processes $B\to K^{(*)}\ell^{+}\ell^{-}$ started in  1998~\cite{Skwarnicki:1998ph,CDF:1999uew,BaBar:2000jlq}. The first observation of $B\to{K}\ell^{+}\ell^{-}$ was made by the Belle collaboration in 2001 with a statistical significance of  $5.3$ \cite{Belle:2001oey}. From 2001 to now, the $B\to{K}^{(*)}\ell^{+}\ell^{-}$ with $\ell^{+}\ell^{-}$ being either an $e^{+}e^{-}$ or $\mu^{+}\mu^{-}$ pair has been observed or measured by the Belle~\cite{Belle:2001oey,Belle:2003ivt,Belle:2009zue,Belle:2016fev,BELLE:2019xld,Belle:2019oag}, the $BABAR$~\cite{BaBar:2003szi,BaBar:2008jdv,BaBar:2012mrf}, the CDF~\cite{CDF:2011buy}, the CMS~\cite{CMS:2015bcy}, and the LHCb collaborations~\cite{LHCb:2012juf,LHCb:2013ghj,LHCb:2014vgu,LHCb:2016ykl,LHCb:2017avl,LHCb:2021trn}. In particular, the LHCb collaboration measured the form-factor-independent observable $P_{5}^{\prime}$~\cite{LHCb:2013ghj}, and found a $2.5$ standard deviation ($\sigma$) discrepancy to the SM prediction~\cite{Egede:2008uy} after integrating over $1.0<q^{2}<6.0$ $\text{GeV}^{2}$. In addition, the LHCb collaboration recently reported the most precise measurement of the ratio of branching fractions for $B^{+}\to{K}^{+}\mu^{+}\mu^{-}$ and $B^{+}\to{K}^{+}e^{+}e^{-}$ decays in $1.1<q^{2}<6.0~\text{GeV}^{2}$ as $R^{{\mu}{e}}_{K}=0.846^{+0.044}_{-0.041}$~\cite{LHCb:2021trn}, indicating a $3.1$$\sigma$ discrepancy with the SM prediction~\cite{Descotes-Genon:2015uva,Bordone:2016gaq}, and providing evidence for the violation of lepton flavor universality (LFU). For the $B_{s}$ decays, there have been some experiments, such as the CDF~\cite{CDF:2001yrm,CDF:2008zhr} and the $\mathrm{D}\O$ experiments~\cite{D0:2006pmq}, to search for the $B_{s}\to\phi\ell^{+}\ell^{-}$ mode. In 2011, the $B_{s}\to\phi\mu^{+}\mu^{-}$ mode was first observed in the CDF experiment~\cite{CDF:2011grz}, and then measured by the CDF~\cite{CDF:2011buy} and the LHCb collaborations~\cite{LHCb:2013tgx,LHCb:2015wdu,LHCb:2021zwz}. The electron mode is still missing in the  experiment. Moreover, in Ref.~\cite{LHCb:2021zwz} the LHCb collaboration also reported their measurement of the $B_{s}\to{f}_{2}^{\prime}(1525)\mu^{+}\mu^{-}$ process. Compared to the dielectronic and dimuonic modes, the ditauic mode is less studied. There is a Belle experiment, which focused on the $B^{0}\to{K}^{*0}\tau^{+}\tau^{-}$ process, and determined the upper limit of the branching fraction $\mathcal{B}(B^{0}\to K^{*0}\tau^{+}\tau^{-})<3.1\times10^{-3}$ at $90\%$ confidence level~\cite{Belle:2021ecr}.

The FCNC decay of bottom(-stranged) mesons has also been studied by various theoretical approaches, such as the lattice QCD (LQCD)~\cite{Bouchard:2013eph,Horgan:2013hoa,Bailey:2015dka}, the light-cone sum rule~\cite{Ball:2004rg,Ball:2004ye,Wu:2006rd,Bartsch:2009qp,Bharucha:2015bzk,Cheng:2017bzz,Gao:2019lta,Wang:2015vgv,Wang:2017jow,Lu:2018cfc,Gao:2021sav,Cui:2022zwm,Cui:2023bzr}, the QCD factorization~\cite{Bobeth:2008ij}, the perturbative QCD (pQCD)~\cite{Li:2009tx,Wang:2007an,Li:2009rc,Wang:2012ab,Wang:2013ix,Xiao:2013lia} and its combination with LQCD data~\cite{Jin:2020jtu,Jin:2020qfp}, as well as various quark models~\cite{Deandrea:2001qs,Geng:2003su,Chen:2010aq,Li:2010ra,Dubnicka:2016nyy,Soni:2020bvu,Issadykov:2022imz}, and so on \cite{Lu:2011jm,Ahmady:2019hag,Rajeev:2020aut}. On the other hand, in order to understand the discrepancy of the value of $R$ with the SM prediction, the effects beyond the SM are considered. Following this line of thought, the extensions of the SM via the extended Higgs-boson~\cite{Li:2018rax,Barman:2018jhz,DelleRose:2019ukt,Ordell:2019zws,Marzo:2019ldg,Iguro:2018qzf,Iguro:2023jju}, supersymmetry~\cite{Aslam:2009cv,Trifinopoulos:2019lyo}, and extra dimensions~\cite{Shaw:2019fin} have been used. At the same time, some NP models with an additional heavy neutral boson~\cite{Altmannshofer:2014cfa,Bhattacharya:2014wla,Crivellin:2015lwa,Celis:2015ara,Falkowski:2015zwa,Bhattacharya:2016mcc,Chiang:2017hlj,King:2017anf,Falkowski:2018dsl,Allanach:2019mfl,Dwivedi:2019uqd,Capdevila:2020rrl,Sheng:2021tom} or leptoquarks~\cite{Hiller:2014yaa,Gripaios:2014tna,deMedeirosVarzielas:2015yxm,Becirevic:2017jtw,DiLuzio:2017vat,Becirevic:2018afm,Angelescu:2018tyl,Cornella:2019hct,Popov:2019tyc,DaRold:2019fiw,Hati:2019ufv,Datta:2019bzu,Balaji:2019kwe,Crivellin:2019dwb,Saad:2020ihm,Babu:2020hun,Iguro:2021kdw} were also considered.

Although great progress has been made both experimentally and theoretically in the rare semileptonic decays of bottom(-strange) mesons in recent decades, those of bottom-charmed mesons have been less studied. Compared to the $B_{(s)}$ mesons, the $B_{c}$ meson is difficult to produce at the Belle experiment because the $B_{c}\bar{B}_{c}$ is close to 12.5 GeV, which is far from the energy region of $\Upsilon(4S)$. Moreover, according to $f_{c}/f_{u}=(7.5\pm1.8)\times10^{-3}$ measured by the LHCb collaboration~\cite{LHCb:2019tea}, the $B_{c}$ meson is also underproductivity in the $pp$ experiment. Here, the $f_{c}$ and $f_{u}$ are the fragmentation fractions of $B_{c}$ and $B$ meson, respectively, in $pp$ collisions. As a result, the $B_c$ meson decay has received less experimental attention in the past. Recently, the LHCb collaboration reported the result of the $B_{c}^{+}\to{D}_{s}^{+}\mu^{+}\mu^{-}$ process~\cite{LHCb:2023lyb}. Using the $pp$ collision data collected by the LHCb experiment at the center-of-mass energies of 7, 8, and 13 TeV, corresponding to a total integrated luminosity of 9 $\text{fb}^{-1}$, the LHCb collaboration did not observe significant signals in the nonresonant $\mu^{+}\mu^{-}$ modes, but set an upper limit as $f_{c}/f_{u}\times \mathcal{B}(B_{c}^{+}\to{D}_{s}^{+}\mu^{+}\mu^{-})<9.6\times10^{-8}$ at the $95\%$ confidence level. {Moreover, considering that the $B_{c}\to D_{s}^{(*)}\ell^{+}\ell^{-}$ channels have similar amounts of branching fractions~\cite{Wang:2014yia}, and the $D_{s}^{*}$ needs to be reconstructed by the $D_{s}$ meson in the experiment, the measurement of $B_{c}\to D_{s}^{*}\ell^{+}\ell^{-}$ will be more difficult. This indicates that the search for rare semileptonic decays of $B_{c}$ is difficult for the present experiment. However, with the high-luminosity upgrade of the Large Hadron Collider (LHC), this situation is likely to improve. In any case, with the accumulation of data in the experiment, we expect the LHCb experiment to search for these rare semileptonic decays of the $B_{c}$ meson.}

In the theoretical sector, the rare semileptonic decays of $B_{c}$ have been studied by the light-front quark model (LFQM)~\cite{Geng:2001vy}, the pQCD~\cite{Wang:2014yia}, the QCD sum rule~\cite{Kiselev:2002vz,Azizi:2008vv}, the constituent quark model (CQM)~\cite{Geng:2001vy}, {\it et al}. The branching fractions of $B_{c}\to{D_{s}^{*}}\ell^{+}\ell^{-}$ with $\ell=e$ or $\mu$ are predicted to be  approximately $10^{-7}$. In Refs. \cite{Dutta:2019wxo,Mohapatra:2021ynn,Zaki:2023mcw}, the $B_{c}\to{D_{s}^{*}}\mu^{+}\mu^{-}$ process was been studied within the SM and beyond. In this work, we also focus on the $B_{c}\to{D}_{s}^{*}\ell^{+}\ell^{-}$ process, where the necessary form factors are calculated via the covariant LFQM approach. To provide more physical observables, we present the angular distribution of the quasi-four-body process $B_{c}\to{D_{s}^{*}}(\to{D_{s}}\pi)\ell^{+}\ell^{-}$.

The applications of the standard and(or) covariant LFQM have proved successful in the study of the meson~\cite{Jaus:1989au,Jaus:1996np,Cheng:1996if,Cheng:1997au,Jaus:1999zv,Cheng:2003sm,Chua:2003ac,Cheng:2004yj,Wang:2007sxa,Wang:2008ci,Shen:2008zzb,Wang:2008xt,Wang:2009mi,Cheng:2009ms,Chen:2009qk,Choi:2010zb,Choi:2010be,Li:2010bb,Ke:2011mu,Verma:2011yw,Ke:2013yka,Xu:2014mqa,Shi:2016gqt,Cheng:2017pcq,Chen:2017vgi,Kang:2018jzg,Chang:2018zjq,Chang:2019mmh,Chang:2019xtj,Chang:2019obq,Chang:2020xvu,Chang:2020wvs,Chen:2021ywv,Choi:2021mni,Choi:2021qza,Arifi:2022qnd,Zhang:2023ypl,Shi:2023qnw,Hazra:2023zno,Zhang:2020dla} and baryon weak decays~\cite{Ke:2007tg,Ke:2012wa,Wang:2017mqp,Ke:2017eqo,Zhu:2018jet,Zhao:2018zcb,Zhao:2018mrg,Xing:2018lre,Chua:2018lfa,Ke:2019smy,Chua:2019yqh,Ke:2019lcf,Hu:2020mxk,Geng:2020fng,Hsiao:2020gtc,Geng:2021nkl,Li:2021qod,Ke:2021pxk,Hsiao:2021mlp,Li:2021kfb,Li:2022nim,Geng:2022xpn,Wang:2022ias,Zhao:2022vfr,Li:2022hcn,Lu:2023rmq,Zhao:2023yuk,Liu:2023zvh}. The $B_{c}\to{D}_{s}^{*}$ weak transition form factors deduced by (axial)-vector currents have been calculated in Ref. \cite{Zhang:2023ypl} with the covariant LFQM. Probably in the series of papers~\cite{Cheng:2003sm,Chua:2003ac,Wang:2007sxa,Wang:2008ci,Shen:2008zzb,Wang:2008xt,Wang:2009mi,Chen:2009qk,Cheng:2009ms,Choi:2010zb,Li:2010bb,Verma:2011yw,Ke:2013yka,Xu:2014mqa,Shi:2016gqt,Chang:2018zjq,Chang:2019xtj,Chang:2019obq,Chang:2019mmh,Chang:2020xvu,Chang:2020wvs,Chen:2021ywv,Choi:2021mni,Choi:2021qza,Arifi:2022qnd,Shi:2023qnw,Ke:2007tg,Ke:2012wa,Ke:2017eqo,Wang:2017mqp,Zhu:2018jet,Zhao:2018zcb,Xing:2018lre,Chua:2018lfa,Zhao:2018mrg,Chua:2019yqh,Ke:2019lcf,Ke:2019smy,Hu:2020mxk,Geng:2020fng,Hsiao:2020gtc,Hsiao:2021mlp,Geng:2021nkl,Ke:2021pxk,Zhao:2022vfr,Geng:2022xpn,Wang:2022ias,Liu:2023zvh,Lu:2023rmq,Zhao:2023yuk}, the hadron wave function was taken as a Gaussian-like form with phenomenal parameter $\beta$, which represents the hadron structure. To fix the phenomenal parameter, the corresponding decay constant was used. However, as we all know, the decay constant is only associated with the zero-point wave function. This indicates that the oversimplified Gaussian-form wave function is not able to depict the behavior far away from the zero point. For this object, we propose to directly adopt the numerical spatial wave function by solving the Schr\"{o}dinger equation with the modified Godfrey-Isgur (GI) model. By fitting the mass spectrum of the observed heavy flavor mesons, the parameters of the potential model can be fixed. This strategy avoids the $\beta$ dependence, and can also reduce the corresponding uncertainty. {We note that in Ref. \cite{Faustov:2022ybm}, the authors used a relativistic quark model based on the quasipotential approach in QCD to study the semileptonic decay of bottom mesons. In their approach, the numerical wave functions of the mesons are obtained, thus avoiding the corresponding uncertainty.}

This paper is organized as follows. After the Introduction, we illustrate the angular distributions of the quasi-four-body decays $B_{c}\to{D}_{s}^{*}(\to{D}_{s}\pi)\ell^{+}\ell^{-}$ ($\ell$= $e$, $\mu$, $\tau$) in Sec. \ref{sec02}. In Sec. \ref{sec03}, we introduce the covariant LFQM and derive the formula of the weak transition form factors. Then in Sec. \ref{sec04}, the numerical results, including the from factors of $B_{c}\to{D}_{s}^{*}$ and physical observables of $B_{c}\to{D}_{s}^{*}(\to{D}_{s}\pi)\ell^{+}\ell^{-}$ processes, are presented. Finally, this paper ends with a short summary.

\section{The angular distribution of $B_{c}\to{D}_{s}^{*}(\to{D}_{s}\pi)\ell^{+}\ell^{-}$}\label{sec02}

\subsection{The effective Hamiltonian for $b\to{s}\ell^{+}\ell^{-}$}

The effective Hamiltonian associated with $b\to{s}\ell^{+}\ell^{-}$ is \cite{Buchalla:1995vs}
\begin{equation}
\begin{split}
\mathcal{H}{\!}=&{\!}-{\!}\frac{4G_{F}}{\sqrt{2}}\Big{\{}V_{tb}V_{ts}^{*}
\bigg{[}{C}_{1}(\mu)\mathcal{O}_{1}^{c}(\mu)
{\!}+{\!}{C}_{2}(\mu)\mathcal{O}_{2}^{c}(\mu)
{\!}+{\!}\sum_{i=3}^{10}{C}_{i}(\mu)\mathcal{O}_{i}(\mu)\bigg{]}\\
&{\!}+{\!}V_{ub}V_{us}^{*}\big{[}{C}_{1}(\mu)(\mathcal{O}_{1}^{c}(\mu){\!}-{\!}\mathcal{O}_{1}^{u}(\mu)){\!}+{\!}{C}_{2}(\mu)(\mathcal{O}_{2}^{c}(\mu){\!}-{\!}\mathcal{O}_{2}^{u}(\mu))\big{]}\Big{\}},
\label{eq:Hamiltonian1}
\end{split}
\end{equation}
where $V_{ij}$ are the Cabibbo-Kobayashi-Maskawa (CKM) matrix elements and $G_{F}=1.16637\times10^{-5}~\text{GeV}^{-2}$ \cite{ParticleDataGroup:2022pth} is the Fermi constant. Also, the ${C}_{i}(\mu)$ are Wilson coefficients and the $\mathcal{O}_{i}(\mu)$ are four fermion operators. They all depend on the QCD renormalization scale $\mu$. More specifically, the $\mathcal{O}_{1,2}^{c,u}$ are current-current operations, the $\mathcal{O}_{3-6}$ are QCD penguin operators, the $\mathcal{O}_{7,8}$ are electromagnetic and chromomagnetic penguin operators, and the $\mathcal{O}_{9,10}$ are semileptonic operators, respectively.

\begin{figure}[htbp]\centering
  \includegraphics[width=70mm]{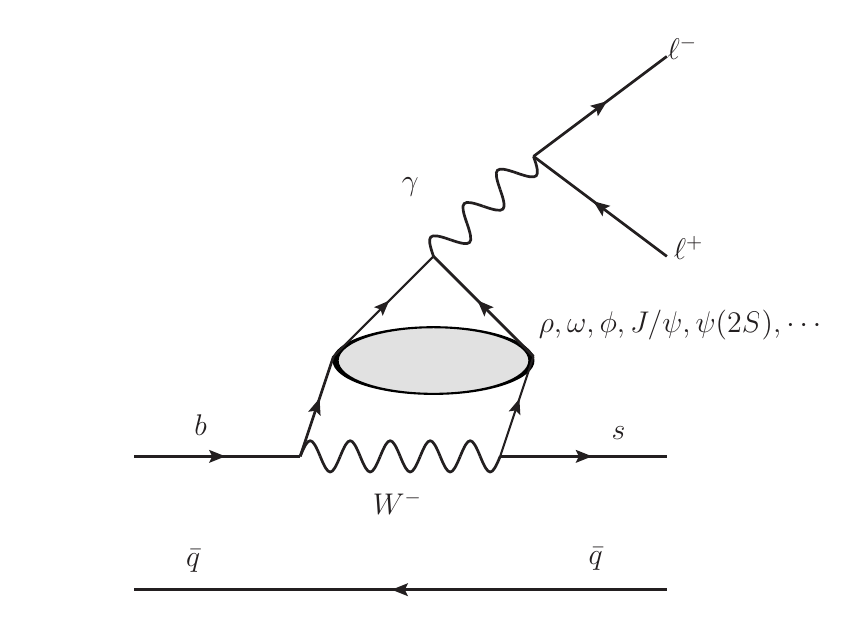}\\
  \caption{The contributions of the intermediate vector states $(\rho,\omega,\phi.J/\psi,\psi(2S),\dots)$ to the $b\to{s}\ell^{+}\ell^{-}$ process resulting from the current-current operators $\mathcal{O}_{1,2}^{c,u}$.}
  \label{fig:Resonance}
\end{figure}

Apart from the $\gamma$ and $Z$ penguin diagrams, and the $W^{+}W^{-}$ box diagram, the long distance contribution, via the intermediate vector states $(\rho,\omega,\phi.J/\psi,\psi(2S),\dots)$ (see Fig. \ref{fig:Resonance}) also shows an unignorable influence. By adding the factorable quark-loop contributions from $\mathcal{O}_{1-6,8}$ to the effective Wilson coefficients ${C}_{7,9}^{\text{eff}}$, the effective Hamiltonian in Eq.~\eqref{eq:Hamiltonian1} can be simplified. In the calculation, we have adopted the following effective Hamiltonian, i.e.,
\begin{equation}
\begin{split}
\mathcal{H}^{\text{eff}}(b\to{s}\ell^{+}\ell^{-})=&-\frac{4G_{F}}{\sqrt{2}}V_{tb}V_{ts}^{*}\frac{\alpha_{e}}{4\pi}
\bigg{\{}\bar{s}\big{[}{C}_{9}^{\text{eff}}(q^{2},\mu)\gamma^{\mu}P_{L}\\
&-\frac{2m_{b}}{q^{2}}{C}_{7}^{\text{eff}}(\mu)i\sigma^{\mu\nu}q_{\nu}P_{R}\big{]}b(\bar{\ell}\gamma_{\mu}\ell)\\
&+{C}_{10}(\mu)(\bar{s}\gamma^{\mu}P_{L}b)(\bar{\ell}\gamma_{\mu}\gamma_{5}\ell)\bigg{\}},
\label{eq:Hamiltonian2}
\end{split}
\end{equation}
where $P_{L(R)}=(1\mp\gamma_{5})/2$, $\sigma^{\mu\nu}=i(\gamma^{\mu}\gamma^{\nu}-\gamma^{\nu}\gamma^{\mu})/2$, and the electromagnetic coupling constant $\alpha_{e}=1/137$. The $\mathcal{C}_{7}^{\text{eff}}$ and $\mathcal{C}_{9}^{\text{10}}$ are the effective Wilson coefficients, defined as \cite{Chen:2001zc}
\begin{equation}
\begin{split}
{C}_{7}^{\text{eff}}(\mu)=&{C}_{7}(\mu)+{C}^{\prime}_{b\to s\gamma}(\mu),\\
{C}_{9}^{\text{eff}}(q^{2},\mu)=&{C}_{9}(\mu)+Y_{\text{pert}}(q^{2},\mu)+Y_{\text{res}}(q^{2},\mu),
\end{split}
\end{equation}
where the term ${C}^{\prime}_{b\to{s}{\gamma}}$ is the absorptive part of the $b\to{s}{c}{\bar{c}}\to{s}{\gamma}$ rescattering~\cite{Asatrian:1996as,Chen:2001zc,Aslam:2008hp,Aslam:2009cv,Wang:2012ab,Soni:2020bvu,Jin:2020jtu,Jin:2020qfp}:
\begin{equation}
\begin{split}
{C}^{\prime}_{b\to s\gamma}(\mu)=&
i\alpha_{s}\Big{\{}\frac{2}{9}\eta^{\frac{14}{23}}
\Big{[}\frac{x_{t}(x_{t}^{2}-5x_{t}-2)}{8(x_{t}-1)^{3}}+\frac{3x_{t}^{2}\ln x_{t}}{4(x_{t}-1)^{4}}-0.1687\Big{]}\\
&-0.03\mathcal{C}_{2}(\mu)\Big{\}}
\end{split}
\end{equation}
with $x_{t}=m_{t}^{2}/m_{W}^{2}$, $\eta=\alpha_{s}(m_{W})/\alpha_{s}(\mu)$, and $\alpha_{s}$ being adopted as $\alpha_{s}(m_{b})=0.217$ in our calculation. The short-distance contributions from the soft-gluon emission and the one-loop contributions of the four fermion operators $\mathcal{O}_{1}-\mathcal{O}_{6}$, and the long-distance contributions from the intermediate vector meson states are also taken into account, and have been included in the $Y_{\text{pert}}$ and $Y_{\text{res}}$ terms, respectively.
The $Y_{\text{pert}}$ can be written as \cite{Buras:1994dj}
\begin{equation}
\begin{split}
Y_{\text{pert}}(\hat{s},\mu)=&0.124\omega(\hat{s})+g(\hat{m}_{c},\hat{s}){C}(\mu)\\
&+\lambda_{\mu}\Big{[}g(\hat{m}_{c},\hat{s})-g(0,\hat{s})\Big{]}(3{C}_{1}(\mu)+{C}_{2}(\mu))\\
&-\frac{1}{2}g(0,\hat{s})({C}_{3}(\mu)+3{C}_{4}(\mu))\\
&-\frac{1}{2}g(1,\hat{s})(4{C}_{3}(\mu)+4{C}_{4}(\mu)+3{C}_{5}(\mu)+{C}_{6}(\mu))\\
&+\frac{2}{9}(3{C}_{3}(\mu)+{C}_{4}(\mu)+3{C}_{5}(\mu)+{C}_{6}(\mu)),
\end{split}
\end{equation}
where $\hat{s}=q^{2}/m_{b}^{2}$ and $\hat{m}_{c}=m_{c}/m_{b}$ with $m_{b}=4.8~\text{GeV}$ and $m_{c}=1.6~\text{GeV}$, and ${C}(\mu)=3{C}_{1}(\mu)+{C}_{2}(\mu)+3{C}_{3}(\mu)+{C}_{4}(\mu)+3{C}_{5}(\mu)+{C}_{6}(\mu)$. At the next leading order, the Wilson coefficients at the QCD renormalization scale $\mu=m_{b}$ are chosen as $C_{1}=-0.175$, $C_{2}=1.076$, $C_{3}=1.258\%$, $C_{4}=-3.279\%$, $C_{5}=1.112\%$, $C_{6}=-3.634\%$, $C_{7}=-0.302$, $C_{8}=-0.148$, $C_{9}=4.232$, and $C_{10}=-4.410$ \cite{Buchalla:1995vs}.

In the Wolfenstein representation, the $\lambda_{u}=V_{ub}V_{us}^{*}/(V_{tb}V_{ts}^{*})$ can be expressed as
\begin{equation}
\lambda_{u}\approx-\lambda^{2}(\rho-i\eta),
\end{equation}
approximately, which is a small value suppressed by $\lambda^{2}$ with $\lambda=0.22500\pm0.00067$~\cite{ParticleDataGroup:2022pth}.

In addition, the term $\Omega(\hat{s})$ is the one-gluon correction to the matrix element of the operator $\mathcal{O}_{9}$, represented as \cite{Buras:1994dj,Jin:2020jtu}
\begin{equation}
\begin{split}
\omega(\hat{s})=&-\frac{2}{9}\pi^{2}+\frac{4}{3}\int_{0}^{\hat{s}}\frac{\ln{1-u}}{u}du-\frac{2}{3}\ln{(\hat{s})}\ln{(1-\hat{s})}\\
&-\frac{5+4\hat{s}}{3(1+2\hat{s})}\ln{(1-\hat{s})}-\frac{2\hat{s}(1+\hat{s})(1-2\hat{s})}{3(1-\hat{s})^{2}(1+2\hat{s})}\ln{(\hat{s})}\\
&+\frac{5+9\hat{s}-6\hat{s}^{2}}{6(1-\hat{s})(1+2\hat{s})},
\end{split}
\end{equation}
and the $g$ terms~\cite{Buras:1994dj,Aslam:2008hp,Aslam:2009cv,Wang:2012ab,Soni:2020bvu}:
\begin{equation}
\begin{split}
g(z,\hat{s})=&-\frac{8}{9}\ln{z}+\frac{8}{27}+\frac{4}{9}x-\frac{2}{9}(2+x)\sqrt{\vert1-x\vert}\\
&\times
\begin{array}{ll}
\ln\vert\frac{1+\sqrt{1-x}}{1-\sqrt{1-x}}\vert-i\pi &\text{for}\ x\equiv4z^2/\hat{s}<1\\
2\arctan{\frac{1}{\sqrt{x-1}}} &\text{for}\ x\equiv4z^2/\hat{s}>1
\end{array},\\
g(0,\hat{s})=&\frac{8}{27}-\frac{8}{9}\ln{\frac{m_{b}}{\mu}}-\frac{4}{9}\ln{\hat{s}}+\frac{4}{9}i\pi,
\end{split}
\end{equation}
come from the one-loop contributions of the $\mathcal{O}_{1-6}$.

The $Y_{\text{res}}$ term, which describes the long-distance contributions associated with the intermediate light vector mesons (such as $\rho$, $\omega$, and $\phi$) and vector charmonium states [such as $J/\psi$, $\psi(2S)$, etc.] (see the Fig. \ref{fig:Resonance}), is adopted as \cite{Jin:2020jtu} \footnote{This is a phenomenological method, and for more details on the charm-loop contribution, one can refer to Refs.~\cite{Khodjamirian:2010vf,Khodjamirian:2012rm,Qin:2022rlk}.}
\begin{equation}
\begin{split}
Y_{\text{res}}(q^{2},\mu)=&-\frac{3\pi}{\alpha_{e}^{2}}\bigg{[}{C}(\mu)
\sum_{V_{i}=J/\psi,\psi(2S),\dots}\frac{m_{V_{i}}\mathcal{B}(V_{i}\to\ell^{+}\ell^{-})\Gamma_{V_{i}}}{q^{2}-m_{V_{i}}^{2}+im_{V_{i}}\Gamma_{V_{i}}}\\
&-\lambda_{u}g(0,\hat{s})(3{C}_{1}(\mu)+{C}_{2}(\mu))\\
&\times \sum_{V_{j}=\rho, \omega,\phi}\frac{m_{V_{j}}\mathcal{B}(V_{j}\to\ell^{+}\ell^{-})\Gamma_{V_{j}}}{q^{2}-m_{V_{j}}^{2}+im_{V_{j}}\Gamma_{V_{j}}}\bigg{]},
\label{eq:Yres}
\end{split}
\end{equation}
where $m_{V_{i}}$ and $\Gamma_{V_{i}}$ are the mass and total width of the intermediate vector meson $V_{i}$ respectively, and the $\Gamma(V_{i}\to\ell^{+}\ell^{-})$ is the corresponding dilepton width. These input values are collected in Table \ref{tab:VectorMeson}. In addition, the nonvanished branching fraction for the $\tau$ channel, i.e., $\mathcal{B}(\psi(2S)\to\tau^{+}\tau^{-})=3.1\times10^{-3}$ \cite{ParticleDataGroup:2022pth}, is also used.

For the $J/\psi$ and $\psi(2S)$ states, the small widths and the large dilepton width will have a large influence on the decay width. However, the narrow widths are also used to reject them in the experimental analysis. One the other hand, for those above the $D\bar{D}$ threshold, such as $\psi(3770)$, $\psi(4040)$ and $\psi(4160)$, the board widths and mutual overlap make things difficult. Also, for the charmless vector mesons ($\rho$, $\omega$ and $\phi$), their contributions are suppressed by the $\lambda_{u}$ factor.

\begin{table}[htbp]\centering
\caption{The masses, total widths and dilepton widths of the intermediate vector mesons used in Eq.~\eqref{eq:Yres}.
These values are quoted from the PDG \cite{ParticleDataGroup:2022pth}.}
\label{tab:VectorMeson}
\renewcommand\arraystretch{1.15}
\begin{tabular*}{86mm}{c@{\extracolsep{\fill}}ccc}
\toprule[1pt]
\toprule[0.5pt]
\multirow{2}*{\shortstack{$V_{i}$}}     &\multirow{2}*{\shortstack{$m_{V_{i}}$ (GeV)}}   &\multirow{2}*{\shortstack{$\Gamma_{V_{i}}$ (MeV)}}
&$\mathcal{B}(V_{i}\to\ell^{+}\ell^{-})$   \\
&&&where $\ell=e,\mu$\\
\midrule[0.5pt]
$\rho$   &$0.775$    &$149$     &$4.635\times10^{-5}$   \\
$\omega$   &$0.783$    &$8.68$     &$7.380\times10^{-5}$   \\
$\phi$   &$1.019$    &$4.249$     &$2.915\times10^{-4}$   \\
$J/\psi$   &$3.097$    &$0.093$     &$5.966\times10^{-2}$   \\
$\psi(2S)$   &$3.686$    &$0.294$     &$7.965\times10^{-3}$\\
$\psi(3770)$   &$3.774$    &$27.2$     &$9.6\times10^{-6}$\\
$\psi(4040)$   &$4.039$    &$80$     &$1.07\times10^{-5}$\\
$\psi(4160)$   &$4.191$    &$70$     &$6.9\times10^{-6}$\\
\bottomrule[0.5pt]
\bottomrule[1pt]
\end{tabular*}
\end{table}

\subsection{The angular distributions and physical observables in the $B_{c}\to D_{s}^{*}(\to D_{s}\pi)\ell^{+}\ell^{-}$ decay}

{
In this subsection, we will drive the formula of the quasi-four-body decay $B_{c}\to D_{s}^{*}(\to D_{s}\pi)\ell^{+}\ell^{-}$. The differential decay width of this process is
\begin{equation}
    d\Gamma=\frac{\vert \mathcal{M} \vert^{2}}{2m_{B_{c}}}d\Phi_{4}(p;k_{1},k_{2},q_{1},q_{2}),
\end{equation}
where $p$ is the four momentum of the initial $B_{c}$ meson, $k_{1}(k_{2})$ and $q_{1}(q_{2})$ are the momenta of the mesons $D_{s}(\pi)$ and the lepton $\ell^{-}(\ell^{+})$, respectively, and $d\Phi_{4}$ is the four-body phase space. Taking into account the width of the $D_{s}^{*}$ meson, but treating it as narrow ($\Gamma_{D_{s}^{*}}\ll m_{D_{s}^{*}}$), the width can be obtained by doing the integration as
\begin{equation}
    \begin{split}
        \int d\Phi_{4}\frac{\vert \mathcal{M} \vert^{2}}{2m_{B_{c}}}
        \stackrel{\Gamma_{D_{s}^{*}}\ll m_{D_{s}^{*}}}{\longrightarrow}&
        \frac{1}{2^{15}\pi^{5}m_{B_{c}}m_{D_s^*}\Gamma_{D_{s}^{*}}}
        \int dq^{2} d\cos\theta d\cos\theta_{\ell} d\phi\\
        &\times\frac{\sqrt{\lambda(k^{2},k_{1}^{2},k_{2}^{2})}}{k^{2}}
        \frac{\sqrt{\lambda(q^{2},q_{1}^{2},q_{2}^{2})}}{q^{2}}\\
        &\times\frac{\sqrt{\lambda(p^{2},k^{2},q^{2})}}{p^{\prime2}}
        (k^{2}-m_{D_s^*}^{2})^{2}\vert \mathcal{M} \vert^{2}\Big|_{k^2=m_{D_s^*}^{2}}
    \end{split}
\end{equation}
with $\lambda(x,y,z)=x^{2}+y^{2}+z^{2}-2(xy+xz+yz)$.

The invariant amplitude $\mathcal{M}$ can be calculated from
\begin{widetext}
\begin{equation}
\begin{split}
\mathcal{M}(s_{\ell^{+}},s_{\ell^{-}})=&
\big{\langle}{D_{s}\pi;\ell^{+}(s_{\ell^{+}})\ell^{-}(s_{\ell^{-}})}{\vert}{\mathcal{H}^{\text{eff}}}{\vert}{B_{c}}\big{\rangle}\\
=&\sum_{s_{V}}\frac{i}{k^{2}-m_{D_{s}^{*}}^{2}}\mathcal{M}_{D_{s}^{*}\to{D_{s}\pi}}(s_{V})
\big{\langle}{D_{s}^{*}(s_{V})}{\ell^{+}(s_{\ell^{+}})\ell^{-}(s_{\ell^{-}})}{\vert}\mathcal{H}^{\text{eff}}{\vert}B_{c}\big{\rangle}\\
=&\sum_{s_{V}}\frac{iN}{2(k^{2}-m_{D_{s}^{*}}^{2})}\mathcal{M}_{D_{s}^{*}\to{D_{s}\pi}}(s_{V})
\bigg{\{}{C}_{9}^{\text{eff}}H^{V-A}(s_{V},t)L^{V}(s_{\ell^{+}},s_{\ell^{-}},t)
-\frac{2m_{b}}{q^{2}}{C}_{7}^{\text{eff}}H^{T+T\text{5}}(s_{V},t)L^{V}(s_{\ell^{+}},s_{\ell^{-}},t)\\
&+{C}_{10}H^{V-A}(s_{V},t)L^{A}(s_{\ell^{+}},s_{\ell^{-}},t)
-\sum_{\lambda=\pm1,0}\bigg{[}{C}_{9}^{\text{eff}}H^{V-A}(s_{V},\lambda)L^{V}(s_{\ell^{+}},s_{\ell^{-}},\lambda)\\
&-\frac{2m_{b}}{q^{2}}{C}_{7}^{\text{eff}}H^{T+T\text{5}}(s_{V},\lambda)L^{V}(s_{\ell^{+}},s_{\ell^{-}},\lambda)
+{C}_{10}H^{V-A}(s_{V},\lambda)L^{A}(s_{\ell^{+}},s_{\ell^{-}},\lambda)\bigg{]}
\bigg{\}},
\end{split}
\end{equation}
\end{widetext}
where $N=\frac{4G_{F}}{\sqrt{2}}V_{tb}V_{ts}^{*}\frac{\alpha_{e}}{4\pi}$, and the factor $1/2$ comes from the $P_{L(R)}$ in the effective Hamiltonian in Eq.~\eqref{eq:Hamiltonian2}.

For the amplitude $\mathcal{M}_{D_{s}^{*}\to D_{s}\pi}$, it can be evaluated by the effective Lagrangian approach. The concerned effective Lagrangian is
\begin{equation}
\mathcal{L}=g_{D_{s}^{*}D_{s}\pi}D_{s}^{\dagger}D_{s\mu}^{*}\partial^{\mu}\pi,
\end{equation}
where $g_{D_{s}^{*}D_{s}\pi}$ is the corresponding coupling constant. So we have the decay width of $D_{s}^{*}\to D_{s}\pi$ as
\begin{equation}
    \Gamma_{D_{s}^{*}}\times\mathcal{B}(D_{s}^{*}\to D_{s}\pi)=\frac{g_{D_{s}^{*}D_{s}\pi}^{2}}{48\pi}m_{D_{s}^{*}}\beta^{3}
\end{equation}
with $\beta=\sqrt{\lambda(m_{D_s^*}^{2},m_{D_s}^{2},m_{\pi}^{2})}/m_{D_s^*}^{2}$. Obviously, the coupling constant $g_{D_{s}^{*}D_{s}\pi}$ can be canceled between the vertex factor and the decay width.
}

\begin{figure}[htbp]\centering
  \includegraphics[width=76mm]{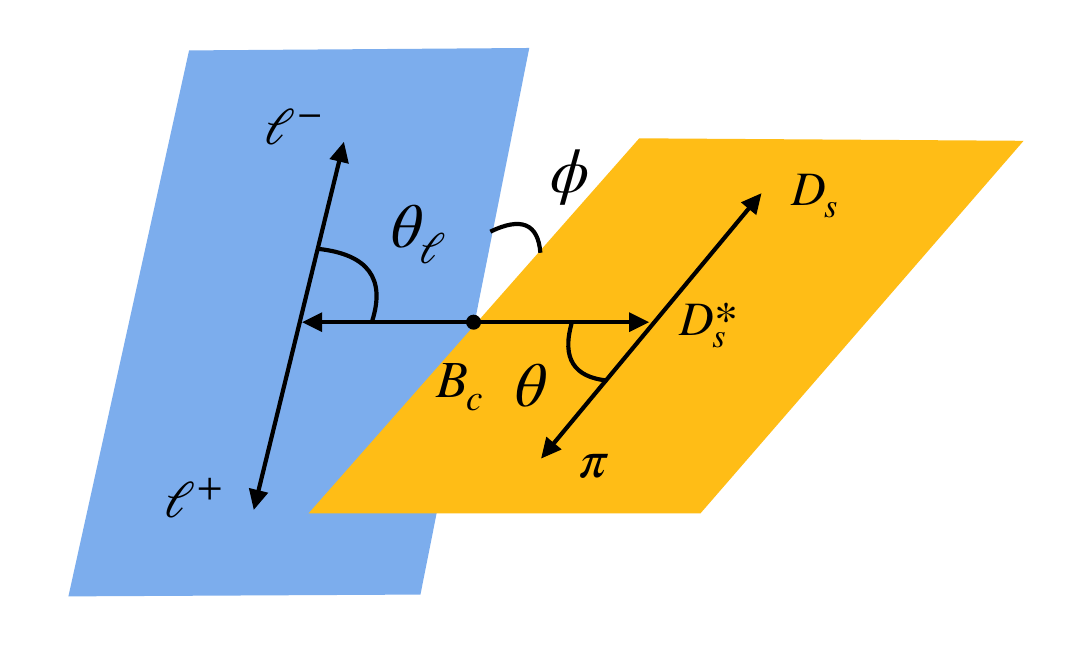}\\
  \caption{{Kinematics of the quasi-four-body decay $B_{c}\to D_{s}^{*}(\to D_{s}\pi)\ell^{+}\ell^{-}$.}}
  \label{fig:kinematics}
\end{figure}

Finally, with the effective Hamiltonian in Eq.~\eqref{eq:Hamiltonian2}, we can calculate the quasi-four-body decay $B_{c}^{-}\to D_{s}^{*-}(\to D_{s}^{-}\pi^{0})\ell^{+}\ell^{-}$. As deduced in Ref. \cite{Altmannshofer:2008dz}, the corresponding angular distributions can be simplified as
\begin{equation}
\frac{d^{4}\Gamma}{dq^{2} d\cos\theta d\cos\theta_{\ell} d\phi}=\frac{9}{32\pi}\sum_{i}I_{i}(q^{2})f_{i}(\theta, \theta_{\ell}, \phi),
\label{eq:fi}
\end{equation}
where the explicit expressions of $I_{i}(q^{2})$ and $f_{i}(\theta, \theta_{\ell}, \phi)$ are shown in Table \ref{tab:Iifi}.
Compared to Ref. \cite{Altmannshofer:2008dz}, the term $I_{6c}$ is neglected since it depends on the scalar operator. As shown in Fig. \ref{fig:kinematics}, the $\theta$ is the angle between the $-\hat{z}$ direction and pion-emitted direction in the rest frame of the $D_{s}^{*}$ meson, the $\theta_{\ell}$ is the angle made by the $\ell^{-}$ with the $+\hat{z}$ direction in the $\ell^{+}\ell^{-}$ center of mass system, and the $\phi$ is the angle between the decay planes, i.e., the $D_{s}^{*}\to D_{s}\pi$ plane and the $\text{virtual~boson}\to\ell^{+}\ell^{-}$ plane.

\begin{table*}[htbp]\centering
\caption{The explicit expressions of the angular coefficients $I_{i}$ and $f_{i}$ \cite{Altmannshofer:2008dz,Jin:2020jtu,Jin:2020qfp} in Eq.~\eqref{eq:fi},
where $\hat{m}_{\ell}^{2}=m_{\ell}^{2}/q^{2}$ and $\beta_{\ell}=\sqrt{1-4\hat{m}_{\ell}^{2}}$.}
\label{tab:Iifi}
\renewcommand\arraystretch{1.25}
\begin{tabular*}{160mm}{c@{\extracolsep{\fill}}cc}
\toprule[1pt]
\toprule[0.5pt]
$i$          &$I_{i}(q^{2})$      &$f_{i}(\theta, \theta_{\ell}, \phi)$\\
\midrule[0.5pt]
$1s$       &$\Big{(}\frac{3}{4}-\hat{m}_{\ell}^{2}\Big{)}
\Big{(}\vert\mathscr{A}_{\|}^{L}\vert^{2}+\vert\mathscr{A}_{\bot}^{L}\vert^{2}
+\vert\mathscr{A}_{\|}^{R}\vert^{2}+\vert\mathscr{A}_{\bot}^{R}\vert^{2}\Big{)}
+4\hat{m}_{\ell}^{2}\text{Re}\big{[}\mathscr{A}_{\bot}^{L}\mathscr{A}_{\bot}^{R*}
+\mathscr{A}_{\|}^{L}\mathscr{A}_{\|}^{R*}\big{]}$           &$\sin^{2}\theta$\\
$1c$       &$\vert\mathscr{A}_{0}^{L}\vert^{2}+\vert\mathscr{A}_{0}^{R}\vert^{2}
+4\hat{m}_{\ell}^{2}\Big{(}\vert\mathscr{A}_{t}\vert^{2}+2\text{Re}\big{[}\mathscr{A}_{0}^{L}\mathscr{A}_{0}^{R*}\big{]}\Big{)}$           &$\cos^{2}\theta$\\
$2s$       &$\beta_{\ell}^{2}\Big{(}\vert\mathscr{A}_{\|}^{L}\vert^{2}+\vert\mathscr{A}_{\bot}^{L}\vert^{2}+\vert\mathscr{A}_{\|}^{R}\vert^{2}+\vert\mathscr{A}_{\bot}^{R}\vert^{2}\Big{)}/4$           &$\sin^{2}\theta\cos2\theta_{\ell}$\\
$2c$       &$-\beta_{\ell}^{2}\Big{(}\vert\mathscr{A}_{0}^{L}\vert^{2}+\vert\mathscr{A}_{0}^{R}\vert^{2}\Big{)}$           &$\cos^{2}\theta\cos2\theta_{\ell}$\\
$3$         &$\beta_{\ell}^{2}\Big{(}\vert\mathscr{A}_{\bot}^{L}\vert^{2}-\vert\mathscr{A}_{\|}^{L}\vert^{2}+\vert\mathscr{A}_{\bot}^{R}\vert^{2}-\vert\mathscr{A}_{\|}^{R}\vert^{2}\Big{)}/2$           &$\sin^{2}\theta\sin^{2}\theta_{\ell}\cos2\phi$\\
$4$         &$\beta_{\ell}^{2}\text{Re}\big{[}\mathscr{A}_{0}^{L}\mathscr{A}_{\|}^{L*}+\mathscr{A}_{0}^{R}\mathscr{A}_{\|}^{R*}\big{]}/\sqrt{2}$           &$\sin2\theta\sin2\theta_{\ell}\cos\phi$\\
$5$         &$\sqrt{2}\beta_{\ell}\text{Re}\big{[}\mathscr{A}_{0}^{L}\mathscr{A}_{\bot}^{L*}-\mathscr{A}_{0}^{R}\mathscr{A}_{\bot}^{R*}\big{]}$           &$\sin2\theta\sin\theta_{\ell}\cos\phi$\\
$6s$         &$2\beta_{\ell}\text{Re}\big{[}\mathscr{A}_{\|}^{L}\mathscr{A}_{\bot}^{L*}-\mathscr{A}_{\|}^{R}\mathscr{A}_{\bot}^{R*}\big{]}$           &$\sin^{2}\theta\cos\theta_{\ell}$\\
$7$         &$\sqrt{2}\beta_{\ell}\text{Im}\big{[}\mathscr{A}_{0}^{L}\mathscr{A}_{\|}^{L*}-\mathscr{A}_{0}^{R}\mathscr{A}_{\|}^{R*}\big{]}$           &$\sin2\theta\sin\theta_{\ell}\sin\phi$\\
$8$         &$\beta_{\ell}^{2}\text{Im}\big{[}\mathscr{A}_{0}^{L}\mathscr{A}_{\bot}^{L*}+\mathscr{A}_{0}^{R}\mathscr{A}_{\bot}^{R*}\big{]}/\sqrt{2}$           &$\sin2\theta\sin2\theta_{\ell}\sin\phi$\\
$9$         &$\beta_{\ell}^{2}\text{Im}\big{[}\mathscr{A}_{\|}^{L*}\mathscr{A}_{\bot}^{L}+\mathscr{A}_{\|}^{R*}\mathscr{A}_{\bot}^{R}\big{]}$           &$\sin^{2}\theta\sin^{2}\theta_{\ell}\sin2\phi$\\
\bottomrule[0.5pt]
\bottomrule[1pt]
\end{tabular*}
\end{table*}

The amplitudes $\mathscr{A}_{0,\|,\bot}^{L,R}$ and $\mathscr{A}_{t}$ are the functions of the transferred momentum square $q^{2}$, and the seven independent form factors $V$, $A_{0,1,2}$ and $T_{1,2,3}$, i.e., \cite{Altmannshofer:2008dz,Jin:2020jtu,Jin:2020qfp}
\begin{widetext}
\begin{equation}
\begin{split}
\mathscr{A}_{\bot}^{L,R}(q^{2})=&-N_{\ell}\sqrt{2N_{D_{s}^{*}}}\sqrt{\lambda(M^{\prime2},M^{\prime\prime2},q^{2})}
\bigg{[}\Big{(}{C}_{9}^{\text{eff}}\mp{C}_{10}\Big{)}\frac{V(q^{2})}{M^{\prime}+M^{\prime\prime}}
+2\hat{m}_{b}{C}_{7}^{\text{eff}}T_{1}(q^{2})\bigg{]},\\
\mathscr{A}_{\|}^{L,R}(q^{2})=&N_{\ell}\sqrt{2N_{D_{s}^{*}}}\bigg{[}
\Big{(}{C}_{9}^{\text{eff}}\mp{C}_{10}\Big{)}(M^{\prime}+M^{\prime\prime})A_{1}(q^{2})
+2\hat{m}_{b}{C}_{7}^{\text{eff}}\Big{(}M^{\prime2}-M^{\prime\prime2}\Big{)}T_{2}(q^{2})\bigg{]},\\
\mathscr{A}_{0}^{L,R}(q^{2})=&\frac{N_{\ell}\sqrt{N_{D_{s}^{*}}}}{2M^{\prime\prime}\sqrt{q^{2}}}
\bigg{\{}\Big{(}{C}_{9}^{\text{eff}}\mp{C}_{10}\Big{)}
\bigg{[}(M^{\prime2}-M^{\prime\prime2}-q^{2})(M^{\prime}+M^{\prime\prime})A_{1}(q^{2})-\frac{\lambda(M^{\prime2},M^{\prime\prime2},q^{2})}{M^{\prime}+M^{\prime\prime}}A_{2}(q^{2})\bigg{]}\\
&+2m_{b}{C}_{7}^{\text{eff}}\bigg{[}(M^{\prime2}+3M^{\prime\prime2}-q^{2})T_{2}(q^{2})-\frac{\lambda(M^{\prime2},M^{\prime\prime2},q^{2})}{M^{\prime2}-M^{\prime\prime2}}T_{3}(q^{2})\bigg{]}
\bigg{\}},\\
\mathscr{A}_{t}(q^{2})=&2N_{\ell}\sqrt{N_{D_{s}^{*}}}\sqrt{\frac{\lambda(M^{\prime2},M^{\prime\prime2},q^{2})}{q^{2}}}{C}_{10}A_{0}(q^{2}),
\end{split}
\end{equation}
\end{widetext}
where $M^{\prime}(M^{\prime\prime})$ is the mass of the $B_{c}$ ($D_{s}^{*}$) meson and $\hat{m}_{b}=m_{b}/q^{2}$, and
\begin{equation}
\begin{split}
N_{\ell}=&\frac{i\alpha_{e}G_{F}}{4\sqrt{2}\pi}V_{tb}V_{ts}^{*},\\
N_{D_{s}^{*}}=&\frac{8\sqrt{\lambda}q^{2}}{3\times256\pi^{3}M^{\prime3}}
\sqrt{1-\frac{4m_{\ell}^{2}}{q^{2}}}\mathcal{B}\big{(}D_{s}^{*}\to D_{s}\pi\big{)}.
\end{split}
\end{equation}

For the $CP$-conjugated mode $B_{c}^{+}\to D_{s}^{*+}(\to D_{s}^{+}\pi^{0})\ell^{+}\ell^{-}$, we have
\begin{equation}
\frac{d^{4}\bar{\Gamma}}{q^{2}{d\cos\theta}{d\cos\theta_{\ell}}{d\phi}}=\sum_{i}\frac{9}{32\pi}\bar{I}_{i}(q^{2})f_{i}(\theta, \theta_{\ell}, \phi),
\end{equation}
where $\bar{I}_{i}$ can be obtained by doing the conjugation for the weak phases of the CKM matrix elements in $I_{i}$ in Table \ref{tab:Iifi}. In addition, we should also do the following substitutions as
\begin{equation}
\begin{split}
I_{1(c,s),2(c,s),3,4,7}\to&\bar{I}_{1(c,s),2(c,s),3,4,7},\\
I_{5,6s,8,9}\to&-\bar{I}_{5,6s,8,9}.
\label{eq:replacementCP}
\end{split}
\end{equation}
This is the result of the operations of $\theta_{\ell}\to\theta_{\ell}-\pi$ and $\phi\to-\phi$.

To separate the $CP$-conserving and the $CP$-violating effects, we define the normalized $CP$-averaged angular coefficients $S_{i}$ and the $CP$ asymmetry angular coefficients $A_{i}$ as
\begin{equation}
\begin{split}
S_{i}=&\frac{I_{i}+\bar{I}_{i}}{d(\Gamma+\bar{\Gamma})/dq^{2}},\\
A_{i}=&\frac{I_{i}-\bar{I}_{i}}{d(\Gamma+\bar{\Gamma})/dq^{2}},\\
\label{eq:SiAi}
\end{split}
\end{equation}
respectively.
To reduce both the experimental and theoretical uncertainties, the $S_{i}$ and $A_{i}$ have been normalized to the $CP$-averaged differential decay width. The other physical observables, such as the forward-backward asymmetry parameter $A_{FB}$, the $CP$-violation $\mathcal{A}_{CP}$, and the longitudinal (transverse) polarization fractions of $D_{s}^{*}$ meson $F_{L}(F_{T})$, can thus be easily expressed in terms of these normalized angular coefficients. With the above preparations, we continue to study the physical observables.
\begin{enumerate}
\item[{(a)}]
By integrating over the angles in the regions $\theta\in[0,\pi]$, $\theta_{\ell}\in[0,\pi]$, and $\phi\in[0,2\pi]$, the $q^{2}$-dependent differential decay width becomes
\begin{equation}
\frac{d\Gamma}{dq^{2}}=\frac{1}{4}\big{(}3I_{1c}+6I_{1s}-I_{2c}-2I_{2s}\big{)},
\label{eq:Br}
\end{equation}
and that of the $CP$-conjugated mode $d\bar{\Gamma}/dq^{2}$ is analogous and can be obtained with the replacement in Eq.~\eqref{eq:replacementCP}.
So the $CP$-averaged differential decay width of $B_{c}\to D_{s}^{*}(\to D_{s}\pi)\ell^{+}\ell^{-}$ can be evaluated by
\begin{equation}
\frac{d\Gamma}{dq^{2}}=\frac{1}{2}\Big{(}\frac{d\Gamma}{dq^{2}}+\frac{d\bar{\Gamma}}{dq^{2}}\Big{)}.
\label{eq:BrCP}
\end{equation}
In this work, we focus on the $CP$-averaged decay width.
\item[{(b)}]
The $CP$ violation of the decay width can thus be estimated by
\begin{equation}
\mathcal{A}_{CP}(q^{2})=\frac{(d\Gamma-d\bar{\Gamma})/dq^{2}}{(d\Gamma+d\bar{\Gamma})/dq^{2}}=\frac{1}{4}\big{(}3A_{1c}+6A_{1s}-A_{2c}-2A_{2s}\big{)}.
\end{equation}
\item[{(c)}]
The $CP$ asymmetry lepton forward-backward asymmetry is
\begin{equation}
\begin{split}
A_{FB}^{CP}(q^{2})=&\frac{\big{(}\int_{-1}^{0}-\int_{0}^{1}\big{)}d\cos\theta_{\ell}\int_{-1}^{1}d\cos\theta\int_{0}^{2\pi}d\phi\frac{d^{4}(\Gamma+\bar{\Gamma})}{dq^{2}{d\cos\theta}{d\cos\theta_{\ell}}{d\phi}}}
{d(\Gamma+\bar{\Gamma})/dq^{2}}\\
=&\frac{3}{4}A_{6},
\end{split}
\end{equation}
and the $CP$-averaged lepton forward-backward asymmetry is
\begin{equation}
A_{FB}(q^{2})=\frac{3}{4}S_{6}.
\end{equation}
\item[{(d)}]
The longitudinal and transverse $D_{s}^{*}$ polarization fractions are
\begin{equation}
\begin{split}
F_{L}=&\frac{1}{4}(3S_{1c}-S_{2c}),\\
F_{T}=&\frac{1}{2}(3S_{1s}-S_{2s}),
\end{split}
\end{equation}
respectively.
\end{enumerate}

Furthermore, the clean angular observables $P_{1,2,3}$ and $P_{4,5,6,8}^{\prime}$ ({more details can be found in Refs.~\cite{Matias:2012xw,Descotes-Genon:2013vna}}) are associated with the $CP$-averaged angular coefficients:
\begin{equation}
\begin{split}
P_{1}=&\frac{S_{3}}{2S_{2s}},\\
P_{2}=&\frac{\beta_{\ell}S_{6s}}{8S_{2s}},\\
P_{3}=&-\frac{S_{9}}{4S_{2s}},\\
\end{split}
\end{equation}
\begin{equation}
\begin{split}
P_{4}^{\prime}=&\frac{S_{4}}{\sqrt{S_{1c}S_{2s}}},\\
P_{5}^{\prime}=&\frac{\beta_{\ell}S_{5}}{2\sqrt{S_{1c}S_{2s}}},\\
P_{6}^{\prime}=&-\frac{\beta_{\ell}S_{7}}{2\sqrt{S_{1c}S_{2s}}},\\
P_{8}^{\prime}=&-\frac{S_{8}}{\sqrt{S_{1c}S_{2s}}}.
\end{split}
\end{equation}
{As pointed out in Refs.~\cite{Matias:2012xw,Descotes-Genon:2013vna,LHCb:2013ghj,Jin:2020jtu}, in the large-recoiled limit, these observables are largely free of form factor uncertainties.}

Finally, we also focus on the ratios, i.e.,
\begin{equation}
\begin{split}
R^{e\mu}=&\frac{\int_{4m_{\mu}^{2}}^{(M^{\prime}-M^{\prime\prime})^{2}}\frac{d\Gamma[B_{c}\to D_{s}^{*}(\to D_{s}\pi)e^{+}e^{-}]}{dq^{2}}dq^{2}}
{\int_{4m_{\mu}^{2}}^{(M^{\prime}-M^{\prime\prime})^{2}}\frac{d\Gamma[B_{c}\to D_{s}^{*}(\to D_{s}\pi)\mu^{+}\mu^{-}]}{dq^{2}}dq^{2}},\\
R^{\tau\mu}=&\frac{\int_{4m_{\tau}^{2}}^{(M^{\prime}-M^{\prime\prime})^{2}}\frac{d\Gamma[B_{c}\to D_{s}^{*}(\to D_{s}\pi)\tau^{+}\tau^{-}]}{dq^{2}}dq^{2}}
{\int_{4m_{\mu}^{2}}^{(M^{\prime}-M^{\prime\prime})^{2}}\frac{d\Gamma[B_{c}\to D_{s}^{*}(\to D_{s}\pi)\mu^{+}\mu^{-}]}{dq^{2}}dq^{2}},
\end{split}
\end{equation}
which reflect the LFU. We would like to emphasize that the lower limit of the integral of the electron mode is chosen as $4m_{\mu}^{2}$ instead of the kinematic limit $4m_{e}^{2}$ in order to exclude the large enhancement dominated by the photon pole in the small $q^{2}$ region due to the $C_{7}^{\text{eff}}$-associated factor $1/q^{2}$. In the $B\to K\ell^{+}\ell^{-}$ process, the experimental measurements of the ratio $R_{K}^{e\mu}$ by Belle \cite{Belle:2009zue,Belle:2016fev,BELLE:2019xld,Belle:2019oag} and $BABAR$ \cite{BaBar:2012mrf} are in agreement with the SM prediction, while the LHCb result \cite{LHCb:2014vgu,LHCb:2017avl,LHCb:2021trn} shows a clear deviation from the SM expectation (see Fig. 4 of Ref. \cite{LHCb:2021trn}) with $3.1$$\sigma$. {We note that in Ref.~\cite{Alok:2023yzg}, the authors used the ratios $R_{K^{(*)}}^{\tau\mu}$ to study the LFU violation, and found that they can deviate from the SM prediction even if the NP couplings are universal. Therefore, in order to use these ratios to study the LFU violation, we should compare the allowed ranges, considering both the solutions with only universal couplings and those with universal and nonuniversal components.} Whatever, the ratio in the $B_{c}\to D_{s}^{*}\ell^{+}\ell^{-}$ sector is also interesting to investigate whether it is consistent with the SM expectation or not. The breaking of the LFU may require an expansion of the gauge structure of the SM, and of course probes the NP effects~\cite{Li:2018lxi}.

\section{weak transition form factors}\label{sec03}

The standard and(or) covariant LFQMs have been widely used to study the decays of mesons~\cite{Jaus:1989au,Jaus:1996np,Cheng:1996if,Cheng:1997au,Jaus:1999zv,Cheng:2003sm,Chua:2003ac,Cheng:2004yj,Wang:2007sxa,Wang:2008ci,Shen:2008zzb,Wang:2008xt,Wang:2009mi,Cheng:2009ms,Chen:2009qk,Choi:2010zb,Choi:2010be,Li:2010bb,Ke:2011mu,Verma:2011yw,Ke:2013yka,Xu:2014mqa,Shi:2016gqt,Cheng:2017pcq,Chen:2017vgi,Kang:2018jzg,Chang:2018zjq,Chang:2019mmh,Chang:2019xtj,Chang:2019obq,Chang:2020xvu,Chang:2020wvs,Chen:2021ywv,Choi:2021mni,Choi:2021qza,Arifi:2022qnd,Zhang:2023ypl,Shi:2023qnw,Hazra:2023zno} and baryons~\cite{Ke:2007tg,Ke:2012wa,Wang:2017mqp,Ke:2017eqo,Zhu:2018jet,Zhao:2018zcb,Zhao:2018mrg,Xing:2018lre,Chua:2018lfa,Ke:2019smy,Chua:2019yqh,Ke:2019lcf,Hu:2020mxk,Geng:2020fng,Hsiao:2020gtc,Geng:2021nkl,Li:2021qod,Ke:2021pxk,Hsiao:2021mlp,Li:2021kfb,Li:2022nim,Geng:2022xpn,Wang:2022ias,Zhao:2022vfr,Li:2022hcn,Lu:2023rmq,Zhao:2023yuk,Liu:2023zvh}.
In the conventional LFQM framework, the consistent quark (or antiquark) of the meson is required to be on its mass shell, and thus the initial (or final) meson is offshell. This procedure misses the zero-mode effects and makes the matrix element noncovariant. To avoid this shortcoming, Jaus \cite{Jaus:1989au,Jaus:1999zv} proposed a covariant framework for the $S$-waved pseudoscalar and vector meson decays in which the zero-mode contributions are systematically taken into account. Cheng {\it et al.} \cite{Cheng:2003sm,Cheng:2009ms} extended this approach to the case of the $P$-wave meson (such as scalar, axial-vector and tensor mesons). The physical quantities, such as the decay constant and the form factor of the weak transition, are obtained in terms of the Feynman loop integration. Unlike the conventional LFQM, the covariant LFQM requires the initial (or final) meson to be on its mass shell. For more details on the difference, see Refs. \cite{Cheng:2003sm,Chang:2019obq}. In this section, we will use the covariant LFQM to calculate the $B_{c}\to{D}_{s}^{*}$ form factors.

Following Ref.~\cite{Zhang:2023ypl}, the $B_{c}\to{D_{s}^{*}}$ weak transition form factors deduced by (axial-)vector currents are defined as
\begin{widetext}
\begin{equation}
\begin{split}
\langle D_{s}^{*}(p^{\prime\prime})\vert \bar{s}\gamma_{\mu}b \vert B_{c}(p^{\prime})\rangle=&
\epsilon_{\mu\nu\alpha\beta}\varepsilon^{*\nu}P^{\alpha}q^{\beta}g(q^{2}),\\
\langle D_{s}^{*}(p^{\prime\prime})\vert \bar{s}\gamma_{\mu}\gamma_{5}b \vert B_{c}(p^{\prime})\rangle=&
-i\Big{\{}\varepsilon^{*}_{\mu}f(q^{2})+\varepsilon^{*}\cdot P(P_{\mu}a_{+}(q^{2})+q_{\mu}a_{-}(q^{2}))\Big{\}},
\label{eq:ffs1}
\end{split}
\end{equation}
where we use the convention $\epsilon_{0123}=+1$ and define $P_{\mu}=p_{\mu}^{\prime}+p_{\mu}^{\prime\prime}$ and $q_{\mu}=p_{\mu}^{\prime}-p_{\mu}^{\prime\prime}$, and
$\varepsilon$ is the polarization vector of the $D_{s}^{*}$ meson. These amplitudes can also be parametrized as the Bauer-Stech-Wirbel (BSW) form \cite{Wirbel:1985ji}, i.e.,
\begin{equation}
\begin{split}
\langle D_{s}^{*}(p^{\prime\prime})\vert \bar{s}\gamma_{\mu}b \vert B_{c}(p^{\prime})\rangle=&
-\frac{1}{M^{\prime}+M^{\prime\prime}}\epsilon_{\mu\nu\alpha\beta}\varepsilon^{*\nu}P^{\alpha}q^{\beta}V(q^{2}),\\
\langle D_{s}^{*}(p^{\prime\prime})\vert \bar{s}\gamma_{\mu}\gamma_{5}b \vert B_{c}(p^{\prime})\rangle=&
i\Bigg{\{}(M^{\prime}+M^{\prime\prime})\varepsilon_{\mu}^{*}A_{1}(q^{2})-\frac{\varepsilon^{*}\cdot P}{M^{\prime}+M^{\prime\prime}}P_{\mu}A_{2}(q^{2})-2M^{\prime\prime}\frac{\varepsilon^{*}\cdot P}{q^{2}}q_{\mu}\Big{[}A_{3}(q^{2})-A_{0}(q^{2})\Big{]}\Bigg{\}}
\label{eq:ffs2}
\end{split}
\end{equation}
with $M^{\prime}(M^{\prime\prime})$ being the mass of the parent (daughter) meson. These two definitions are related by the relations \cite{Zhang:2023ypl}
\begin{equation}
\begin{split}
V(q^{2})=&-(M^{\prime}+M^{\prime\prime})g(q^{2}),~~~\quad
A_{1}(q^{2})=-\frac{f(q^{2})}{M^{\prime}+M^{\prime\prime}},\\
A_{2}(q^{2})=&(M^{\prime}+M^{\prime\prime})a_{+}(q^{2}),~~\quad
A_{3}(q^{2})-A_{0}(q^{2})=\frac{q^{2}}{2M^{\prime\prime}}a_{-}(q^{2}),\\
A_{3}(q^{2})=&\frac{M^{\prime}+M^{\prime\prime}}{2M^{\prime\prime}}A_{1}(q^{2})-\frac{M^{\prime}-M^{\prime\prime}}{2M^{\prime\prime}}A_{2}(q^{2}).
\end{split}
\end{equation}

In addition, the (pseudo)tensor current amplitudes can be defined as \cite{Ball:1998kk,Ali:1999mm}
\begin{equation}
\begin{split}
\langle D_{s}^{*}(p^{\prime\prime})\vert \bar{s}i\sigma_{\mu\nu}q^{\nu}b \vert B_{c}(p^{\prime})\rangle=&T_{1}(q^{2})\epsilon_{\mu\nu\alpha\beta}\varepsilon^{*\nu}P^{\alpha}q^{\beta},\\
\langle D_{s}^{*}(p^{\prime\prime})\vert \bar{s}i\sigma_{\mu\nu}q^{\nu}\gamma_{5}b \vert B_{c}(p^{\prime})\rangle=&iT_{2}(q^{2})
\Big{[}(M^{\prime2}-M^{\prime\prime2})\varepsilon^{*}_{\mu}-\varepsilon^{*}\cdot q P_{\mu}\Big{]}
+iT_{3}(q^{2})\varepsilon^{*}\cdot q\Big{[}q_{\mu}-\frac{q^{2}}{M^{\prime2}-M^{\prime\prime2}}P_{\mu}\Big{]},
\label{eq:ffs3}
\end{split}
\end{equation}
\end{widetext}
where, we have $T_{1}(0)=T_{2}(0)$ since the identity $2\sigma_{\mu\nu}\gamma_{5}=-i\epsilon_{\mu\nu\alpha\beta}\sigma^{\alpha\beta}$.

The form factors require a nonperturbative calculation. In this work, we use the covariant LFQM to calculate the relevant form factors for the weak transition. In this approach, the constituent quark and the antiquark inside a meson are off shell.
We define the incoming (outgoing) meson to  have the momentum $P^{\prime}=p_{1}^{\prime}+p_{2}(P^{\prime\prime}=p_{1}^{\prime\prime}+p_{2})$, where $p_{1}^{\prime(\prime\prime)}$ and $p_{2}$ are the off-shell momenta of the quark and the antiquark, respectively.
These momenta can be expressed in terms of the internal variables ($x_{i},\vec{k}_{\bot}^{\prime}$) ($i=1,2$), defined by
\begin{equation}
p_{1}^{\prime+}=x_{1}P^{\prime+},~~\quad p_{1}^{+}=x_{2}P^{\prime+},~~\quad \vec{p}_{1\bot}^{\prime}=x_{1}\vec{P}_{\bot}^{\prime}+\vec{k}_{\bot}^{\prime}.
\end{equation}
They must also satisfy $x_{1}+x_{2}=1$.

\begin{figure}[htbp]\centering
  \includegraphics[width=76mm]{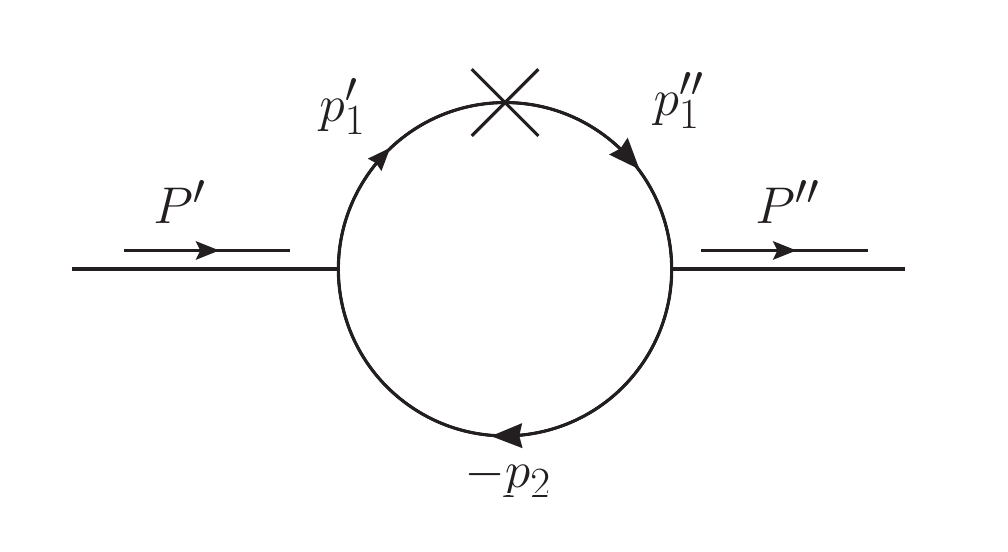}\\
  \caption{The one-loop Feynman diagram for meson weak transition amplitude, where $P^{\prime}(P^{\prime\prime})$ is the momentum of the incoming (outgoing) meson,
  $p_{1}^{\prime(\prime\prime)}$ and $p_{2}$ are the momenta of the quark and antiquark, respectively. {The symbol ``cross" denotes the weak interaction vertex.}}
  \label{fig:loop}
\end{figure}

According to Refs. \cite{Cheng:2003sm,Zhang:2023ypl}, the corresponding weak transition matrix element at the one-loop level can be calculated in terms of the Feynman loop integral, as shown in Fig.~\ref{fig:loop}. Then the form factors can be extracted from the corresponding matrix element. To write down the transition amplitude, we need the meson-quark antiquark vertices for the initial meson as $i\Gamma^{\prime}=H_{P}^{\prime}\gamma_{5}$,
and that of the outgoing meson as $i(\gamma_{0}\Gamma^{\prime\prime\dagger}\gamma_{0})$ with $\Gamma^{\prime\prime}=H_{V}^{\prime\prime}\big{[}\gamma_{\mu}-(p_{1}^{\prime\prime}-p_{2})_{\mu}/W_{V}^{\prime\prime}\big{]}$ \cite{Cheng:2003sm,Zhang:2023ypl},
where the subscripts $P$ and $V$ denote the pseudoscalar and vector meson, respectively.

For Fig. \ref{fig:loop}, the concrete expression of the transition amplitude for $P \to V$ can be expressed as
\begin{equation}
\begin{split}
\mathcal{B}_{\mu}^{V(A,T,T\text{5})}=-i^{3}\frac{N_{c}}{(2\pi)^{4}}\int d^{4}p_{1}^{\prime}\frac{iH_{P}^{\prime}H_{V}^{\prime\prime}}{N_{1}^{\prime}N_{1}^{\prime\prime}N_{2}}S_{\mu\nu}^{V(A,T,T\text{5})}\epsilon_{V}^{*\nu},
\end{split}
\end{equation}
where $N_{1}^{\prime(\prime\prime)}=p_{1}^{\prime(\prime\prime)2}-m_{1}^{\prime(\prime\prime)2}$ and $N_{2}=p_{2}^{2}-m_{2}^{2}$ come from the propagators of the quarks.
The superscripts $V$, $A$, $T$, and $T\text{5}$ represent the vector, axial-vector, tensor, and pseudotensor currents, respectively.
The traces $S_{\mu\nu}^{V}$ are written as
\begin{equation}
\begin{split}
S_{\mu\nu}^{V}{\!}=&\text{Tr}\Big{[}\Big{(}\gamma_{\nu}{\!}-{\!}\frac{(p_{1}^{\prime\prime}{\!}-{\!}p_{2})_{\nu}}{W_{V}^{\prime\prime}}\Big{)}(\slashed{p}_{1}^{\prime\prime}+m_{1}^{\prime\prime})\gamma_{\mu}(\slashed{p}_{1}^{\prime}+m_{1}^{\prime})\gamma_{5}(-\slashed{p}_{2}+m_{2})\Big{]}\\
=&-2i\epsilon_{\mu\nu\alpha\beta}\bigg{[}p_{1}^{\prime\alpha}P^{\beta}(m_{1}^{\prime\prime}-m_{1}^{\prime})
+p_{1}^{\prime\alpha}q^{\beta}(m_{1}^{\prime\prime}+m_{1}^{\prime}-2m_{2})\\
&+q^{\alpha}P^{\beta}m_{1}^{\prime}\bigg{]}+\frac{1}{W_{V}^{\prime\prime}}(4p_{1\nu}^{\prime}-3q_{\nu}-P_{\nu})i\epsilon_{\mu\alpha\beta\rho}p_{1}^{\prime\alpha}q^{\beta}P^{\rho}.
\label{eq:SV}
\end{split}
\end{equation}
To make reading easier, the relevant expressions of the traces $S_{\mu\nu}^{A,T,T5}$ are collected in Appendix \ref{app01}.

Following Refs.~\cite{Cheng:2003sm,Cheng:2004yj,Cheng:2009ms}, the execution of the $p_{1}^{\prime-}$ integration went to the replacement:
\begin{equation}
\begin{split}
N_{1}^{\prime(\prime\prime)}&\to\hat{N}_{1}^{\prime(\prime\prime)}=x_{1}\big{(}M^{\prime(\prime\prime)2}-M_{0}^{\prime(\prime\prime)2}\big{)},\\
H_{P(V)}^{\prime(\prime\prime)}&\to h_{P(V)}^{\prime(\prime\prime)},\\
W_{V}^{\prime\prime}&\to\omega_{V}^{\prime\prime},\\
\int\frac{d^{4}p_{1}^{\prime}}{(2\pi)^{4}}H_{P}^{\prime}H_{V}^{\prime\prime}S_{\mu\nu}\epsilon^{*\nu}&\to -i\pi\int\frac{dx_{2}d^{2}\vec{k}_{\bot}^{\prime}}{x_{2}\hat{N}_{1}^{\prime}\hat{N}_{1}^{\prime\prime}}h_{P}^{\prime}h_{V}^{\prime\prime}\hat{S}_{\mu\nu}\epsilon^{*\nu},
\label{eq:Bc2P}
\end{split}
\end{equation}
where we define
\begin{equation}
M_{0}^{\prime(\prime\prime)2}=\frac{\vec{k}_{\bot}^{\prime(\prime\prime)2}+m_{1}^{\prime(\prime\prime)2}}{x_{1}}+\frac{\vec{k}_{\bot}^{\prime(\prime\prime)2}+m_{2}^{\prime(\prime\prime)2}}{x_{2}},
\end{equation}
with $\vec{k}_{\bot}^{\prime\prime}=\vec{k}_{\bot}^{\prime}-x_{2}\vec{q}_{\bot}$ and $\omega_{V}^{\prime\prime}=M_{0}^{\prime\prime}+m_{1}^{\prime\prime}+m_{2}$.

To write down the concrete expression of $\hat{S}_{\mu\nu}$, we should take into account the so-called zero-mode contribution. As shown in Refs.~\cite{Cheng:2003sm,Chen:2017vgi}, after doing the integration in Eq.~\eqref{eq:Bc2P} we have $p_{2}=\hat{p}_{2}$, and
\begin{equation}
\begin{split}
\hat{p}_{1}^{\prime\mu}&=(P^{\prime}-\hat{p}_{2})^{\mu}\\
&=x_{1}P^{\prime\mu}+(0,0,\vec{k}_{\bot}^{\prime})^{\mu}+\frac{1}{2}\Big{(}x_{2}P^{\prime-}-\frac{\vec{p}_{2\bot}^{2}+m_{2}^{2}}{x_{2}P^{\prime+}}\Big{)}\tilde{\omega}^{\mu},
\end{split}
\end{equation}
where $\tilde{\omega}=(2,0,\vec{0}_{\bot})$ is a lightlike four vector in the light-front coordinate. Following the discussions in a series of papers~\cite{Cheng:2003sm,Cheng:2004yj,Cheng:2009ms,Chen:2017vgi}, for avoiding the $\tilde{\omega}$ dependence, we need to do the following replacements \cite{Cheng:2003sm,Cheng:2004yj,Cheng:2009ms}:
\begin{equation}
\begin{split}
\hat{p}_{1\mu}^{\prime}{\!}\doteq&P_{\mu}A_{1}^{(1)}+q_{\mu}A_{2}^{(1)},\\
\hat{p}_{1\mu}^{\prime}\hat{p}_{1\nu}^{\prime}{\!}\doteq&g_{\mu\nu}A_{1}^{(2)}{\!}+{\!}P_{\mu}P_{\nu}A_{2}^{(2)}{\!}+{\!}(P_{\mu}q_{\nu}{\!}+{\!}P_{\nu}q_{\mu})A_{3}^{(2)}{\!}+{\!}q_{\mu}q_{\nu}A_{4}^{(2)},\\
N_{2}\doteq&Z_{2},\\
\hat{p}_{1\mu}^{\prime}\hat{N}_{2}{\!}\doteq&q_{\mu}\Big{(}A_{2}^{(1)}Z_{2}+\frac{P\cdot q}{q^{2}}A_{1}^{(2)}\Big{)},\\
\hat{p}_{1\mu}^{\prime}\hat{p}_{1\nu}^{\prime}\hat{N}_{2}{\!}\doteq&g_{\mu\nu}A_{1}^{(2)}Z_{2}
+q_{\mu}q_{\nu}\Big{[}A_{4}^{(2)}Z_{2}+2\frac{P\cdot q}{q^{2}}A_{2}^{(1)}A_{1}^{(2)}\Big{]},
\label{eq:replacement}
\end{split}
\end{equation}
in Eqs.~\eqref{eq:SV}, ~\eqref{eq:SA}, ~\eqref{eq:ST1}, and ~\eqref{eq:ST2}. Here, $Z_{2}=\hat{N}_{1}^{\prime}+m_{1}^{\prime2}-m_{2}^{2}+(1-2x_{1})M^{\prime2}+(q^{2}+P\cdot q)\frac{\vec{k}_{\bot}^{\prime}\cdot\vec{q}_{\bot}}{q^{2}}$, $P\cdot q=M^{\prime2}-M^{\prime\prime2}$, and
\begin{equation}
\begin{split}
A_{1}^{(1)}&=\frac{x_{1}}{2},~~~A_{2}^{(1)}=A_{1}^{(1)}-\frac{\vec{k}_{\bot}^{\prime}\cdot\vec{q}_{\bot}}{q^{2}},\\
A_{1}^{(2)}&=-\vec{k}_{\bot}^{\prime2}-\frac{(\vec{k}_{\bot}^{\prime}\cdot\vec{q}_{\bot})^{2}}{q^{2}},~~~A_{2}^{(2)}=(A_{1}^{(1)})^{2},\\
A_{3}^{(2)}&=A_{1}^{(1)}A_{2}^{(1)},~~~A_{4}^{(2)}=(A_{2}^{(1)})^{2}-\frac{1}{q^{2}}A_{1}^{(2)}.
\end{split}
\end{equation}

After performing the replacements \eqref{eq:replacement} in the decay amplitudes \eqref{eq:SV} and \eqref{eq:SA}, the form factors $g$, $f$, $a_{+}$, and $a_{-}$ can be obtained from the terms proportional to the $\epsilon_{\mu\nu\alpha\beta}P^{\alpha}q^{\beta}$, $g_{\mu\nu}$, $P_{\mu}P_{\nu}$ and $P_{\mu}q_{\nu}$, and $q_{\mu}P_{\nu}$ and $q_{\mu}q_{\nu}$, respectively.
The $\epsilon_{V}^{*\mu}P^{\prime\prime}_{\mu}=0$ is used here. Finally, the expressions of these form factors in covariant LFQMs can be written as~\cite{Jaus:1999zv,Cheng:2003sm,Zhang:2023ypl}
\begin{widetext}
\begin{equation}
\begin{split}
g(q^{2})=&-\frac{N_{c}}{16\pi^{3}}\int dx_{2}d^{2}\vec{k}_{2\bot}\frac{2h_{P}^{\prime}h_{V}^{\prime\prime}}{x_{2}\hat{N}_{1}^{\prime}\hat{N}_{1}^{\prime\prime}}
\Bigg{\{}x_{2}m_{1}^{\prime}+x_{1}m_{2}+(m_{1}^{\prime}-m_{1}^{\prime\prime})\frac{k_{\bot}^{\prime}\cdot q_{\bot}}{q^{2}}+\frac{2}{\omega_{V}^{\prime\prime}}\Big{[}k_{\bot}^{\prime2}+\frac{(k_{\bot}^{\prime}\cdot q_{\bot})^{2}}{q^{2}}\Big{]}\Bigg{\}},
\end{split}
\label{eq:formfactorVg}
\end{equation}
\begin{equation}
\begin{split}
f(q^{2})=&\frac{N_{c}}{16\pi^{3}}\int dx_{2}d^{2}\vec{k}_{\bot}^{\prime}\frac{h_{P}^{\prime}h_{V}^{\prime\prime}}{x_{2}\hat{N}_{1}^{\prime}\hat{N}_{1}^{\prime\prime}}
\Bigg{\{}2x_{1}(m_{2}-m_{1}^{\prime})(M_{0}^{\prime2}+M_{0}^{\prime\prime2})-4x_{1}m_{1}^{\prime\prime}M_{0}^{\prime2}+2x_{2}m_{1}^{\prime}P\cdot q+2m_{2}q^{2}\\
&-2x_{1}m_{2}(M^{\prime2}+M^{\prime\prime2})+2(m_{1}^{\prime}-m_{2})(m_{1}^{\prime}+m_{1}^{\prime\prime})^{2}
+8(m_{1}^{\prime}-m_{2})\Big{[}k_{\bot}^{\prime2}+\frac{(k_{\bot}^{\prime}\cdot q_{\bot})^{2}}{q^{2}}\Big{]}\\
&+2(m_{1}^{\prime}+m_{1}^{\prime\prime})(q^{2}+P\cdot q)\frac{k_{\bot}^{\prime}\cdot q_{\bot}}{q^{2}}
-4\frac{q^{2}k_{\bot}^{\prime2}+(k_{\bot}^{\prime}\cdot q_{\bot})^{2}}{q^{2}\omega_{V}^{\prime\prime}}
\Big{[}2x_{1}(M^{\prime2}+M_{0}^{\prime2})-q^{2}-P\cdot q\\
&-2(q^{2}+P\cdot q)\frac{k_{\bot}^{\prime}\cdot q_{\bot}}{q^{2}}-2(m_{1}^{\prime}-m_{1}^{\prime\prime})(m_{1}^{\prime}-m_{2})\Big{]}\Bigg{\}},
\label{eq:formfactorVf}
\end{split}
\end{equation}
\begin{equation}
\begin{split}
a_{+}(q^{2})=&\frac{N_{c}}{16\pi^{3}}\int dx_{2}d^{2}\vec{k}_{\bot}^{\prime}
\frac{2h_{P}^{\prime}h_{V}^{\prime\prime}}{x_{2}\hat{N}_{1}^{\prime}\hat{N}_{1}^{\prime\prime}}
\Bigg{\{}(x_{1}-x_{2})(x_{2}m_{1}^{\prime}+x_{1}m_{2})
-\big{[}2x_{1}m_{2}+m_{1}^{\prime\prime}+(x_{2}-x_{1})m_{1}^{\prime}\big{]}\frac{k_{\bot}^{\prime}\cdot q_{\bot}}{q^{2}}\\
&-2\frac{x_{2}q^{2}+k_{\bot}^{\prime}\cdot q_{\bot}}{x_{2}q^{2}\omega_{V}^{\prime\prime}}
\big{[}k_{\bot}^{\prime}\cdot k_{\bot}^{\prime\prime}+(x_{1}m_{2}+x_{2}m_{1}^{\prime})(x_{1}m_{2}-x_{2}m_{1}^{\prime\prime})\big{]}\Bigg{\}},
\label{eq:formfactorVaplus}
\end{split}
\end{equation}
\begin{equation}
\begin{split}
a_{-}(q^{2})=&\frac{N_{c}}{16\pi^{3}}\int dx_{2}d^{2}\vec{k}_{\bot}^{\prime}
\frac{h_{P}^{\prime}h_{V}^{\prime\prime}}{x_{2}\hat{N}_{1}^{\prime}\hat{N}_{1}^{\prime\prime}}
\Bigg{\{}2(2x_{1}-3)(x_{2}m_{1}^{\prime}+x_{1}m_{2})-8(m_{1}^{\prime}-m_{2})\Big{[}\frac{k_{\bot}^{\prime2}}{q^{2}}+2\frac{(k_{\bot}^{\prime}\cdot q_{\bot})^{2}}{q^{4}}\Big{]}\\
&-\big{[}(14-12x_{1})m_{1}^{\prime}-2m_{1}^{\prime\prime}-(8-12x_{1})m_{2}\big{]}\frac{k_{\bot}^{\prime}\cdot q_{\bot}}{q^{2}}
+\frac{4}{\omega_{V}^{\prime\prime}}\Big{(}
\big{[}M^{\prime2}+M^{\prime\prime2}-q^{2}+2(m_{1}^{\prime}-m_{2})(m_{1}^{\prime\prime}+m_{2})\big{]}\\
&\times(A_{3}^{2}+A_{4}^{(2)}-A_{2}^{1})+Z_{2}(3A_{2}^{(1)}-2A_{4}^{(2)}-1)
+\frac{1}{2}P\cdot q(A_{1}^{(1)}+A_{2}^{(1)}-1)\big{[}x_{1}(q^{2}+P\cdot q)-2M^{\prime2}-2k_{\bot}^{\prime}\cdot q_{\bot}\\
&-2m_{1}^{\prime}(m_{1}^{\prime\prime}+m_{2}-2m_{2}(m_{1}^{\prime}-m_{2}))\big{]}
\Big{[}\frac{k_{\bot}^{\prime2}}{q^{2}}+\frac{(k_{\bot}^{\prime}\cdot q_{\bot})^{2}}{q^{4}}\Big{]}(4A_{2}^{(1)}-3)\Big{)}\Bigg{\}},
\label{eq:formfactorVamin}
\end{split}
\end{equation}
The form factors deduced by (axial)vector currents defined in Eq.~\eqref{eq:ffs2} can thus be evaluated by
\begin{equation}
\begin{split}
V(q^{2})=&-(M^{\prime}+M^{\prime\prime})g(q^{2}),\\
A_{0}(q^{2})=&-\frac{1}{2M^{\prime\prime}}f(q^{2})
-\frac{M^{\prime2}-M^{\prime\prime2}}{2M^{\prime\prime}}a_{+}(q^{2})
-\frac{q^{2}}{2M^{\prime\prime}}a_{-}(q^{2}),\\
A_{1}(q^{2})=&-f(q^{2})/(M^{\prime}+M^{\prime\prime}),~~~A_{2}(q^{2})=(M^{\prime}+M^{\prime\prime})a_{+}(q^{2}).
\label{eq:formfactorV}
\end{split}
\end{equation}

Analogously, we can obtain the concrete expressions of the (pseudo)tensor form factors defined in Eq.~\eqref{eq:ffs3} as~\cite{Cheng:2009ms}
\begin{equation}
\begin{split}
T_{1}(q^{2})=&\frac{N_{c}}{16\pi^{3}}\int dx_{2}d^{2}\vec{k}_{\bot}^{\prime}
\frac{h_{P}^{\prime}h_{V}^{\prime\prime}}{x_{2}\hat{N}_{1}^{\prime}\hat{N}_{1}^{\prime\prime}}
\Bigg{\{}
2A_{1}^{(1)}[M^{\prime2}-M^{\prime\prime2}-2m_{1}^{\prime2}-2\hat{N}_{1}^{\prime}+q^{2}+2(m_{1}^{\prime}m_{2}+m_{1}^{\prime\prime}m_{2}-m_{1}^{\prime}m_{1}^{\prime\prime})]\\
&-8A_{1}^{(2)}+(m_{1}^{\prime}+m_{1}^{\prime\prime})^{2}+\hat{N}_{1}^{\prime}+\hat{N}_{1}^{\prime\prime}-q^{2}+4(M^{\prime2}-M^{\prime\prime2})(A_{2}^{(2)}-A_{3}^{(2)})
+4q^{2}(-A_{1}^{(1)}+A_{2}^{(1)}+A_{3}^{(2)}-A_{4}^{(2)})\\
&-\frac{4}{\omega_{V}^{\prime\prime}}(m_{1}^{\prime}+m_{1}^{\prime\prime})A_{1}^{(2)}
\Bigg{\}},
\label{eq:formfactorT1}
\end{split}
\end{equation}

\begin{equation}
\begin{split}
T_{2}(q^{2})=&T_{1}(q^{2})+\frac{q^{2}}{M^{\prime2}-M^{\prime\prime2}}\frac{N_{c}}{16\pi^{3}}\int dx_{2}d^{2}\vec{k}_{\bot}^{\prime}
\frac{h_{P}^{\prime}h_{V}^{\prime\prime}}{x_{2}\hat{N}_{1}^{\prime}\hat{N}_{1}^{\prime\prime}}
\Bigg{\{}
2A_{2}^{(1)}[M^{\prime2}-M^{\prime\prime2}-2m_{1}^{\prime2}-2\hat{N}_{1}^{\prime}+q^{2}\\
&+2(m_{1}^{\prime}m_{2}+m_{1}^{\prime\prime}m_{2}-m_{1}^{\prime}m_{1}^{\prime\prime})]-8A_{1}^{(2)}-2M^{\prime2}+2m_{1}^{\prime2}+(m_{1}^{\prime}+m_{1}^{\prime\prime})^{2}
+2(m_{2}-2m_{1}^{\prime})m_{2}+3\hat{N}_{1}^{\prime}+\hat{N}_{1}^{\prime\prime}\\
&-q^{2}+2Z_{2}+4(q^{2}-2M^{\prime2}-2M^{\prime\prime2})(A_{2}^{(2)}-A_{3}^{(2)})-4(M^{\prime2}-M^{\prime\prime2})(-A_{1}^{(1)}+A_{2}^{(1)}+A_{3}^{(2)}-A_{4}^{(2)})\\
&-\frac{4}{\omega_{V}^{\prime\prime}}(m_{1}^{\prime\prime}-m_{1}^{\prime}+2m_{2})A_{1}^{(2)}
\Bigg{\}},
\label{eq:formfactorT2}
\end{split}
\end{equation}

\begin{equation}
\begin{split}
T_{3}(q^{2})=&\frac{N_{c}}{16\pi^{3}}\int dx_{2}d^{2}\vec{k}_{\bot}^{\prime}
\frac{h_{P}^{\prime}h_{V}^{\prime\prime}}{x_{2}\hat{N}_{1}^{\prime}\hat{N}_{1}^{\prime\prime}}
\Bigg{\{}
-2A_{2}^{(1)}[M^{\prime2}-M^{\prime\prime2}-2m_{1}^{\prime2}-2\hat{N}_{1}^{\prime}+q^{2}+2(m_{1}^{\prime}m_{2}+m_{1}^{\prime\prime}m_{2}-m_{1}^{\prime}m_{1}^{\prime\prime})]\\
&+8A_{1}^{(2)}+2M^{\prime2}-2m_{1}^{\prime2}-(m_{1}^{\prime}+m_{1}^{\prime\prime})^{2}-2(m_{2}-2m_{1}^{\prime})m_{2}-3\hat{N}_{1}^{\prime}-\hat{N}_{1}^{\prime\prime}+q^{2}
-2Z_{2}-4(q^{2}-M^{\prime2}-3M^{\prime\prime2})\\
&\times(A_{2}^{(2)}-A_{3}^{(2)})+\frac{4}{\omega_{V}^{\prime\prime}}\Big{(}(m_{1}^{\prime\prime}-m_{1}^{\prime}+2m_{2})[A_{1}^{(2)}+(M^{\prime2}-M^{\prime\prime2})(A_{2}^{(2)}+A_{3}^{(2)}-A_{1}^{(1)})]\\
&+(m_{1}^{\prime}+m_{1}^{\prime\prime})(M^{\prime2}-M^{\prime\prime2})(A_{2}^{(1)}-A_{3}^{(2)}-A_{4}^{(2)})+m_{1}^{\prime}(M^{\prime2}-M^{\prime\prime2})(A_{1}^{(1)}+A_{2}^{(1)}-1)\Big{)}
\Bigg{\}}.
\label{eq:formfactorT3}
\end{split}
\end{equation}
\end{widetext}

Following the treatment in Ref. \cite{Cheng:2003sm}, $h_{M}$ is taken as
\begin{equation}
\begin{split}
h_{P}^{\prime}=&(M^{\prime2}-M_{0}^{\prime2})\sqrt{\frac{x_{1}x_{2}}{N_{c}}}\frac{1}{\sqrt{2}\tilde{M}_{0}^{\prime}}\phi_{s}(x_{2},\vec{k}_{\bot}^{\prime}),\\
h_{V}^{\prime\prime}=&(M^{\prime\prime2}-M_{0}^{\prime\prime2})\sqrt{\frac{x_{1}x_{2}}{N_{c}}}\frac{1}{\sqrt{2}\tilde{M}_{0}^{\prime\prime}}\phi_{s}(x_{2},\vec{k}_{\bot}^{\prime\prime}),
\end{split}
\end{equation}
where $\tilde{M}_{0}^{\prime(\prime\prime)}=\sqrt{M_{0}^{\prime(\prime\prime)2}-(m_{1}^{\prime(\prime\prime)}-m_{2})^{2}}$, and $\phi_{s}$ is the space wave function of the pseudoscalar or vector meson.

In the previous theoretical work \cite{Cheng:2003sm,Zhang:2023ypl}, the phenomenological Gaussian-type wave functions
\begin{equation}
\phi_{s}(x_{2},\vec{k}_{\bot}^{\prime(\prime\prime)})=4\Big{(}\frac{\pi}{\beta^{\prime(\prime\prime)2}}\Big{)}^{3/4}{\!}
\sqrt{\frac{e_{1}^{\prime(\prime\prime)}e_{2}}{x_{1}x_{2}M_{0}^{\prime(\prime\prime)}}}
\exp\Big{(}{\!}-{\!}\frac{\vec{k}_{\bot}^{\prime(\prime\prime)2}+k_{z}^{\prime(\prime\prime)2}}{2\beta^{\prime(\prime\prime)2}}\Big{)},
\label{eq:wavefunctionbeta}
\end{equation}
with
\begin{equation}
\begin{split}
k_{z}^{\prime(\prime\prime)}&=\frac{x_{2}M_{0}^{\prime(\prime\prime)}}{2}-\frac{m_{2}^{2}+\vec{k}_{\bot}^{\prime(\prime\prime)2}}{2x_{2}M_{0}^{\prime(\prime\prime)}},\\
e_{1}^{\prime(\prime\prime)}&=\sqrt{m_{1}^{\prime(\prime\prime)2}+\vec{k}_{\bot}^{\prime(\prime\prime)2}+k_{z}^{\prime(\prime\prime)2}},\\
e_{2}&=\sqrt{m_{2}^{2}+\vec{k}_{\bot}^{\prime2}+k_{z}^{\prime2}},
\end{split}
\end{equation}
are widely used. It inevitably introduces the dependence of the parameter $\beta$. The phenomenological parameter $\beta$ can be fixed by the decay constant \cite{Cheng:2003sm,Cheng:2004yj,Cheng:2009ms}. However, as we all know, the decay constant is only associated with the meson wave function at the end point $q^{2}=0$. This indicates that the simple wave function Eq. \eqref{eq:wavefunctionbeta} deviating from the $q^{2}=0$ region may be unreliable.

Taking advantage of the modified GI model \cite{Li:2023wgq}, we can obtain the numerical spatial wave functions of the mesons concerned. By replacing the form in Eq.~\eqref{eq:wavefunctionbeta} with
\begin{equation}
\begin{split}
\phi_{l}(x_{2},\vec{k}_{\bot}^{\prime(\prime\prime)})&=\sqrt{4}\pi\sum_{n=1}^{N_{\text{max}}}c_{n}\sqrt{\frac{e_{1}^{\prime(\prime\prime)}e_{2}}{x_{1}x_{2}M_{0}^{\prime(\prime\prime)}}}
R_{nl}\Big{(}\sqrt{\vec{k}_{\bot}^{\prime(\prime\prime)2}+k_{z}^{\prime(\prime\prime)2}}\Big{)},\\
\phi_{s}(x_{2},\vec{k}_{\bot}^{\prime(\prime\prime)})&\equiv\phi_{l=0}(x_{2},\vec{k}_{\bot}^{\prime(\prime\prime)}),
\label{eq:wavefunctionSHO}
\end{split}
\end{equation}
where $c_{n}$ are the expansion coefficients of the corresponding eigenvectors and $l$ is the orbital angular momentum of the meson, we can avoid the corresponding uncertainty. In Table~\ref{tab:cn}, we collect the expansion coefficients $c_{n}$ of the meson wave functions involved. In addition, the factor $\sqrt{4}\pi$ is needed to satisfy the normalization:
\begin{equation}
\int\frac{dx_{2}d\vec{k}_{\bot}}{2(2\pi)^{3}}\phi_{l}^{*}(x_{2},\vec{k}_{\bot})\phi_{l}(x_{2},\vec{k}_{\bot})=1.
\end{equation}
Besides, the $R_{nl}$ is the simple harmonic oscillator wave function as
\begin{equation}
R_{nl}(\vert p \vert){\!}={\!}\frac{(-1)^{n-1}}{\beta^{3/2}}{\!}\sqrt{\frac{2(n-1)!}{\Gamma(n{\!}+{\!}l{\!}+{\!}1/2)}}
\Bigg{(}\frac{p}{\beta}\Bigg{)}^{l}\exp\Bigg{(}{\!}-{\!}\frac{p^{2}}{2\beta^{2}}\Bigg{)}L_{n{\!}-{\!}1}^{l}\Bigg{(}\frac{p^{2}}{\beta^{2}}\Bigg{)}.
\label{eq:SHO}
\end{equation}
The parameter $\beta=0.5\ \text{GeV}$ in the above equation is consistent with Ref. \cite{Li:2023wgq}.

\begin{table*}[htbp]\centering
\caption{The calculated masses and the expansion coefficients $c_{n}$ of the wave function of the mesons involved \cite{Li:2023wgq}. The masses are given in units of MeV.}
\label{tab:cn}
\renewcommand\arraystretch{1.05}
\begin{tabular*}{170mm}{c@{\extracolsep{\fill}}ccc}
\toprule[1pt]
\toprule[0.5pt]
States    &Masses \cite{Li:2023wgq} &Experiments \cite{ParticleDataGroup:2022pth} &Eigenvector coefficients $c_{n}$ \cite{Li:2023wgq}\\
\midrule[0.5pt]
\multirow{3}*{\shortstack{$B_{c}$}}   &\multirow{3}*{\shortstack{$6271$}}  &\multirow{3}*{\shortstack{$6274.47\pm0.32$}}  &$\Big{\{}0.7877, 0.4410, 0.2857, 0.1991, 0.1470, 0.1132, 0.0900, 0.0734, 0.0611, 0.0517,$\\
&&&$0.0444, 0.0385, 0.0338, 0.0299, 0.0266, 0.0238, 0.0215, 0.0195, 0.0177, 0.0162,$\\
&&&$0.0148, 0.0136, 0.0125, 0.0116, 0.0107, 0.0099, 0.0092, 0.0084, 0.0081, 0.0066, 0.0081\Big{\}}$\\
\specialrule{0em}{2pt}{2pt}
\multirow{3}*{\shortstack{$D_{s}^{*}$}} &\multirow{3}*{\shortstack{$2112$}}  &\multirow{3}*{\shortstack{$2112.2\pm0.4$}}    &$\Big{\{}0.9708, 0.16203, 0.1515, 0.0605, 0.0518, 0.0286, 0.0240, 0.0156, 0.0130, 0.0093,$\\
&&&$0.0078, 0.0059, 0.0050, 0.0039, 0.0033, 0.0027, 0.0023, 0.0019, 0.0016, 0.0013$\\
&&&$0.0012, 0.0010, 0.0008, 0.0007, 0.0006, 0.0005, 0.0005, 0.0004, 0.0003, 0.0003, 0.0003\Big{\}}$\\
\bottomrule[0.5pt]
\bottomrule[1pt]
\end{tabular*}
\end{table*}

\section{Numerical results and discussions}\label{sec04}

\subsection{The form factors}

With the input of the numerical wave functions, and the concrete expressions of the seven form factors in Eqs.~\eqref{eq:formfactorVg}-\eqref{eq:formfactorT3}, we present in this subsection the numerical results of $B_{c}\to D_{s}^{*}$ form factors.

Following the approach described in Refs.~\cite{Jaus:1999zv,Cheng:2003sm}, we assume the condition $q^{+}=0$. This implies that our form factor calculations are performed in the spacelike region ($q^{2}<0$), and therefore we need to extrapolate them to the timelike region ($q^{2}>0$). To perform the analytical continuation, we utilize the $z$-series parametrization~\cite{Chen:2017vgi}
\begin{equation}
\begin{split}
\mathcal{F}(q^{2}){\!}=&{\!}\frac{\mathcal{F}(0)}{1{\!}-{\!}q^{2}/m_{\text{pole}}^{2}}\Bigg{\{}
1{\!}+{\!}a_{1}\Big{(}z(q^{2}){\!}-{\!}z(0){\!}-{\!}\frac{1}{3}\big{[}z(q^{2})^{2}{\!}-{\!}z(0)^{2}\big{]}\Big{)}\\
&+a_{2}\Big{(}z(q^{2}){\!}-{\!}z(0){\!}+{\!}\frac{2}{3}\big{[}z(q^{2})^{2}{\!}-{\!}z(0)^{2}\big{]}\Big{)}
\Bigg{\}},
\end{split}
\label{eq:fittingffs}
\end{equation}
where $a_{i}$ ($i$ = 1, 2) are free parameters needed to fit in the $q^{2}<0$ region, and the $z(q^{2})$ is taken as
\begin{equation}
z(q^{2})=\frac{\sqrt{t_{+}-q^{2}}-\sqrt{t_{+}-t_{0}}}{\sqrt{t_{+}-q^{2}}+\sqrt{t_{+}-t_{0}}}
\end{equation}
with $t_{\pm}=(M^{\prime}\pm M^{\prime\prime})^{2}$ and $t_{0}=t_{+}\Big{(}1-\sqrt{1-t_{-}/t_{+}}\Big{)}$.

\begin{figure*}[htbp]\centering
  \begin{tabular}{cc}
  \includegraphics[width=68mm]{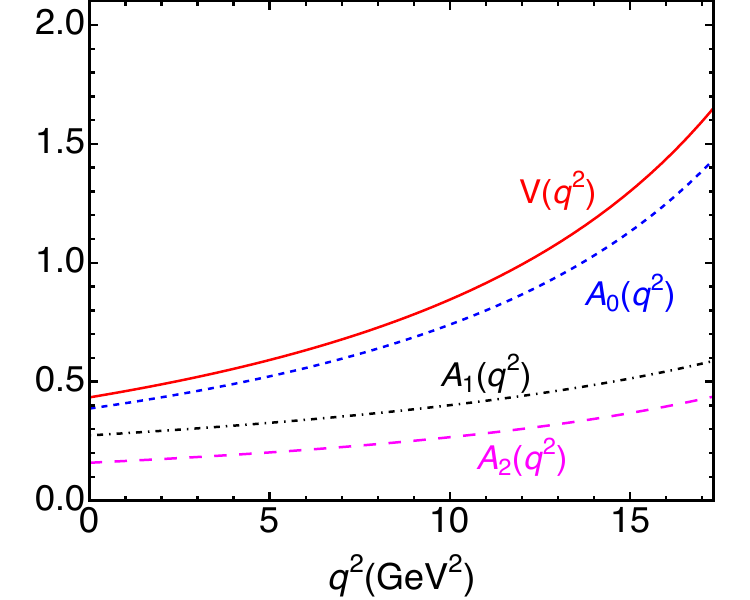}
  \includegraphics[width=68mm]{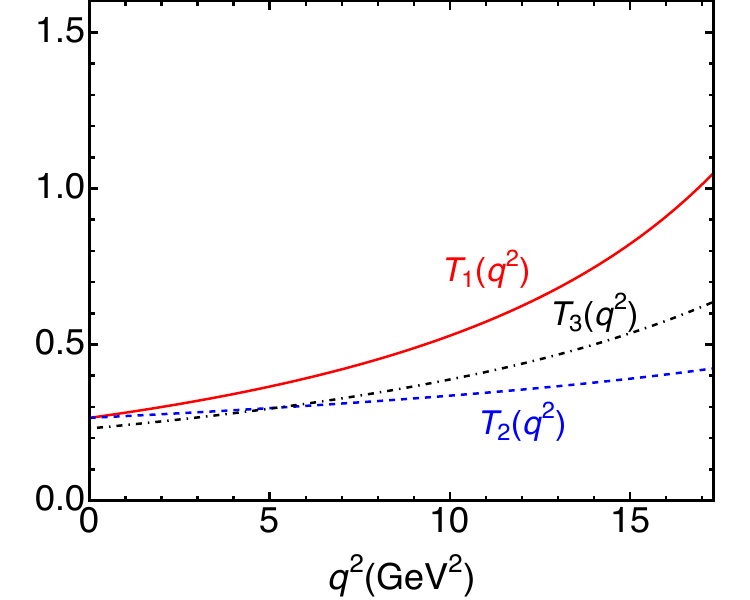}
  \end{tabular}
  \caption{The $q^{2}$ dependence of the $B_{c}\to{D}_{s}^{*}$ weak transition form factors. Here, the four dependent form factors deduced by (axial)vector are presented in the left panel, while the three dependent ones deduced by (pseudo)tensor are shown in the right panel.}
\label{fig:ffs}
\end{figure*}

To determine the values of the free parameters $a_{i}$, as given in Eq.~\eqref{eq:fittingffs}, we perform numerical calculations at 200 equally spaced points for each form factor, ranging from $-20$ to $-0.1\ \text{GeV}^{2}$, using Eqs.~\eqref{eq:formfactorVg}-\eqref{eq:formfactorT3}. The calculated points are then fitted using Eq.~\eqref{eq:fittingffs}. The fitted values of the free parameters, as well as $\mathcal{F}(0)$, $\mathcal{F}(q_{\text{max}}^{2})$, and the pole masses, are listed in Table~\ref{tab:ffs}. Additionally, the $q^{2}$ dependence of the transition form factors $B_{c}\to D_{s}^{*}$ is shown in Fig.~\ref{fig:ffs}.

\begin{table}[htbp]\centering
\caption{Our results of the weak transition form factors of $B_{c}\to D_{s}^{*}$ by using the covariant LFQM.}
\label{tab:ffs}
\renewcommand\arraystretch{1.15}
\begin{tabular*}{86mm}{c@{\extracolsep{\fill}}ccccc}
\toprule[1pt]
\toprule[0.5pt]
                        &$\mathcal{F}(0)$   &$\mathcal{F}(q_{\text{max}}^{2})$   &$m_{\text{pole}}$ (GeV)   &$a_{1}$   &$a_{2}$\\
\midrule[0.5pt]
$V^{B_c\to B_s^{*}}$       &$0.434$    &$1.652$    &$5.415$    &$-7.909$    &$15.667$\\
$A_0^{B_c\to B_s^{*}}$   &$0.387$    &$1.436$    &$5.367$    &$-6.790$    &$9.427$\\
$A_1^{B_c\to B_s^{*}}$   &$0.274$    &$0.588$    &$5.829$    &$-0.721$    &$-4.299$\\
$A_2^{B_c\to B_s^{*}}$   &$0.159$    &$0.438$    &$5.829$    &$-4.942$    &$5.168$\\
$T_1^{B_c\to B_s^{*}}$   &$0.265$    &$1.050$    &$5.415$    &$-8.821$    &$19.272$\\
$T_2^{B_c\to B_s^{*}}$   &$0.265$    &$0.424$    &$5.829$    &$3.067$     &$-9.950$\\
$T_3^{B_c\to B_s^{*}}$   &$0.231$    &$0.637$    &$5.829$    &$-4.985$    &$4.566$\\
\bottomrule[0.5pt]
\bottomrule[1pt]
\end{tabular*}
\end{table}

In Table~\ref{tab:Compareffs}, we compare our results for the $B_{c}\to{D}_{s}^{*}$ weak transition form factors at the end point $q^{2}=0$ with other approaches, in which Refs.~\cite{Kiselev:2002vz,Azizi:2008vv} calculated the concerned form factors with the QCD sum rule,  Refs.~\cite{Geng:2001vy,Wang:2008xt,Zhang:2023ypl} used the covariant LFQM, and Ref.~\cite{Dhir:2008hh} used the Bauer-Stech-Wirbel model and considered the effects of flavor dependence on the form factors caused by possible variation of the average transverse quark momentum ($\omega$) inside the meson. In addition, Ref.~\cite{Geng:2001vy} also used the CQM, and Ref.~\cite{Wang:2014yia} used the pQCD approach. References \cite{Geng:2001vy,Wang:2014yia,Azizi:2008vv} contain the results of (pseudo)tensor form factors. Obviously, our results of the (pseudo)tensor form factors, i.e., $T_{1,2,3}$, at the end point $q^{2}=0$ are consistent with the predictions of pQCD~\cite{Wang:2014yia} and the LFQM~\cite{Geng:2001vy}. We expect further theoretical work, especially LQCD, which is useful to constrain the behavior of the form factors in the low-recoiling region, to test our results.

\begin{table*}[htbp]\centering
\caption{Theoretical predictions of the $B_{c}\to{D}_{s}^{*}$ transition form factors at the end point $q^{2}=0$ using different approaches.}
\label{tab:Compareffs}
\renewcommand\arraystretch{1.15}
\begin{threeparttable}
\begin{tabular*}{168mm}{c@{\extracolsep{\fill}}ccccccc}
\toprule[1pt]
\toprule[0.5pt]
                 &$V^{B_c\to{D}_s^{*}}(0)$  &$A_0^{B_c\to{D}_s^{*}}(0)$  &$A_1^{B_c\to{D}_s^{*}}(0)$  &$A_2^{B_c\to{D}_s^{*}}(0)$  &$T_1^{B_c\to{D}_s^{*}}(0)$  &$T_2^{B_c\to{D}_s^{*}}(0)$  &$T_3^{B_c\to{D}_s^{*}}(0)$\\
\midrule[0.5pt]
This work        &$0.434$  &$0.387$  &$0.274$  &$0.159$ &$0.265$  &$0.265$  &$0.231$\\
Ref. \cite{Kiselev:2002vz}       &$2.02$   &$0.47$  &$0.56$  &$0.65$  &$\cdots$  &$\cdots$  &$\cdots$\\
Ref. \cite{Dhir:2008hh}$^{[1]}$  &$0.032$  &$0.016$  &$0.015$  &$0.013$  &$\cdots$  &$\cdots$  &$\cdots$\\
Ref. \cite{Dhir:2008hh}$^{[2]}$  &$0.29^{+0.02}_{-0.03}$  &$0.16^{+0.01}_{-0.01}$  &$0.18^{+0.01}_{-0.02}$  &$0.20^{+0.02}_{-0.03}$  &$\cdots$  &$\cdots$  &$\cdots$\\
Ref. \cite{Wang:2008xt}          &$0.23^{+0.04}_{-0.03}$  &$0.17^{+0.01}_{-0.01}$  &$0.14^{+0.02}_{-0.01}$  &$0.12^{+0.02}_{-0.02}$  &$\cdots$  &$\cdots$  &$\cdots$\\
Ref. \cite{Zhang:2023ypl}        &$0.25^{+0.00}_{-0.00}$  &$0.18^{+0.02}_{-0.03}$  &$0.16^{+0.01}_{-0.02}$  &$0.15^{+0.01}_{-0.01}$  &$\cdots$  &$\cdots$  &$\cdots$\\
Ref. \cite{Geng:2001vy}          &$0.336$  &$0.164$  &$0.118$  &$\cdots$  &$0.214$  &$0.214$  &$\cdots$\\
Ref. \cite{Geng:2001vy}          &$0.262$  &$0.139$  &$0.144$  &$\cdots$  &$0.167$  &$0.167$  &$\cdots$\\
Ref. \cite{Azizi:2008vv}         &$0.54\pm0.018$  &$0.30\pm0.017$  &$0.36\pm0.013$  &$\cdots$  &$0.31\pm0.017$  &$0.33\pm0.016$  &$0.29\pm0.034$\\
Ref. \cite{Wang:2014yia}         &$0.33\pm0.06$  &$0.21\pm0.04$  &$0.23\pm0.04$  &$0.25\pm0.05$  &$0.28\pm0.06$  &$0.28\pm0.06$  &$0.27\pm0.06$\\
\bottomrule[0.5pt]
\bottomrule[1pt]
\end{tabular*}
\begin{tablenotes}
\footnotesize
\item[1] These results, listed in the fourth row, are obtained by using the universe parameter $\omega=0.40$ GeV.
\item[2] These results, listed in the fifth row, are obtained by using different parameters, i.e., $\omega=0.96^{+0.08}_{-0.07}$ GeV for the $B_{c}$ meson and $\omega=0.51$ GeV for the $D_{s}^{*}$ meson.
\end{tablenotes}
\end{threeparttable}
\end{table*}

\subsection{The angular distributions and physical observables}

With the above preparations, in this subsection we present our numerical results of the branching fractions and some angular observables, i.e., the $CP$-averaged normalized angular coefficients $S_{i}$, the lepton's forward-backward asymmetry parameter $A_{FB}$, and the longitudinal (transverse) polarization fractions of the $D_{s}^{*}$ meson $F_{L(T)}$. In addition, we also investigate the clean angular observables $P_{1,2,3}$ and $P^{\prime}_{4,5,6,8}$. The hadron and lepton masses are quoted from the PDG~\cite{ParticleDataGroup:2022pth}, as well as the lifetime $\tau_{B_{c}}=0.510$ps and the branching fraction $\mathcal{B}(D_{s}^{*}\to{D}_{s}\pi)=5\%$.

\begin{figure*}[htbp]\centering
  \begin{tabular}{lccr}
  \includegraphics[width=42mm]{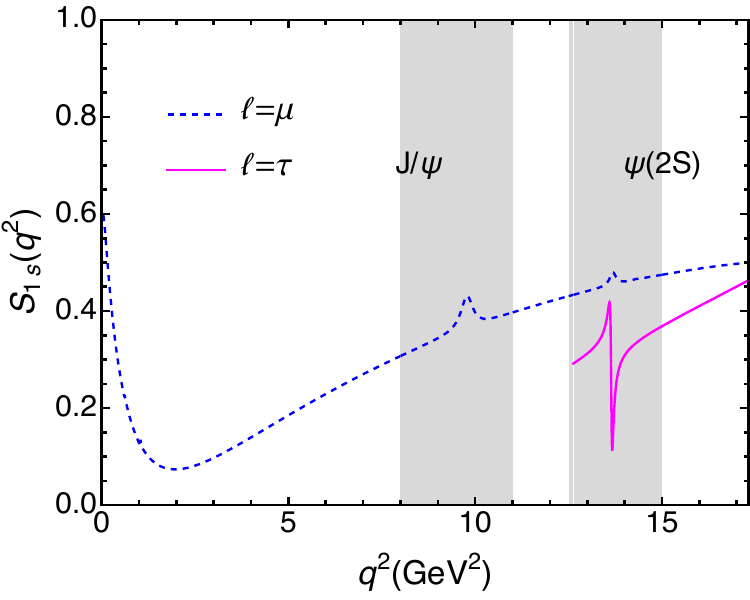}
  \includegraphics[width=42mm]{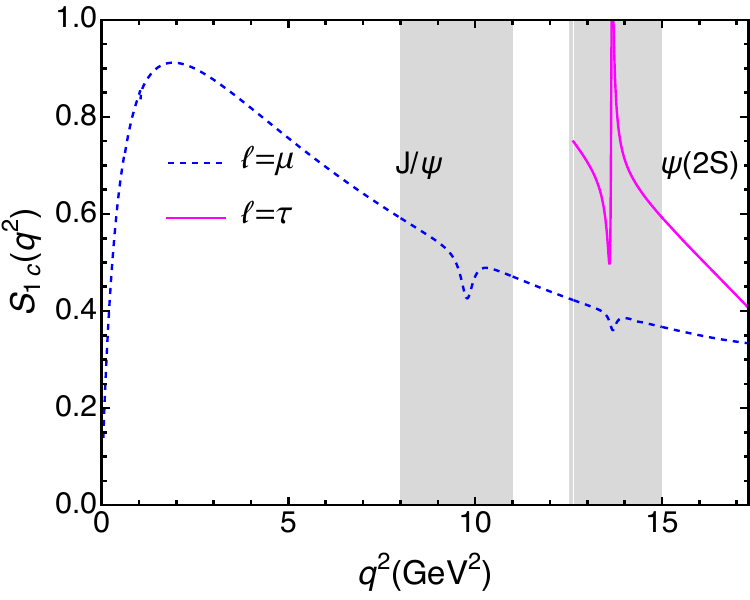}
  \includegraphics[width=42mm]{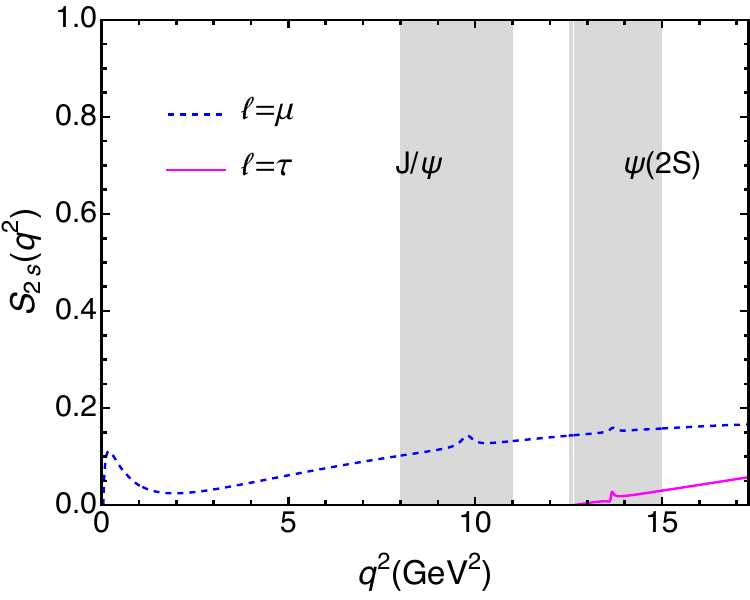}
  \includegraphics[width=42mm]{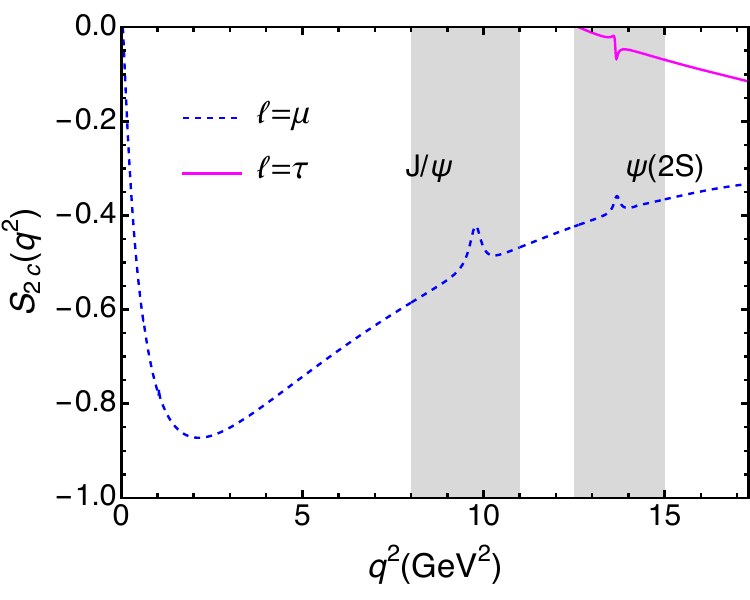}\\
  \includegraphics[width=42mm]{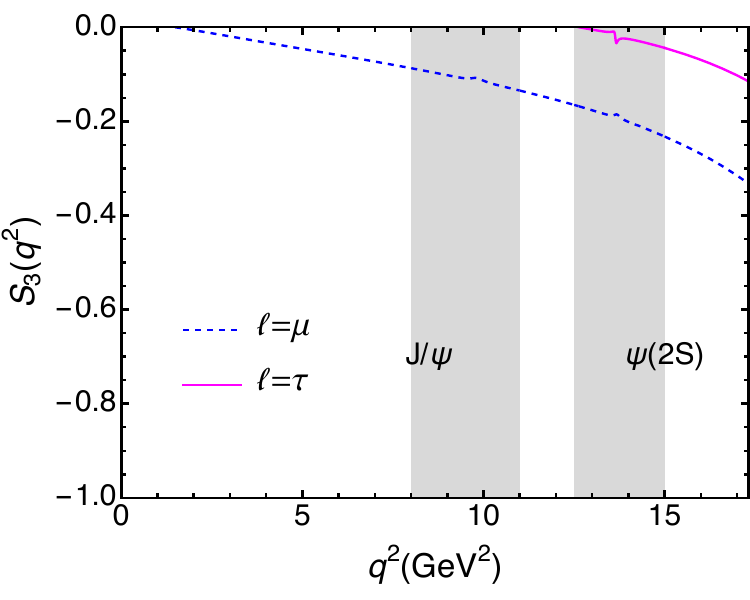}
  \includegraphics[width=42mm]{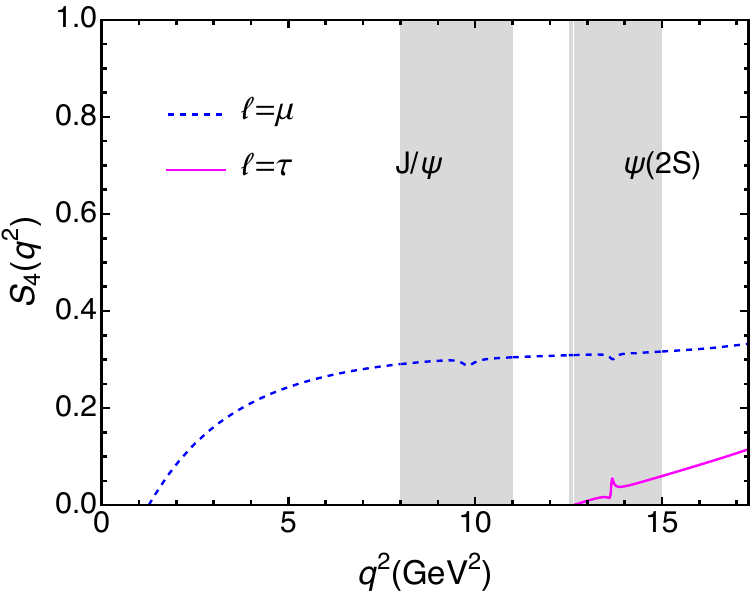}
  \includegraphics[width=42mm]{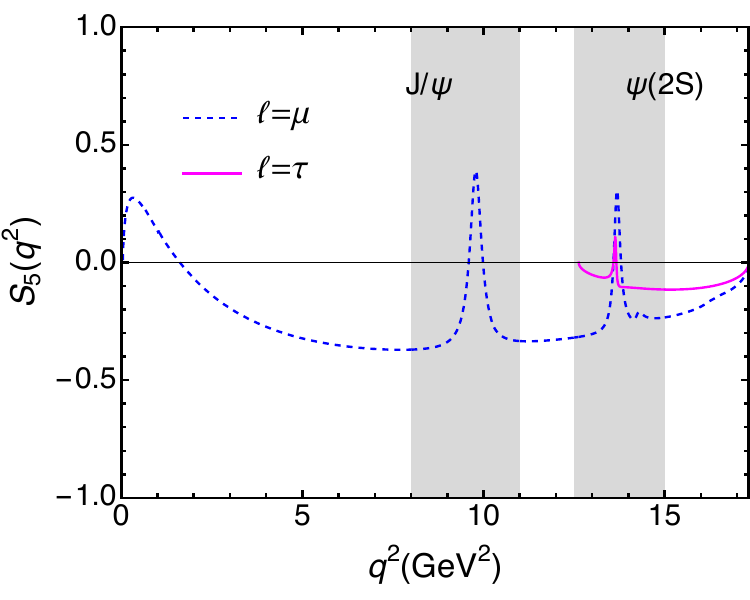}
  \includegraphics[width=42mm]{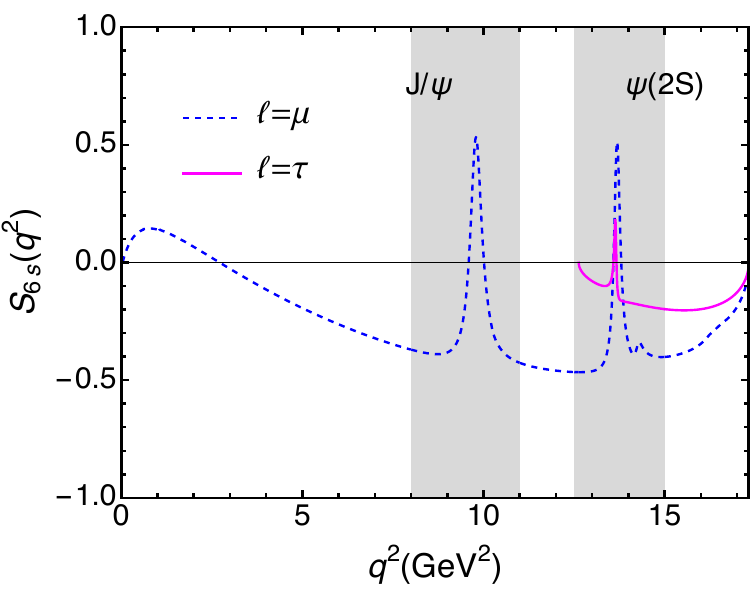}\\
  \includegraphics[width=42mm]{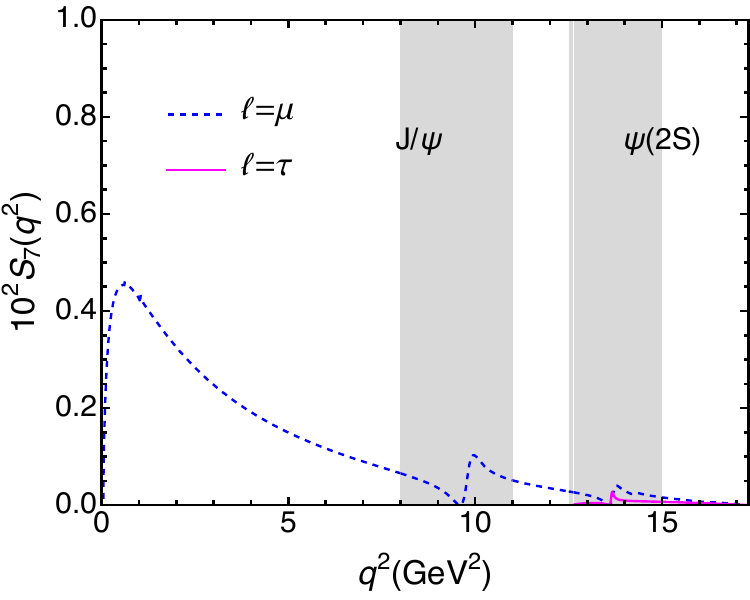}
  \includegraphics[width=42mm]{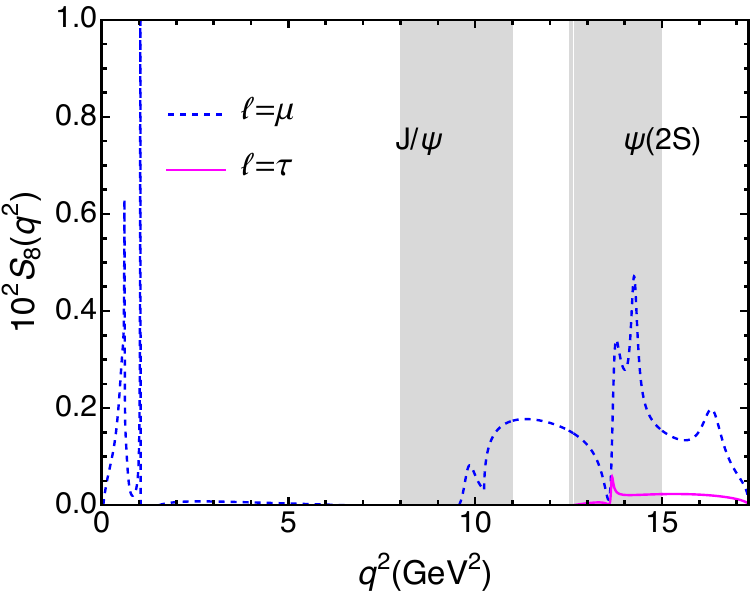}
  \includegraphics[width=42mm]{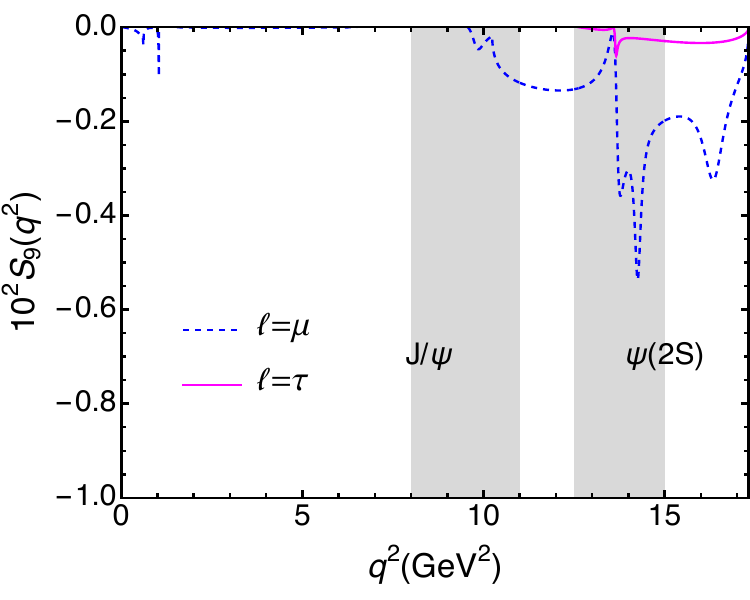}
  \end{tabular}
  \caption{The $q^{2}$ dependence of normalized $CP$-averaged angular coefficients $S_{i}$, where the blue dashed and magenta solid curves are our results for the $\mu$ and $\tau$ modes, respectively.}
\label{fig:Si}
\end{figure*}

\begin{figure*}[htbp]\centering
  \begin{tabular}{lccr}
  \includegraphics[width=42mm]{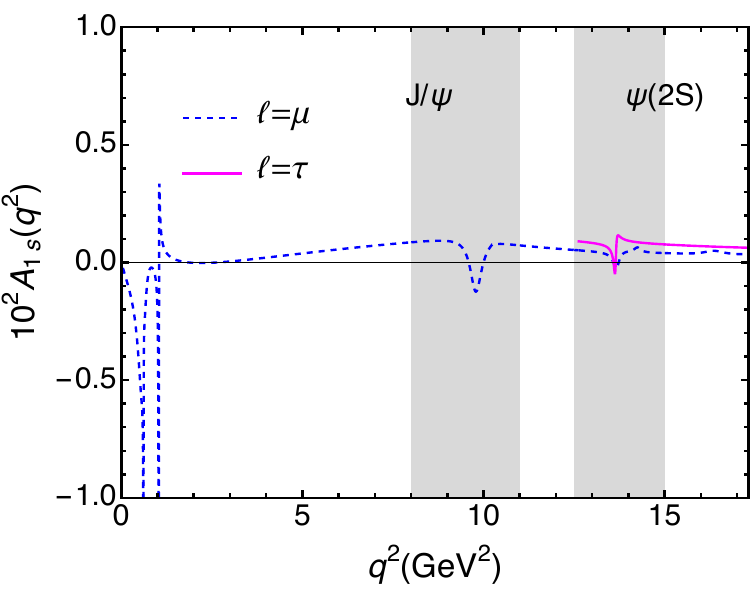}
  \includegraphics[width=42mm]{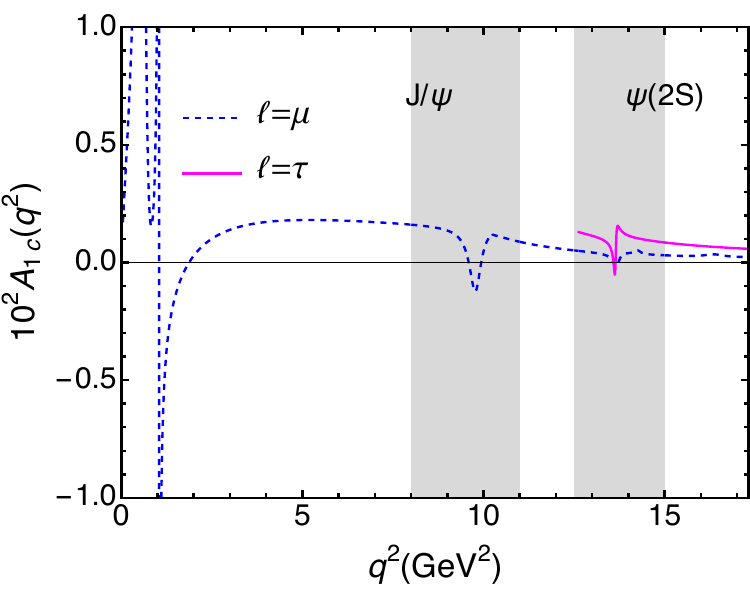}
  \includegraphics[width=42mm]{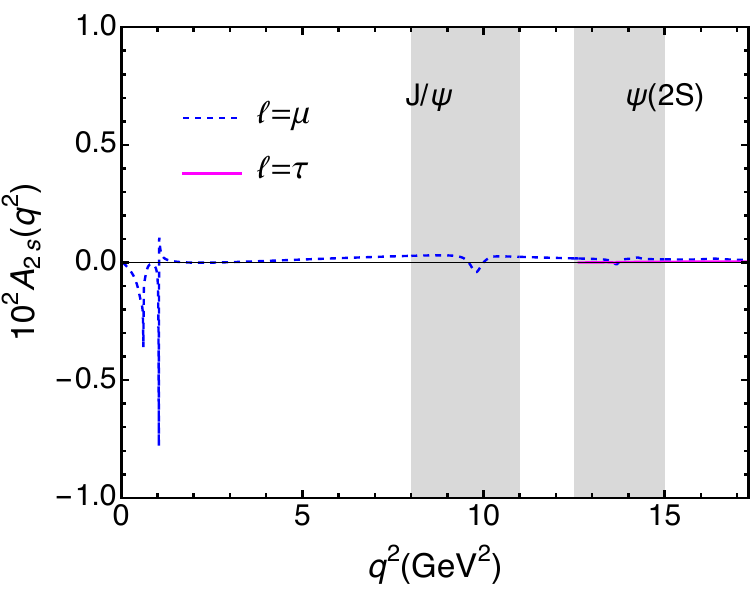}
  \includegraphics[width=42mm]{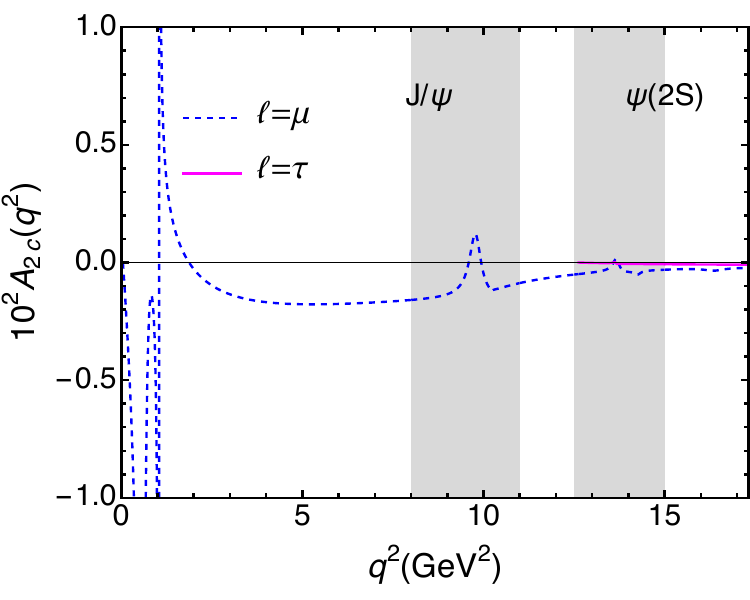}\\
  \includegraphics[width=42mm]{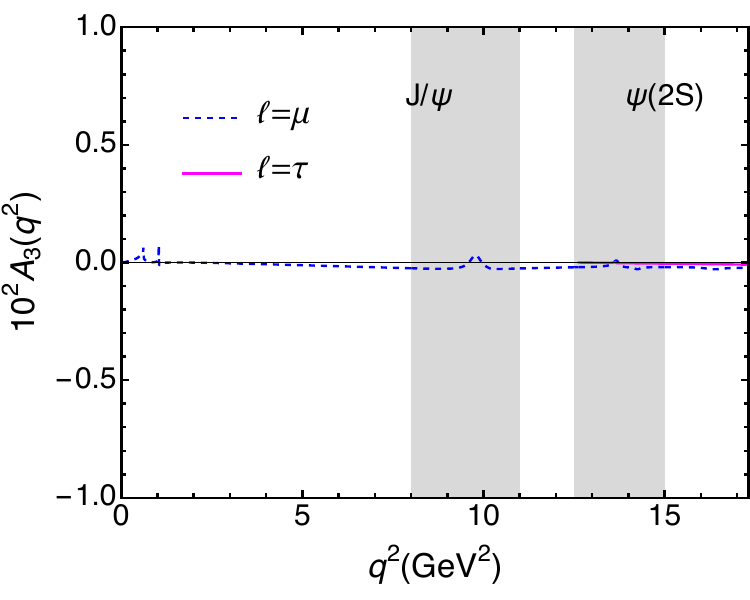}
  \includegraphics[width=42mm]{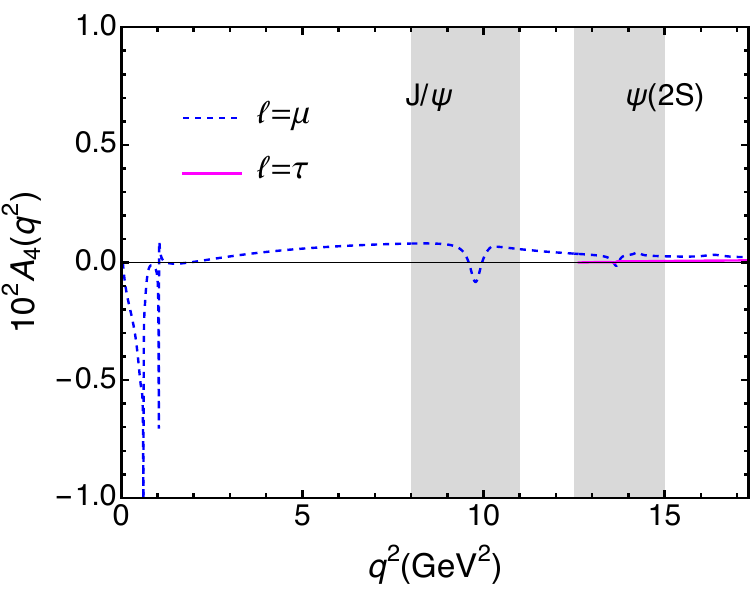}
  \includegraphics[width=42mm]{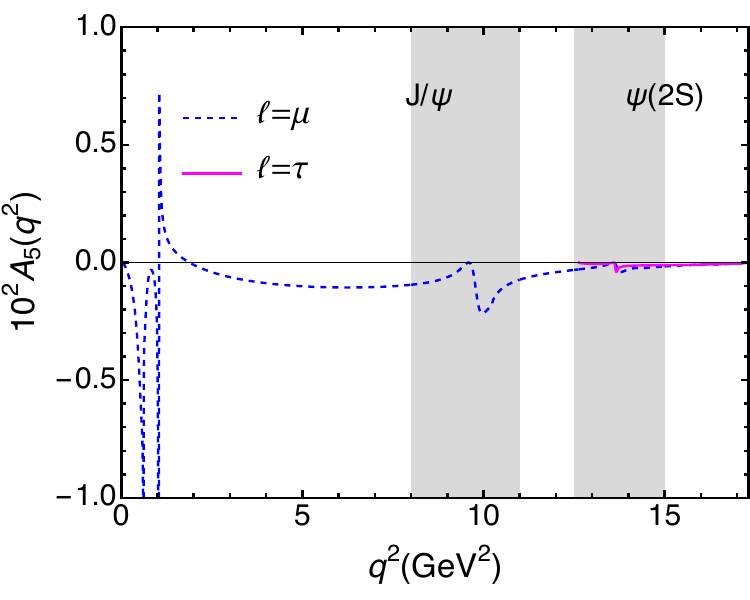}
  \includegraphics[width=42mm]{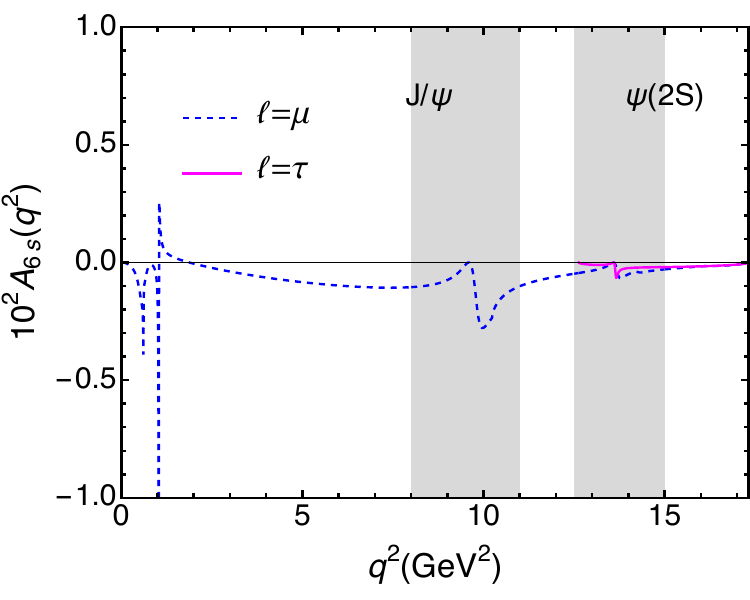}\\
  \includegraphics[width=42mm]{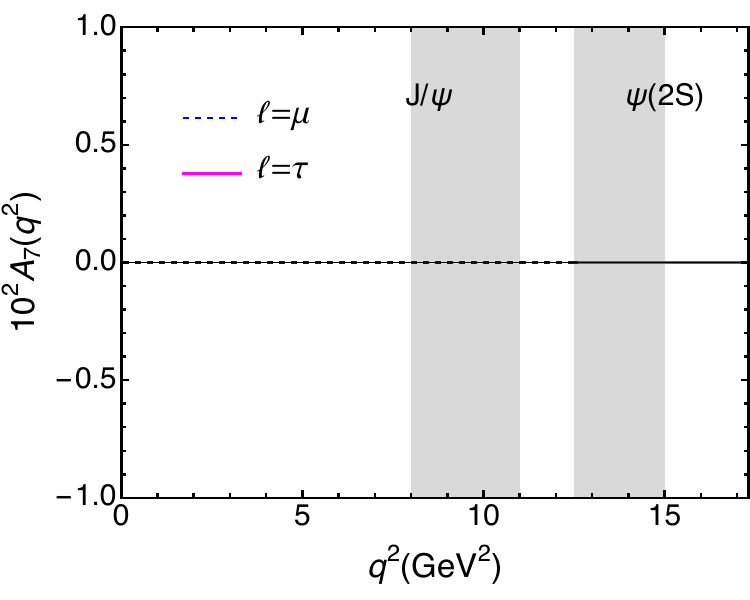}
  \includegraphics[width=42mm]{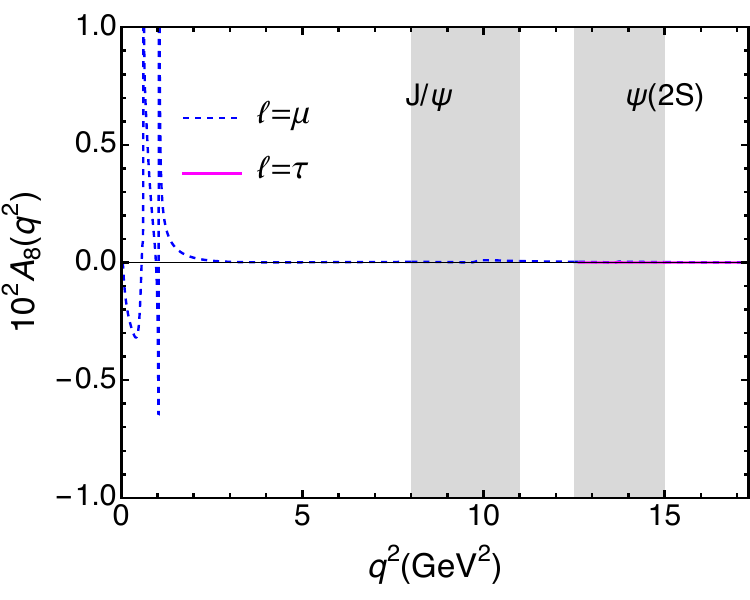}
  \includegraphics[width=42mm]{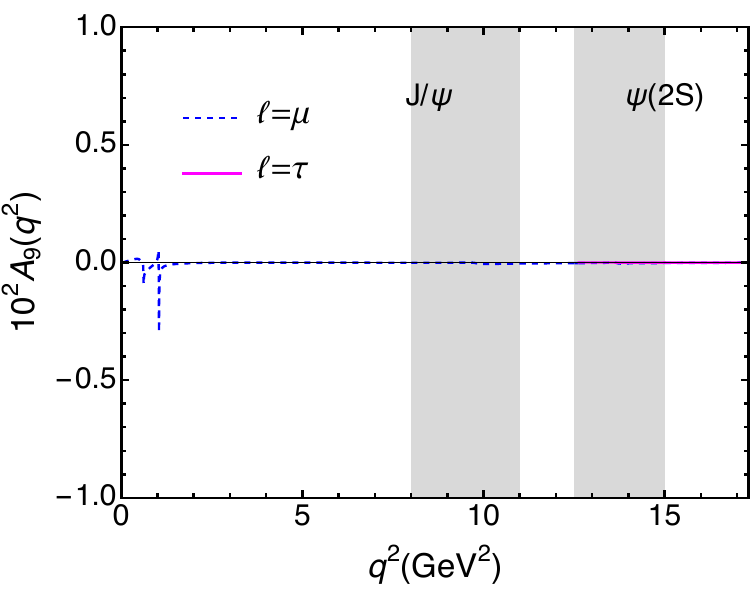}
  \end{tabular}
  \caption{The $q^{2}$ dependence of normalized $CP$ asymmetry angular coefficients $A_{i}$, where the blue dashed and magenta solid curves are our results for the $\mu$ and $\tau$ modes, respectively.}
\label{fig:Ai}
\end{figure*}

First, we focus on the angular coefficients $S_{i}$ and $A_{i}$ defined in Eq.~\eqref{eq:SiAi}. The $q^{2}$ dependence of the normalized $CP$-averaged angular coefficients $S_{i}$ are presented in Fig. \ref{fig:Si}, while the $CP$ asymmetry angular coefficients $A_{i}$ are shown in Fig.~\ref{fig:Ai}. The blue dashed lines and the magenta solid lines represent the muon and the tau channels, respectively. Since the electron channel shows similar behavior to the muon channel, we will only present our results for the muon and the tau channels here. In the energy regions of $8.0<q^{2}<11.0$ and $12.5<q^{2}<15.0~\text{GeV}^{2}$, we use the gray areas to mark the contributions from the charmonium states $J/\psi$ and $\psi(2S)$. {In our calculation, we adopted phenomenological and model-dependent treatment, i.e., the Breit-Wigner ansatz to model the corresponding contribution.} In the experiment, these two regions are generally truncated. The $CP$ asymmetry angular coefficients, $A_{i}$, are shown to be very small in the SM, due to the direct $CP$ violation being proportional to the $\text{Im}[V_{ub}V_{us}^{*}/V_{tb}V_{ts}^{*}]$, which is around $10^{-2}$. This character is very clear in Fig.~\ref{fig:Ai}. Also, the $S_{7,8,9}$ are also very small compared to the other angular coefficients $S_{i}$. These angular coefficients are important physical observables to reveal the underlying decay mechanism, and can be checked by future measurements at LHCb.

\begin{figure*}[htbp]\centering
  \begin{tabular}{cc}
  \includegraphics[width=52mm]{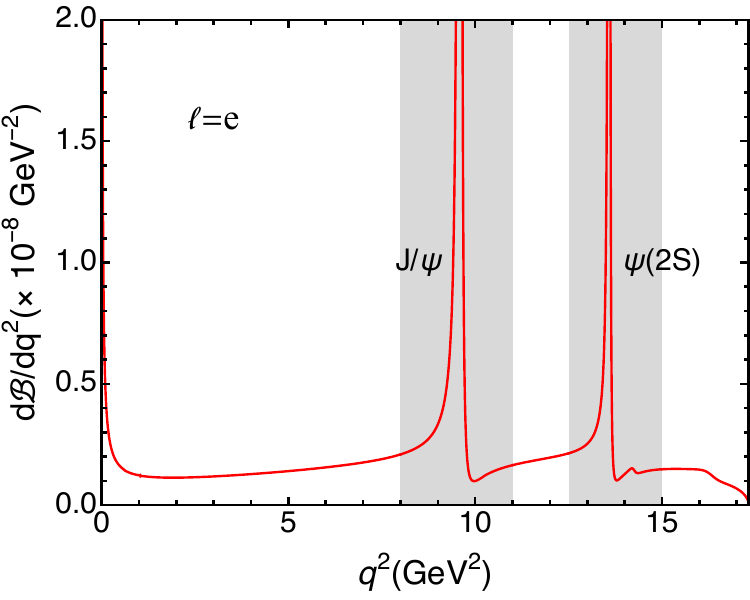}
  \includegraphics[width=52mm]{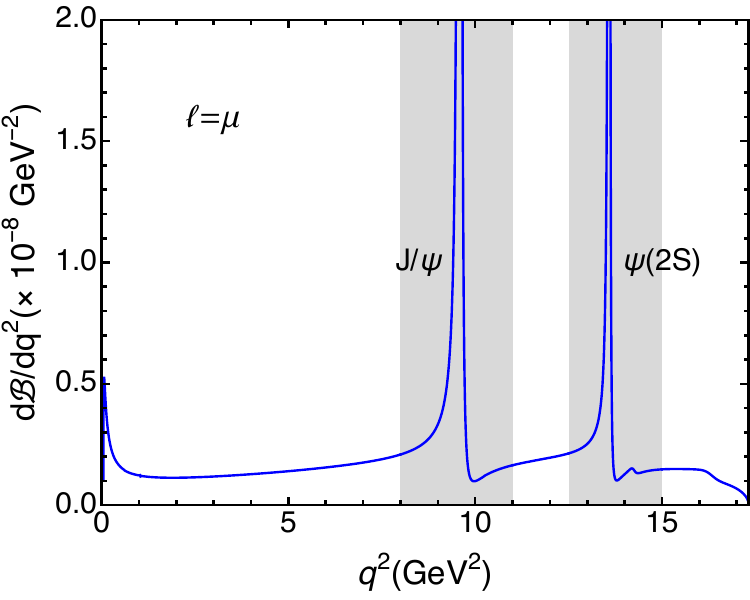}
  \includegraphics[width=52mm]{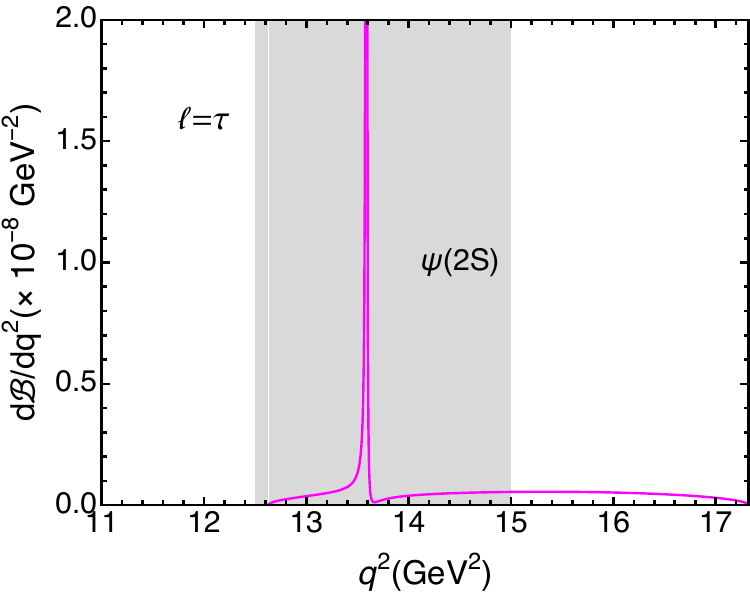}
  \end{tabular}
  \caption{The $q^{2}$ dependence of differential branching fractions $\mathcal{B}(B_{c}\to D_{s}^{*}(\to D_{s}\pi)\ell^{+}\ell^{-})$ [$\ell$=$e$ (left panel), $\mu$ (center panel), and $\tau$ (right panel)], where the red, blue, and magenta curves are our results for the $e$, $\mu$, and $\tau$ modes, respectively.}
\label{fig:BrCP}
\end{figure*}

We further evaluate the $CP$-averaged differential branching fractions by using Eqs.~\eqref{eq:Br} and \eqref{eq:BrCP}. The $q^{2}$ dependence of the differential branching fractions are shown in Fig.~\ref{fig:BrCP}, where the red, blue and magenta curves represent the $e$, $\mu$, and $\tau$ modes, respectively. The gray areas also denote the charm loop contributions from the charmonium states $J/\psi$ and $\psi(2S)$. In Table~\ref{tab:binBr}, we present our result of the branching fractions and their ratios in different $q^{2}$ bins. In the four $q^{2}$ intervals, i.e., $[1.1,6.0]$, $[6.0, 8.0]$, $[11.0, 12.5]$, and $[15.0, 17.0]\ \text{GeV}^{2}$, the branching fractions of the  electron and muon modes can reach up to $10^{-8}$, and the ratio $R^{e\mu}=1$, which is consistent with the SM prediction and reflects the LFU. In the high $q^{2}$ region, that is $[15.0, 17.0]\ \text{GeV}^{2}$, the branching fraction of the tau mode is on the order of magnitude of $10^{-9}$. We also obtain the ratio $R^{\tau\mu}=0.384$. {In the region of $1.1<q^{2}<6.0~\text{GeV}^{2}$, we have the branching fractions as
\begin{eqnarray*}
\mathcal{B}(B_{c}\to{D_{s}^{*}}(\to D_{s}\pi){e}^{+}{e}^{-})_{1.1<q^{2}<6.0\text{GeV}^{2}}&=&0.624\times10^{-8},\\
\mathcal{B}(B_{c}\to{D_{s}^{*}}(\to D_{s}\pi){\mu}^{+}{\mu}^{-})_{1.1<q^{2}<6.0\text{GeV}^{2}}&=&0.622\times10^{-8}.\\
\end{eqnarray*}
In addition, combined with the branching fraction $\mathcal{B}(D_{s}^{*}\to{D_{s}}\pi)=5\%$, we have
\begin{eqnarray*}
\mathcal{B}(B_{c}\to{D_{s}^{*}}{e}^{+}{e}^{-})_{1.1<q^{2}<6.0\text{GeV}^{2}}&=&1.25\times10^{-7},\\
\mathcal{B}(B_{c}\to{D_{s}^{*}}{\mu}^{+}{\mu}^{-})_{1.1<q^{2}<6.0\text{GeV}^{2}}&=&1.24\times10^{-7},\\
\end{eqnarray*}
which may well be tested by the ongoing LHCb experiment.}

\begin{table}[htbp]\centering
\caption{Our results of the branching fractions of $B_{c}\to{D}_{s}^{*}(\to{D}_{s}\pi)\ell^{+}\ell^{-}$ ($\ell$= $e$, $\mu$, $\tau$) (in units of $10^{-8}$) in different $q^{2}$ bins.}
\label{tab:binBr}
\renewcommand\arraystretch{1.05}
\begin{tabular*}{86mm}{c@{\extracolsep{\fill}}ccc}
\toprule[1pt]
\toprule[0.5pt]
$q^{2}$ bins $(\text{GeV}^{2})$  &$\mathcal{B}(\ell=e)$  &$\mathcal{B}(\ell=\mu)$  &$\mathcal{B}(\ell=\tau)$\\
\midrule[0.5pt]
\multirow{2}*{\shortstack{$[1.1, 6.0]$}}   &$0.624$   &$0.622$   &\\
    &\multicolumn{3}{c}{\shortstack{$R^{e\mu}=1.00$}}\\
\specialrule{0em}{3pt}{3pt}
\multirow{2}*{\shortstack{$[6.0, 8.0]$}}   &$0.356$   &$0.355$   &\\
    &\multicolumn{3}{c}{\shortstack{$R^{e\mu}=1.00$}}\\
\specialrule{0em}{3pt}{3pt}
\multirow{2}*{\shortstack{$[11.0, 12.5]$}}   &$0.283$   &$0.283$   &\\
    &\multicolumn{3}{c}{\shortstack{$R^{e\mu}=1.00$}}\\
\specialrule{0em}{3pt}{3pt}
\multirow{2}*{\shortstack{$[15.0, 17.0]$}}   &$0.256$   &$0.256$   &$0.098$\\
    &\multicolumn{3}{c}{\shortstack{$R^{e\mu}=1.00,~~~R^{\tau\mu}=0.384$}}\\
\bottomrule[0.5pt]
\bottomrule[1pt]
\end{tabular*}
\end{table}

\begin{figure*}[htbp]\centering
  \begin{tabular}{cc}
  \includegraphics[width=52mm]{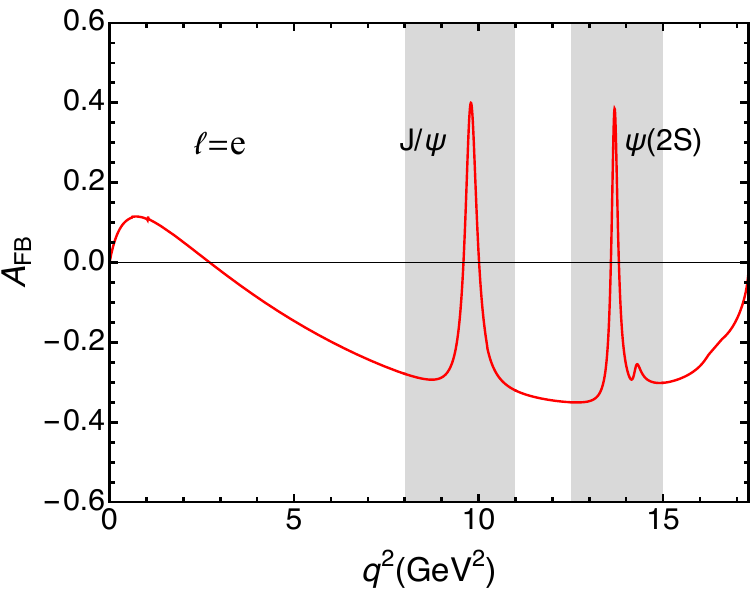}
  \includegraphics[width=52mm]{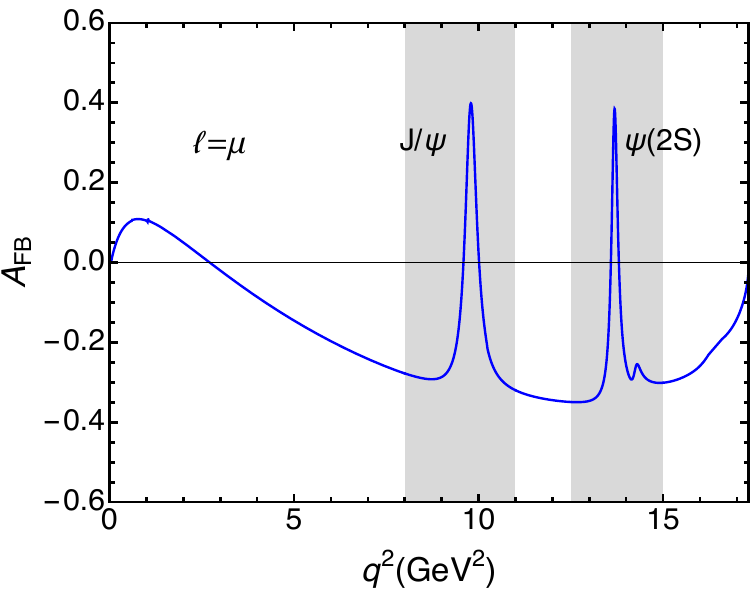}
  \includegraphics[width=52mm]{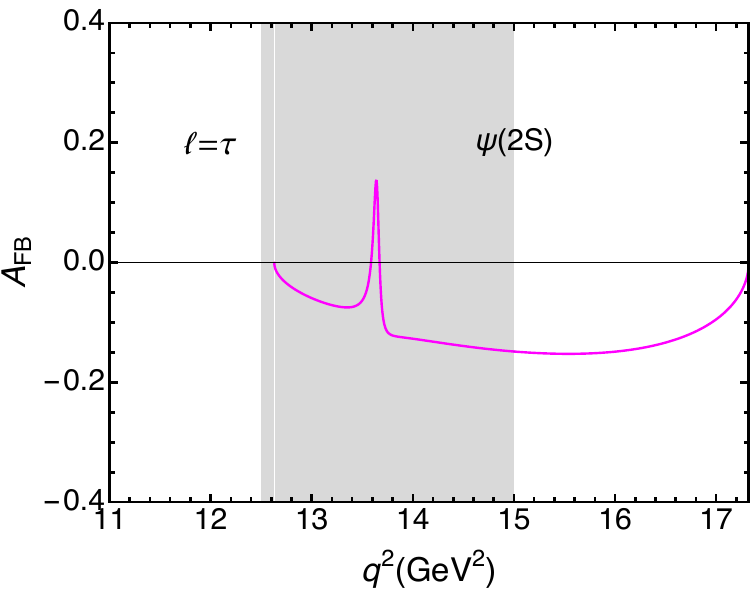}
  \end{tabular}
  \caption{The $q^{2}$ dependence of lepton forward-backward asymmetry parameter $A_{FB}$ in $B_{c}\to D_{s}^{*}(\to D_{s}\pi)\ell^{+}\ell^{-}$ [$\ell$=$e$ (left panel), $\mu$ (center panel), and $\tau$ (right panel)] processes, where the red, the blue, and the magenta curves are our results from the $e$, $\mu$, and $\tau$ modes, respectively.}
\label{fig:AFB}
\end{figure*}

\begin{figure*}[htbp]\centering
  \begin{tabular}{cc}
  \includegraphics[width=52mm]{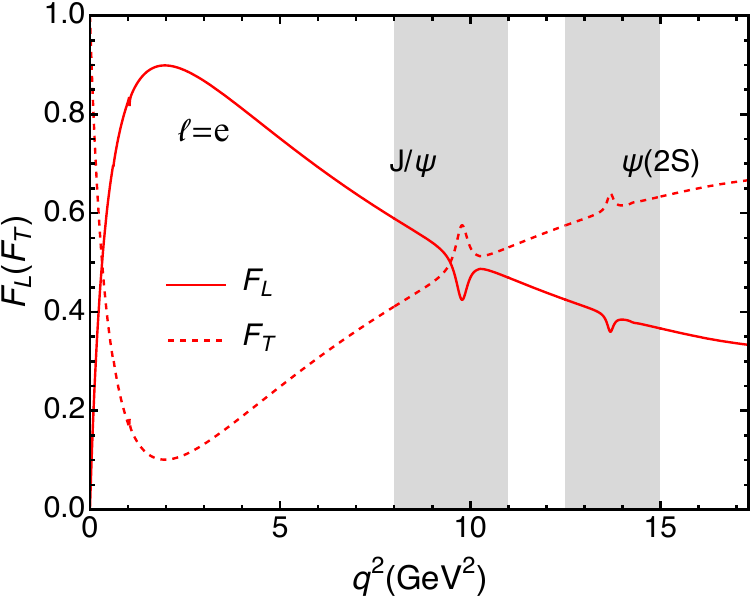}
  \includegraphics[width=52mm]{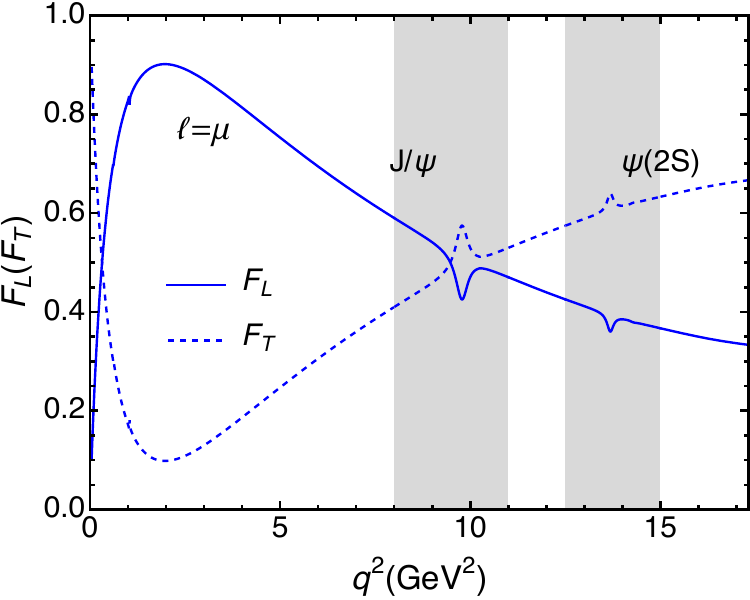}
  \includegraphics[width=52mm]{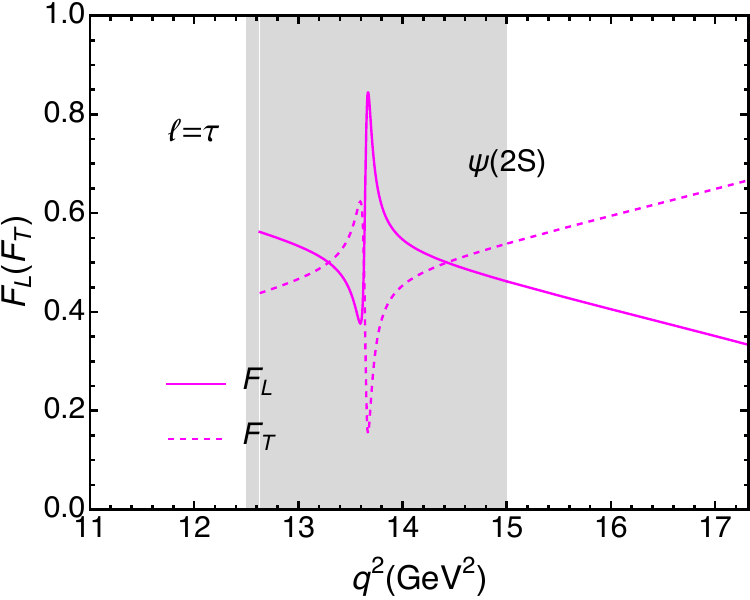}
  \end{tabular}
  \caption{The $q^{2}$ dependence of $D_{s}^{*}$ longitudinal (transverse) polarization fractions $F_{L}(F_{T})$ in $B_{c}\to D_{s}^{*}(\to D_{s}\pi)\ell^{+}\ell^{-}$ [$\ell$=$e$ (left panel), $\mu$ (center panel), and $\tau$ (right panel)] processes, where the red, the blue, and the magenta curves are our results from the $e$, $\mu$, and $\tau$ modes, respectively, and the solid and dashed curves represent the $F_{L}$ and $F_{T}$, respectively.}
\label{fig:FL}
\end{figure*}

We also investigate the physical observables, i.e., the lepton forward-backward asymmetry parameter $A_{FB}$ and the longitudinal (transverse) polarization fractions $F_{L}(F_{T})$. The $q^{2}$ dependence of these observables is presented in Figs.~\ref{fig:AFB} and \ref{fig:FL}, respectively. Their averaged values in different $q^{2}$ bins, defined by
\begin{equation}
\langle{A}\rangle\big{\vert}_{q_{\text{min}}^{2}}^{q_{\text{max}}^{2}}=
\frac{\int_{q_{\text{min}}^{2}}^{q_{\text{max}}^{2}}A[q^{2}]\Big{(}\frac{d\Gamma}{dq^{2}}+\frac{d\bar{\Gamma}}{dq^{2}}\Big{)}dq^{2}}{\int_{q_{\text{min}}^{2}}^{q_{\text{max}}^{2}}\Big{(}\frac{d\Gamma}{dq^{2}}+\frac{d\bar{\Gamma}}{dq^{2}}\Big{)}dq^{2}},
\label{eq:AveragedA}
\end{equation}
where $A=(A_{FB},\ F_{L},\ F_{T})$, are shown in Table \ref{tab:binA}.

\begin{table}[htbp]\centering
\caption{The averaged forward-backward asymmetry $\langle{A_{FB}}\rangle$ and the longitudinal (transverse) polarization fractions $\langle{F_{L}}\rangle(\langle{F_{T}}\rangle)$ in different $q^{2}$ bins.}
\label{tab:binA}
\renewcommand\arraystretch{1.15}
\begin{tabular*}{86mm}{c@{\extracolsep{\fill}}ccc}
\toprule[1pt]
\toprule[0.5pt]
$q^{2}$ bins $(\text{GeV}^{2})$   &$\langle{A_{FB}(\ell=e)}\rangle$   &$\langle{A_{FB}(\ell=\mu)}\rangle$   &$\langle{A_{FB}(\ell=\tau)}\rangle$\\
\midrule[0.5pt]
$[1.1, 6.0]$   &$-0.061$   &$-0.061$   &\\
$[6.0, 8.0]$   &$-0.243$   &$-0.242$   &\\
$[11.0, 12.5]$   &$-0.340$   &$-0.339$   &\\
$[15.0, 17.0]$   &$-0.254$   &$-0.254$   &$-0.143$\\
\midrule[0.5pt]
$q^{2}$ bins $(\text{GeV}^{2})$   &$\langle{F_{L}(\ell=e)}\rangle$   &$\langle{F_{L}(\ell=\mu)}\rangle$   &$\langle{F_{L}(\ell=\tau)}\rangle$\\
\midrule[0.5pt]
$[1.1, 6.0]$   &$0.815$   &$0.817$   &\\
$[6.0, 8.0]$   &$0.637$   &$0.638$   &\\
$[11.0, 12.5]$   &$0.446$   &$0.446$   &\\
$[15.0, 17.0]$   &$0.352$   &$0.352$   &$0.410$\\
\midrule[0.5pt]
$q^{2}$ bins $(\text{GeV}^{2})$   &$\langle{F_{T}(\ell=e)}\rangle$   &$\langle{F_{T}(\ell=\mu)}\rangle$   &$\langle{F_{T}(\ell=\tau)}\rangle$\\
\midrule[0.5pt]
$[1.1, 6.0]$   &$0.185$   &$0.183$   &\\
$[6.0, 8.0]$   &$0.363$   &$0.362$   &\\
$[11.0, 12.5]$   &$0.554$   &$0.554$   &\\
$[15.0, 17.0]$   &$0.648$   &$0.648$   &$0.590$\\
\bottomrule[0.5pt]
\bottomrule[1pt]
\end{tabular*}
\end{table}

\begin{figure*}[htbp]\centering
  \begin{tabular}{lccc}
  \includegraphics[width=42mm]{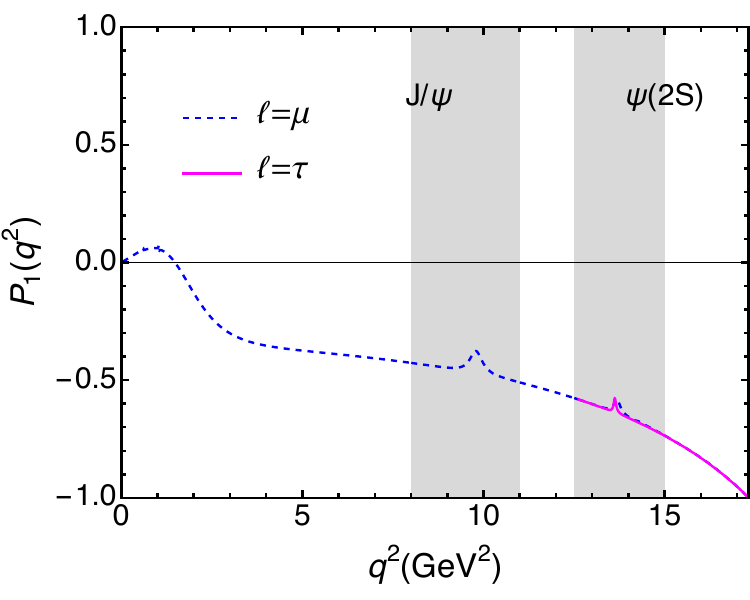}
  \includegraphics[width=42mm]{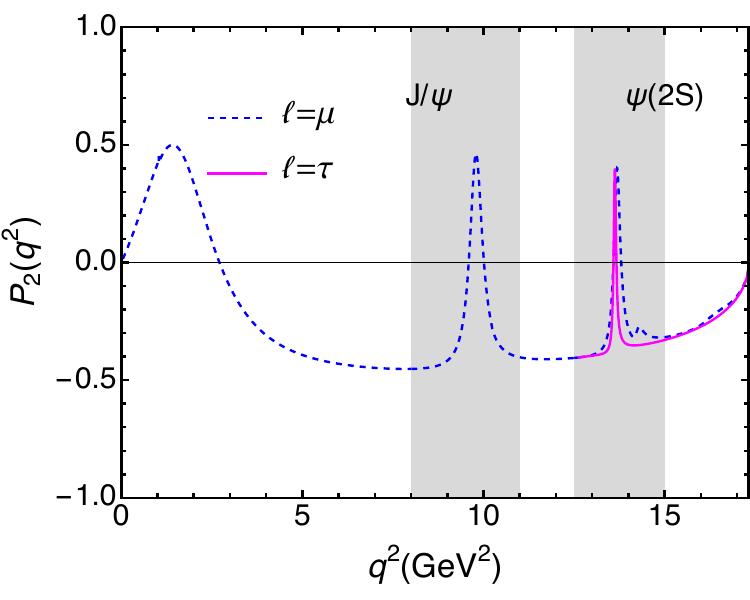}
  \includegraphics[width=42mm]{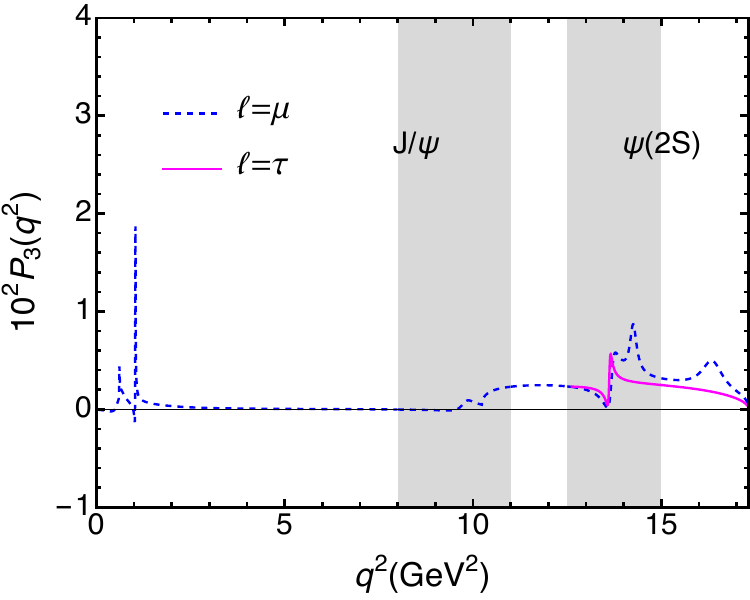}
  \includegraphics[width=42mm]{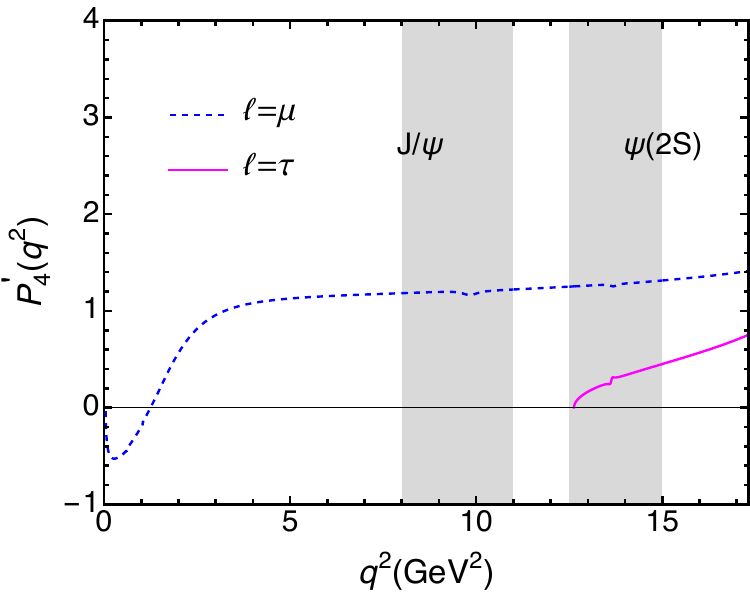}\\
  \includegraphics[width=42mm]{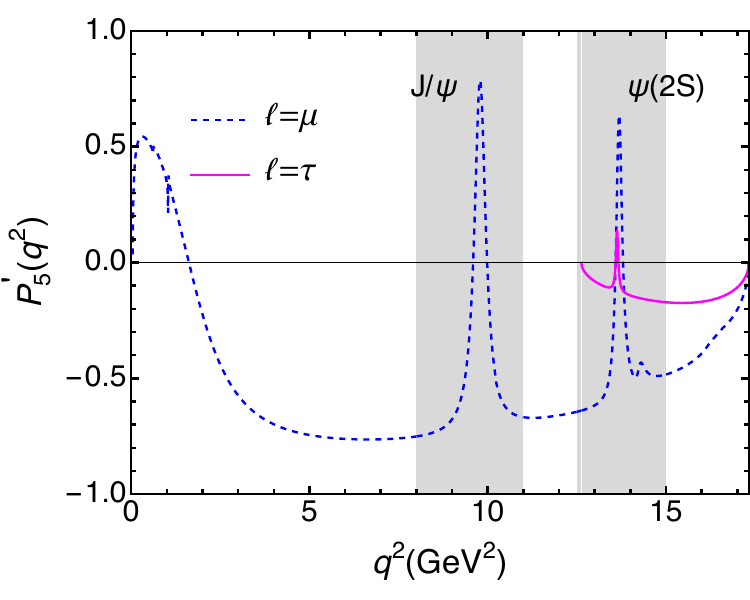}
  \includegraphics[width=42mm]{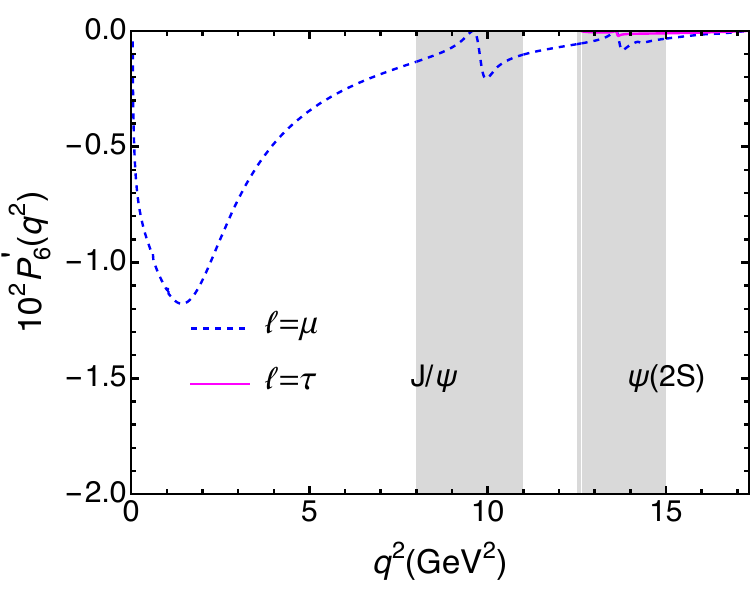}
  \includegraphics[width=42mm]{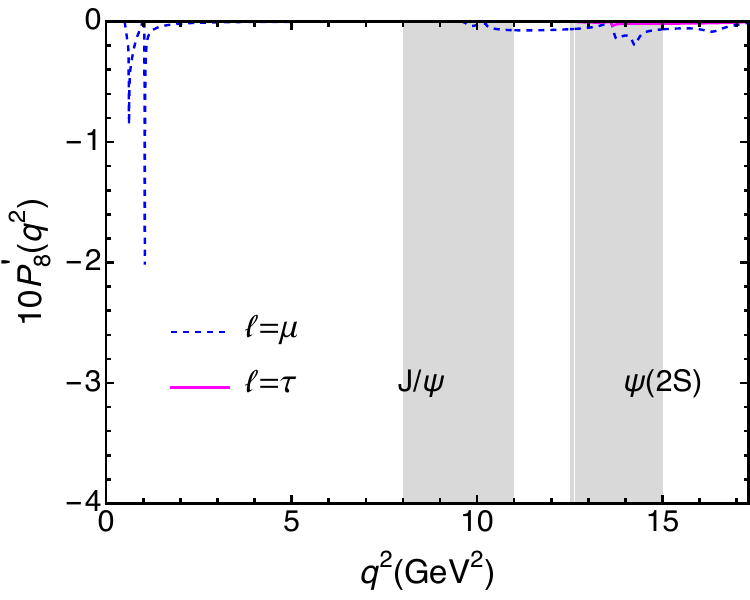}
  \end{tabular}
  \caption{The $q^{2}$ dependence of the clean angular observables $P_{1,2,3}$ and $P_{4,5,6,8}^{\prime}$, where the blue dashed and magenta solid curves represent our results for the $\mu$ and $\tau$ modes, respectively.}
\label{fig:Pi}
\end{figure*}

In addition, we present our results for the $q^{2}$ dependent clean angular observables $P_{1,2,3}$ and $P_{4,5,6,8}^{\prime}$ in Fig.~\ref{fig:Pi}. In Ref.~\cite{LHCb:2013ghj}, the LHCb collaboration reported the measurement of the form-factor-independent observables $P^{\prime}_{4,5,6,8}$ of the $B^{0}\to{K^{*0}}\mu^{+}\mu^{-}$ decay. In particular, in the interval of $4.30<q^{2}<8.68$ $\text{GeV}^{2}$, the observable $P_{5}^{\prime}$ shows $3.7\sigma$ discrepancy with the SM prediction~\cite{Egede:2008uy}. After integration over the energy region $1.0<q^{2}<6.0$ $\text{GeV}^{2}$, the discrepancy is determined to be $2.5\sigma$. So we want to investigate these clean angular observables in the rare semileptonic decay of bottom-charmed meson. In order to exclude the charmonium contributions and make it easy to check  experimentally, we also present the averaged values of these observables in different $q^{2}$ intervals in Table~\ref{tab:binP}. The averaged value in a $q^{2}$ bin is defined by Eq.~\eqref{eq:AveragedA}.

\begin{table*}[htbp]\centering
\caption{The averaged values of the clean angular observables $P_{1,2,3}$ and $P^{\prime}_{4,5,6,8}$ in different $q^{2}$ bins.}
\label{tab:binP}
\renewcommand\arraystretch{1.15}
\begin{tabular*}{162mm}{cccc|cccc}
\toprule[1pt]
\toprule[0.5pt]
$q^{2}$ bins $(\text{GeV}^{2})$   &$\langle{P_{1}(\ell=e)}\rangle$   &$\langle{P_{1}(\ell=\mu)}\rangle$   &$\langle{P_{1}(\ell=\tau)}\rangle$
&$q^{2}$ bins $(\text{GeV}^{2})$   &$\langle{P_{4}^{\prime}(\ell=e)}\rangle$   &$\langle{P_{4}^{\prime}(\ell=\mu)}\rangle$   &$\langle{P_{4}^{\prime}(\ell=\tau)}\rangle$\\
\midrule[0.5pt]
$[1.1, 6.0]$   &$-0.281$   &$-0.281$   &   &$[1.1, 6.0]$  &$0.908$  &$0.898$  &\\
$[6.0, 8.0]$   &$-0.408$   &$-0.408$   &   &$[6.0, 8.0]$  &$1.177$  &$1.169$  &\\
$[11.0, 12.5]$   &$-0.543$   &$-0.543$   &   &$[11.0, 12.5]$  &$1.240$  &$1.236$  &\\
$[15.0, 17.0]$   &$-0.822$   &$-0.822$   &$-0.826$   &$[15.0, 17.0]$  &$1.350$  &$1.347$  &$0.561$\\
\midrule[0.5pt]
$q^{2}$ bins $(\text{GeV}^{2})$   &$\langle{P_{2}(\ell=e)}\rangle$   &$\langle{P_{2}(\ell=\mu)}\rangle$   &$\langle{P_{2}(\ell=\tau)}\rangle$
&$q^{2}$ bins $(\text{GeV}^{2})$   &$\langle{P_{5}^{\prime}(\ell=e)}\rangle$   &$\langle{P_{5}^{\prime}(\ell=\mu)}\rangle$   &$\langle{P_{5}^{\prime}(\ell=\tau)}\rangle$\\
\midrule[0.5pt]
$[1.1, 6.0]$   &$-0.125$   &$-0.125$   &   &$[1.1, 6.0]$  &$-0.540$  &$-0.534$  &\\
$[6.0, 8.0]$   &$-0.446$   &$-0.446$   &   &$[6.0, 8.0]$  &$-0.766$  &$-0.761$  &\\
$[11.0, 12.5]$   &$-0.409$   &$-0.409$   &   &$[11.0, 12.5]$  &$-0.664$  &$-0.662$  &\\
$[15.0, 17.0]$   &$-0.262$   &$-0.262$   &$-0.269$   &$[15.0, 17.0]$  &$-0.390$  &$-0.390$  &$-0.163$\\
\midrule[0.5pt]
$q^{2}$ bins $(\text{GeV}^{2})$   &$10^{3}\langle{P_{3}(\ell=e)}\rangle$   &$10^{3}\langle{P_{3}(\ell=\mu)}\rangle$   &$10^{3}\langle{P_{3}(\ell=\tau)}\rangle$
&$q^{2}$ bins $(\text{GeV}^{2})$   &$10^{3}\langle{P_{6}^{\prime}(\ell=e)}\rangle$   &$10^{3}\langle{P_{6}^{\prime}(\ell=\mu)}\rangle$   &$10^{3}\langle{P_{6}^{\prime}(\ell=\tau)}\rangle$\\
\midrule[0.5pt]
$[1.1, 6.0]$   &$0.184$   &$0.184$   &   &$[1.1, 6.0]$  &$-6.440$  &$-6.320$  &\\
$[6.0, 8.0]$   &$0.007$   &$0.007$   &   &$[6.0, 8.0]$  &$-1.868$  &$-1.856$  &\\
$[11.0, 12.5]$   &$2.421$   &$2.421$   &   &$[11.0, 12.5]$  &$-0.771$  &$-0.769$  &\\
$[15.0, 17.0]$   &$3.442$   &$3.442$   &$2.008$   &$[15.0, 17.0]$  &$-0.188$  &$-0.188$  &$-0.070$\\
\midrule[0.5pt]
   &    &    &
&$q^{2}$ bins $(\text{GeV}^{2})$   &$10^{3}\langle{P_{8}^{\prime}(\ell=e)}\rangle$   &$10^{3}\langle{P_{8}^{\prime}(\ell=\mu)}\rangle$   &$10^{3}\langle{P_{8}^{\prime}(\ell=\tau)}\rangle$\\
\midrule[0.5pt]
                     &   &   &   &$[1.1, 6.0]$  &$-1.196$  &$-1.164$  &\\
                     &   &   &   &$[6.0, 8.0]$  &$-0.029$  &$-0.029$  &\\
                     &   &   &   &$[11.0, 12.5]$  &$-7.054$  &$-7.030$  &\\
                     &   &   &   &$[15.0, 17.0]$  &$-6.147$  &$-6.137$  &$-1.467$\\
\bottomrule[0.5pt]
\bottomrule[1pt]
\end{tabular*}
\end{table*}

In general, this quasi-four-body decay provides a set of  physical observables to study the corresponding weak interaction, and in particular the ratios of the branching fractions $R^{e\mu}$ and $R^{\tau\mu}$, as well as the clean angular coefficients $P_{i}$ and $P_{j}^{\prime}$, can be helpful to search for the NP effects beyond the SM. We call for the ongoing LHCb experiment to search for this process and to measure the corresponding physical observables.

\section{Summary}\label{sec05}

In this work, we have studied the $B_{c}\to{D_{s}^{*}}$ transition form factors deduced by the (axial)vector and (pseudo)tensor currents, and, in the future, investigate the angular distributions of the quasi-four-body processes $B_{c}\to{D}_{s}^{*}(\to{D}_{s}\pi)\ell^{+}\ell^{-}$ ($\ell$=$e$, $\mu$, $\tau$).

To describe the weak process, the relevant seven independent form factors are calculated by utilizing the covariant LFQM approach. The concerned meson wave functions are adopted as the numerical wave functions, which are extracted from the solution of the modified GI model. This treatment avoids the $\beta$ dependence and thus reduces the corresponding uncertainty. Our results of form factors are compared with other approaches. In particular, for the (pseudo)tensor currents deduced form factors $T_{1,2,3}(q^{2}=0)$, our results agree with the pQCD prediction. More theoretical works, especially the LQCD and QCD sum rule (or light-cone sum rule) calculation, are highly appreciated to test our result and to refine the corresponding topic.

With the obtained form factors, the rare semileptonic decays $B_{c}\to{D}_{s}^{*}(\to{D}_{s}\pi)\ell^{+}\ell^{-}$ are studied. Not only the branching fractions, the lepton-side forward-backward asymmetry parameter $A_{FB}$, and the longitudinal and transverse polarization fractions $F_{L}$ and $F_{T}$, but also the angular coefficients $S_{i}$ and $A_{i}$ are investigated. Numerically, the concerned cascade decays with $e$ or $\mu$ final states are around $10^{-8}$, which need to be tested by other approaches and ongoing experiments. {Moreover, the $A_{FB}$ and $F_{L(T)}$ are important physical observables, and they are also feasible observables in the future LHCb experiment, so we look forward to the experimental results.} In addition, the ratios of the branching fractions are also calculated to validate whether or not the LFU violated. Furthermore, the clean coefficient observables $P_{i}$ and $P_{j}^{\prime}$ are presented, which reduce the uncertainty from the form factors and can be a possible signal to search for the NP effects. {Since these observations are largely free of form factor uncertainties in the large-recoiled limit, and are feasible to measure  experimentally, we strongly encourage our experimental colleagues to measure them.}

Overall, in this work we have  systematically studied the angular distribution of $B_{c}\to{D_{s}^{*}}(\to{D_{s}}\pi)\ell^{+}\ell^{-}$ ($\ell$= $e$, $\mu$, $\tau$) with the form factors obtained by the covariant LFQM. We live in the hope that with the completion of the LHCb experiment prepared for the run 3 and run 4 of the LHC and the improvement of the experimental capabilities, this rare semileptonic process can be discovered and we expect that the predicted physical observables can be tested.

\appendix

\section{The weak transition matrix elements deduced by axial-vector and (pseudo)tensor currents}
\label{app01}

\setcounter{equation}{0}
\renewcommand{\theequation}{A.\arabic{equation}}

In this Appendix, we present the concerned expressions of the weak transition matrix elements deduced by axial-vector current , and (pseudo)tensor currents. The expression of the axial-vector current matrix element is
\begin{widetext}
\begin{equation}
\begin{split}
S_{\mu\nu}^{A}=&\text{Tr}\Big{[}\Big{(}\gamma_{\nu}-\frac{(p_{1}^{\prime\prime}-p_{2})_{\nu}}{W_{V}^{\prime\prime}}\Big{)}(\slashed{p}_{1}^{\prime\prime}+m_{1}^{\prime\prime})\gamma_{\mu}\gamma_{5}(\slashed{p}_{1}^{\prime}+m_{1}^{\prime})\gamma_{5}(-\slashed{p}_{2}+m_{2})\Big{]}\\
=&-2g_{\mu\nu}\bigg{[}m_{2}(q^{2}-N_{1}^{\prime}-N_{1}^{\prime\prime}-m_{1}^{\prime2}-m_{1}^{\prime\prime2})
-m_{1}^{\prime}(M^{\prime\prime2}-N_{1}^{\prime\prime}-N_{2}-m_{1}^{\prime\prime2}-m_{2}^{2})\\
&-m_{1}^{\prime\prime}(M^{\prime2}-N_{1}^{\prime}-N_{2}-m_{1}^{\prime2}-m_{2}^{2})-2m_{1}^{\prime}m_{1}^{\prime\prime}m_{2}\bigg{]}
-8p_{1\mu}^{\prime}p_{1\nu}^{\prime}(m_{2}-m_{1}^{\prime})\\
&+2m_{1}^{\prime}(P_{\mu}q_{\nu}+P_{\nu}q_{\mu}+2q_{\mu}q_{\nu})-2p_{1\mu}^{\prime}P_{\nu}(m_{1}^{\prime}-m_{1}^{\prime\prime})-2p_{1\nu}^{\prime}P_{\mu}(m_{1}^{\prime}+m_{1}^{\prime\prime})\\
&-2p_{1\mu}^{\prime}q_{\nu}(3m_{1}^{\prime}-m_{1}^{\prime\prime}-2m_{2})-2p_{1\nu}^{\prime}q_{\mu}(3m_{1}^{\prime}+m_{1}^{\prime\prime}-2m_{2})\\
&-\frac{1}{2W_{V}^{\prime\prime}}\bigg{[}2p_{1\mu}^{\prime}(M^{\prime2}+M^{\prime\prime2}-q^{2}-2N_{2}+2(m_{1}^{\prime}-m_{2})(m_{1}^{\prime\prime}+m_{2}))\\
&+q_{\mu}(q^{2}-2M^{\prime2}+N_{1}^{\prime}-N_{1}^{\prime\prime}+2N_{2}-(m_{1}+m_{1}^{\prime\prime})^{2}+2(m_{1}^{\prime}-m_{2})^{2})\\
&+P_{\mu}(q^{2}-N_{1}^{\prime}-N_{1}^{\prime\prime}-(m_{1}^{\prime}+m_{1}^{\prime\prime})^{2})\bigg{]}(4p_{1\nu}^{\prime}-3q_{\nu}-P_{\nu}),
\label{eq:SA}
\end{split}
\end{equation}
and the expression of the (pseudo)tensor current matrix element is
\begin{equation}
S_{\mu\nu}^{T+T5}=\text{Tr}\Big{[}\Big{(}\gamma_{\nu}-\frac{(p_{1}^{\prime\prime}-p_{2})_{\nu}}{W_{V}^{\prime\prime}}\Big{)}
(\slashed{p}_{1}^{\prime\prime}+m_{1}^{\prime\prime})i\sigma_{\mu\delta}(1+\gamma_{5})q^{\delta}(\slashed{p}_{1}^{\prime}+m_{1}^{\prime})\gamma_{5}(-\slashed{p}_{2}+m_{2})\Big{]}.
\label{eq:ATT5}
\end{equation}
By using the identity $2\sigma_{\mu\delta}\gamma_{5}=-i\epsilon_{\mu\delta\alpha\beta}\sigma^{\alpha\beta}$, the matrix element $S_{\mu\nu}^{T+T5}$ can be decomposed into
\begin{equation}
S_{\mu\nu}^{T+T5}=iq^{\delta}S_{\mu\nu\delta}+\frac{1}{2}\epsilon_{\mu\delta\alpha\beta}q^{\delta}S_{\nu}^{\alpha\beta},
\end{equation}
where $iq^{\delta}S_{\mu\nu\delta}$ and $\frac{1}{2}\epsilon_{\mu\delta\alpha\beta}q^{\delta}S_{\nu}^{\alpha\beta}$ are expressed as
\begin{equation}
\begin{split}
iq^{\delta}S_{\mu\nu\delta}=&\text{Tr}\Big{[}\Big{(}\gamma_{\nu}-\frac{(p_{1}^{\prime\prime}-p_{2})_{\nu}}{W_{V}^{\prime\prime}}\Big{)}
(\slashed{p}_{1}^{\prime\prime}+m_{1}^{\prime\prime})i\sigma_{\mu\delta}q^{\delta}(\slashed{p}_{1}^{\prime}+m_{1}^{\prime})\gamma_{5}(-\slashed{p}_{2}+m_{2})\Big{]}\\
=&i\epsilon_{\mu\nu\alpha\beta}P^{\alpha}p_{1}^{\beta}\big{(}m_{1}^{\prime2}-m_{1}^{\prime\prime2}+N_{1}^{\prime}-N_{1}^{\prime\prime}\big{)}
-\frac{i}{2}\epsilon_{\mu\nu\alpha\beta}P^{\alpha}q^{\beta}\big{(}m_{1}^{\prime2}+4m_{1}^{\prime}m_{1}^{\prime\prime}-m_{1}^{\prime\prime2}+N_{1}^{\prime}-N_{1}^{\prime\prime}+q^{2}\big{)}\\
&-i\epsilon_{\mu\nu\alpha\beta}p_{1}^{\alpha}q^{\beta}\big{(}M^{\prime2}-m_{1}^{\prime2}+4m_{2}(m_{1}^{\prime}+m_{1}^{\prime\prime})-4m_{1}^{\prime}m_{1}^{\prime\prime}-m_{1}^{\prime\prime2}-2m_{2}^{2}+M^{\prime\prime2}-N_{1}^{\prime}-N_{1}^{\prime\prime}-2N_{2}-q^{2}\big{)}\\
&+i\epsilon_{\mu\alpha\beta\gamma}P^{\alpha}p_{1}^{\beta}q^{\gamma}P_{\nu}\big{(}\frac{m_{1}^{\prime}+m_{1}^{\prime\prime}}{W_{V}^{\prime\prime}}\big{)}
+i\epsilon_{\mu\alpha\beta\gamma}P^{\alpha}p_{1}^{\beta}q^{\gamma}p_{1\nu}\bigg{(}2-\frac{4(m_{1}^{\prime}+m_{1}^{\prime\prime})}{W_{V}^{\prime\prime}}\bigg{)}
+i\epsilon_{\mu\alpha\beta\gamma}P^{\alpha}p_{1}^{\beta}q^{\gamma}q_{\nu}\bigg{(}\frac{3(m_{1}^{\prime}+m_{1}^{\prime\prime})}{W_{V}^{\prime\prime}}-1\bigg{)}\\
&-i\epsilon_{\nu\alpha\beta\gamma}P^{\alpha}p_{1}^{\beta}q^{\gamma}q_{\mu}
+2i\epsilon_{\nu\alpha\beta\gamma}P^{\alpha}p_{1}^{\beta}q^{\gamma}p_{1\mu},
\label{eq:ST1}
\end{split}
\end{equation}
\begin{equation}
\begin{split}
\frac{1}{2}\epsilon_{\mu\delta\alpha\beta}q^{\delta}S_{\nu}^{\alpha\beta}=&\text{Tr}\Big{[}\Big{(}\gamma_{\nu}-\frac{(p_{1}^{\prime\prime}-p_{2})_{\nu}}{W_{V}^{\prime\prime}}\Big{)}
(\slashed{p}_{1}^{\prime\prime}+m_{1}^{\prime\prime})\frac{1}{2}\sigma^{\alpha\beta}\epsilon_{\mu\delta\alpha\beta}q^{\delta}(\slashed{p}_{1}^{\prime}+m_{1}^{\prime})\gamma_{5}(-\slashed{p}_{2}+m_{2})\Big{]}\\
=&-2ig_{\mu\nu}\big{\{}M^{\prime2}[m_{1}^{\prime\prime}(m_{1}^{\prime}-m_{1}^{\prime\prime})-N_{1}^{\prime\prime}]+m_{1}^{\prime3}(m_{2}-m_{1}^{\prime\prime})+m_{1}^{\prime2}(m_{2}(m_{1}^{\prime\prime}-m_{2})+M^{\prime\prime2}-N_{2})\\
&+m_{1}^{\prime}(m_{1}^{\prime\prime3}-m_{1}^{\prime\prime2}m_{2}-m_{1}^{\prime\prime}(M^{\prime\prime2}+N_{1}^{\prime}-N_{1}^{\prime\prime})+m_{2}(N_{1}^{\prime}-N_{1}^{\prime\prime}-q^{2}))-m_{1}^{\prime\prime3}m_{2}+m_{1}^{\prime\prime2}m_{2}^{2}\\
&+m_{1}^{\prime\prime2}N_{2}+m_{1}^{\prime\prime}m_{2}N_{1}^{\prime}-m_{1}^{\prime\prime}m_{2}N_{1}^{\prime\prime}+m_{1}^{\prime\prime}m_{2}q^{2}-m_{2}^{2}N_{1}^{\prime}+m_{2}^{2}N_{1}^{\prime\prime}+M^{\prime\prime2}N_{1}^{\prime}-N_{1}^{\prime}N_{2}+N_{1}^{\prime\prime}N_{2}\big{\}}\\
&+iP_{\mu}P_{\nu}\Big{\{}\frac{(m_{1}^{\prime}+m_{1}^{\prime\prime})(m_{1}^{\prime2}-m_{1}^{\prime\prime2}+N_{1}^{\prime}-N_{1}^{\prime\prime})+q^{2}(m_{1}^{\prime\prime}-m_{1}^{\prime})}{2W_{V}^{\prime\prime}}\Big{\}}\\
&+iq_{\mu}q_{\nu}\Big{\{}-2M^{\prime2}-m_{1}^{\prime2}+2m_{1}^{\prime}m_{1}^{\prime\prime}-4m_{1}^{\prime}m_{2}+m_{1}^{\prime\prime2}+2m_{2}^{2}-N_{1}^{\prime}+N_{1}^{\prime\prime}+2N_{2}-q^{2}\\
&+\frac{3}{2W_{V}^{\prime\prime}}\Big{[}2M^{\prime2}m_{1}^{\prime}-m_{1}^{\prime3}+m_{1}^{\prime2}(m_{1}^{\prime\prime}+2m_{2})+m_{1}^{\prime}(m_{1}^{\prime\prime2}-2M^{\prime\prime2}-N_{1}^{\prime}+N_{1}^{\prime\prime}-q^{2})\\
&-(m_{1}^{\prime\prime}+2m_{2})(m_{1}^{\prime\prime2}-N_{1}^{\prime}+N_{1}^{\prime\prime}-q^{2})\Big{]}\Big{\}}\\
&+ip_{1\mu}p_{1\nu}\Big{\{}-4M^{\prime2}+4M^{\prime\prime2}-4q^{2}+\frac{4}{W_{V}^{\prime\prime}}\Big{[}(M^{\prime2}-M^{\prime\prime2})(m_{1}^{\prime}+m_{1}^{\prime\prime})+q^{2}(-m_{1}^{\prime}+m_{1}^{\prime\prime}+2m_{2})\Big{]}\Big{\}}\\
&+iP_{\mu}p_{1\nu}\Big{\{}2m_{1}^{\prime2}-2m_{1}^{\prime\prime2}+2N_{1}^{\prime}-2N_{1}^{\prime\prime}+\frac{2}{W_{V}^{\prime\prime}}\Big{[}q^{2}(m_{1}^{\prime}-m_{1}^{\prime\prime})-(m_{1}^{\prime}+m_{1}^{\prime\prime})(m_{1}^{\prime2}-m_{1}^{\prime\prime2}+N_{1}^{\prime}-N_{1}^{\prime\prime})\Big{]}\Big{\}}\\
&+iP_{\nu}p_{1\mu}\Big{\{}2q^{2}+\frac{1}{W_{V}^{\prime\prime}}\Big{[}q^{2}(m_{1}^{\prime}-m_{1}^{\prime\prime}-2m_{2})-(M^{\prime2}-M^{\prime\prime2})(m_{1}^{\prime}+m_{1}^{\prime\prime})\Big{]}\Big{\}}\\
&+iP_{\mu}q_{\nu}\Big{\{}-2m_{1}^{\prime2}+2m_{1}^{\prime}m_{1}^{\prime\prime}-2N_{1}^{\prime}+\frac{3}{2W_{V}^{\prime\prime}}\Big{[}(m_{1}^{\prime}+m_{1}^{\prime\prime})(m_{1}^{\prime2}-m_{1}^{\prime\prime2}+N_{1}^{\prime}-N_{1}^{\prime\prime})+q^{2}(m_{1}^{\prime\prime}-m_{1}^{\prime})\Big{]}\Big{\}}\\
&+iP_{\nu}q_{\mu}\Big{\{}-m_{1}^{\prime2}+m_{1}^{\prime\prime2}-N_{1}^{\prime}+N_{1}^{\prime\prime}-q^{2}+\frac{1}{2W_{V}^{\prime\prime}}\Big{[}2M^{\prime2}m_{1}^{\prime}-m_{1}^{\prime3}+m_{1}^{\prime2}(m_{1}^{\prime\prime}+2m_{2})\\
&+m_{1}^{\prime}(m_{1}^{\prime\prime2}-2M^{\prime\prime2}-N_{1}^{\prime}+N_{1}^{\prime\prime}-q^{2})-(m_{1}^{\prime\prime}+2m_{2})(m_{1}^{\prime\prime2}-N_{1}^{\prime}+N_{1}^{\prime\prime}-q^{2})\Big{]}\Big{\}}\\
&+ip_{1\mu}q_{\nu}\Big{\{}4M^{\prime2}-4m_{1}^{\prime}m_{1}^{\prime\prime}+4m_{1}^{\prime}m_{2}+4m_{1}^{\prime\prime}m_{2}-4m_{2}^{2}-4N_{2}+2q^{2}\\
&+\frac{3}{W_{V}^{\prime\prime}}\Big{[}q^{2}(m_{1}^{\prime}-m_{1}^{\prime\prime}-2m_{2})-(M^{\prime2}-M^{\prime\prime2})(m_{1}^{\prime}+m_{1}^{\prime\prime})\Big{]}\Big{\}}\\
&+ip_{1\nu}q_{\mu}\Big{\{}2M^{\prime2}+2m_{1}^{\prime2}-2m_{1}^{\prime\prime2}-2M^{\prime\prime2}+2N_{1}^{\prime}-2N_{1}^{\prime\prime}+2q^{2}
-\frac{2}{W_{V}^{\prime\prime}}\Big{[}2M^{\prime2}m_{1}^{\prime}-m_{1}^{\prime3}+m_{1}^{\prime2}(m_{1}^{\prime\prime}+2m_{2})\\
&+m_{1}^{\prime}(m_{1}^{\prime\prime2}-2M^{\prime\prime2}-N_{1}^{\prime}+N_{1}^{\prime\prime}-q^{2})-(m_{1}^{\prime\prime}+2m_{2})(m_{1}^{\prime\prime2}-N_{1}^{\prime}+N_{1}^{\prime\prime}-q^{2})\Big{]}\Big{\}},
\label{eq:ST2}
\end{split}
\end{equation}
respectively.
\end{widetext}

\section*{ACKNOWLEDGMENTS}

This work is supported by the China National Funds for Distinguished Young Scientists under Grant No. 11825503, the National Key Research and Development Program of China under Contract No. 2020YFA0406400, the 111 Project under Grant No. B20063, the National Natural Science Foundation of China under Grant No. 12247101 and No. 12335001, the fundamental Research Funds for the Central Universities.

\end{document}